\documentclass[journal,twoside,web]{IEEEtran}

\usepackage{cite}
\usepackage{amsmath,amssymb,amsfonts}
\usepackage{algorithmic}
\usepackage{graphicx}
\usepackage{textcomp}
\usepackage{xcolor}
\usepackage{subfig}
\usepackage{tabularx}
\usepackage{array}
\usepackage{booktabs}
\usepackage{hyperref}
\usepackage{amsmath}
\usepackage[flushleft]{threeparttable}

\usepackage{bm}
\usepackage{xspace}
\newcommand{\bmath}[1]{\ensuremath{\bm{#1}}\xspace}

\newcommand{\x}{\bmath{x}}
\newcommand{\y}{\bmath{y}}
\newcommand{\z}{\bmath{z}}
\newcommand{\e}{\bmath{e}}
\newcommand{\f}{\bmath{f}}
\newcommand{\g}{\bmath{g}}

\newcommand{\lv}{\bmath{\ell}}

\newcommand{\n}{\bmath{n}}

\newcommand{\uv}{\bmath{u}}

\newcommand{\pv}{\bmath{p}}
\newcommand{\rv}{\bmath{r}}

\newcommand{\0}{\bmath{0}}

\newcommand{\alp}{\bmath{\alpha}}

\newcommand{\tht}{\bmath{\theta}}
\newcommand{\lam}{\bmath{\lambda}}
\newcommand{\ome}{\bmath{\omega}}

\newcommand{\ro}{\bmath{\rho}}

\newcommand{\muv}{\bmath{\mu}}
\newcommand{\etav}{\bmath{\eta}}

\newcommand{\phiv}{\bmath{\phi}}

\newcommand{\A}{\bmath{A}}

\newcommand{\G}{\bmath{G}}
\newcommand{\I}{\bmath{I}}
\newcommand{\K}{\bmath{K}}

\newcommand{\R}{\bmath{R}}

\newcommand{\U}{\bmath{U}}

\newcommand{\Nc}{\mathcal{N}}

\newcommand{\beq}{\begin{equation}}
\newcommand{\eeq}{\end{equation}}
\newcommand{\bea}{\begin{eqnarray}}
\newcommand{\eea}{\end{eqnarray}}
\newcommand{\ba}{\left(\!\!\begin{array}}
\newcommand{\ea}{\end{array}\!\!\right)}
\newcommand{\bc}{\begin{center}}
\newcommand{\ec}{\end{center}}

\newcommand{\diag}{\mathrm{diag}}

\def\BibTeX{{\rm B\kern-.05em{\sc i\kern-.025em b}\kern-.08em
		T\kern-.1667em\lower.7ex\hbox{E}\kern-.125emX}}

\newcommand{\txtc}[1]{\textcolor{black}{#1}}

\newcommand{\txty}[1]{\textcolor{black}{#1}}
\newcommand{\psiv}{\boldsymbol{\psi}}
\newcommand{\txtr}[1]{\textcolor{black}{#1}}

\begin{document}
	\title{{Single-Subject Deep-Learning Image Reconstruction with a Neural Optimization Transfer Algorithm for PET-enabled Dual-Energy CT Imaging}}
	\author{Siqi Li, Yansong Zhu, Benjamin A. Spencer and Guobao Wang
		\thanks{This work was supported in part by National Institutes of Health (NIH) under grant no. R21 EB027346. Part of this work was presented at the 2021 SPIE Medical Imaging Conference.}
		\thanks{S. Q. Li , Y. S. Zhu, and G. B. Wang are with the Department of Radiology, University of California Davis Health, Sacramento, CA 95817, USA. (e-mail: sqlli@ucdavis.edu, yszhu@ucdavis.edu, gbwang@ucdavis.edu).}
		\thanks{B. A. Spencer is with the Department of Biomedical Engineering, University of California at Davis, Davis, CA 95616, USA. (e-mail: benspencer@ucdavis.edu}}

\maketitle
\begin{abstract}
	Combining dual-energy computed tomography (DECT) with positron emission tomography (PET) offers many potential clinical applications but typically requires expensive hardware upgrades or increases radiation doses on PET/CT scanners due to an extra X-ray CT scan. The recent PET-enabled DECT method allows DECT imaging on PET/CT without requiring a second X-ray CT scan. It combines the already existing X-ray CT image with a 511 keV $\gamma$-ray CT (gCT) image reconstructed from time-of-flight PET emission data. A kernelized framework has been developed for reconstructing gCT image but this method has not fully exploited the potential of prior knowledge. Use of deep neural networks may explore the power of deep learning in this application. However, common approaches require a large database for training, which is impractical for a new imaging method like \emph{PET-enabled DECT}. Here, we propose a single-subject method by using neural-network representation as a deep coefficient prior to improving gCT image reconstruction without population-based pre-training. The resulting optimization problem becomes the tomographic estimation of nonlinear neural-network parameters from gCT projection data. This complicated problem can be efficiently solved by utilizing the optimization transfer strategy with quadratic surrogates. Each iteration of the proposed neural optimization transfer algorithm includes: PET activity image update; gCT image update; and least-square neural-network learning in the gCT image domain. This algorithm is guaranteed to monotonically increase the data likelihood. Results from computer simulation, real phantom data \txtr{and real patient data} have demonstrated that the proposed method can significantly improve gCT image quality and consequent multi-material decomposition as compared to other methods.
\end{abstract}
\begin{IEEEkeywords}
	PET-enabled dual-energy CT, PET/CT, image reconstruction, kernel methods, convolutional neural-network
\end{IEEEkeywords}

\section{Introduction}

\IEEEPARstart{C}{onventionally}, standard dual-energy computed tomography (DECT) uses two different X-ray energies to obtain energy-dependent tissue attenuation information to allow quantitative material decomposition \cite{McCollough2015}. 
Integration of DECT with positron emission tomography (PET) offers more accurate attenuation correction for PET \cite{Kinahan2006Dual,Xia2013Dual}, enables multi-modality characterization of disease states in cancer and other diseases \cite{Wu2020} and would open up novel clinical applications. However, it is not trivial to combine DECT with PET because it either requires costly CT hardware upgrade on existing PET/CT or significantly increases CT radiation dose due to the need for the second X-ray CT scan.

A PET-enabled DECT method has been proposed to enable DECT imaging on clinical time-of-flight PET/CT scanners without a change of scanner hardware or adding additional radiation dose or scan time \cite{Wang2020}. In PET-enabled DECT imaging, a high-energy ``$\gamma$-ray CT (gCT)" image at 511 keV is reconstructed from a standard time-of-flight PET emission scan and combined with the already-existing low-energy X-ray CT (usually $<$ 140 keV) to produce a pair of DECT images for multi-material decomposition. 

The gCT image can be reconstructed from PET emission data using the maximum-likelihood attenuation and activity (MLAA) method \cite{Rezaei2012, Nuyts1999}. However, standard MLAA reconstruction can be very noisy due to the limited counting statistics of PET data. Regularization-based MLAA methods \cite{Nuyts1999, Mehranian2015, Mehranian2017MR} can suppress noise, but generally require a more complex optimization algorithm, involve one or more hard-to-tune penalty parameters, and need to run for many iterations for a convergent solution. The kernel MLAA method, or KAA in short in this paper, has been developed by integrating the X-ray CT image prior into the forward model of MLAA attenuation image reconstruction through a kernel framework \cite{Wang2020}. KAA has demonstrated substantial improvements over the MLAA for PET-enabled DECT imaging. Nonetheless the estimated kernel coefficient image may still suffer from noise and result in artifacts in material decomposition, such as in low-count scan cases. The aim of this paper is to improve gCT image reconstruction algorithm by exploring the power of deep neural networks.

For general CT image reconstruction, common ways to explore deep learning include direct end-to-end mapping or unrolled model-based deep-learning reconstruction \cite{McLeavy2021, Wang2020Deep}. However, all these approaches require pre-training using a large population-based database, which is not applicable for a new imaging method like the \emph{PET-enabled DECT} because no existing database of 511 keV gCT images is available yet. An alternative way is the single-subject learning approach that explores neural network representation, for example, in a way similar to the deep image prior (DIP) framework \cite{Ulyanov2020}. Such methods \cite{Gong2019,Gong2019a, Ote2023, Xie2020, Baguer2020, Shu2022} do not require population-based pre-training but is based on the data of single subjects. Direct application of DIP reconstruction to standard MLAA, however, would result in oversmoothing and subsequently induce artifacts in material decomposition, as will be demonstrated later in the evaluation studies.

In this paper, we integrate deep image prior with the KAA framework to develop a neural KAA approach for gCT image reconstruction. This leads to the tomographic estimation of nonlinear neural network parameters from projection data. Different from a possible DIP reconstruction method that would be directly using neural networks for gCT image representation, our method employs the DIP model to represent the kernel coefficient image within the KAA framework, enforcing an implicit regularization on gCT image reconstruction to suppress high noise while preserve image contrast at the same time. 

For all DIP-type CT reconstructions, including the proposed neural KAA for gCT, the optimization problem becomes more challenging due to the coupling of unknown neural-network parameters in the projection domain. One solution involves employing a class of gradient descent algorithms (e.g., \cite{Akcakaya2022Unsupervised, Baguer2020}) by violently computing the derivatives of the projection-based likelihood function with respect to neural-network parameters. However, the computational efficiency is low due to the forward and backward processes containing a large-size system matrix. It is also challenging to select an appropriate step size for algorithm convergence due to the high computational cost. 
Alternatively the alternating direction method of multipliers (ADMM) algorithm can be used \cite{Gong2019, Barutcu2021Limited, Zhao2024J} and may decouple the neural network learning step from the tomographic reconstruction step. However, ADMM often involves one or more hyper-parameters that are known difficult to tune \cite{Zhang2020Review, Ben2020Deep}. 

To overcome these issues, we propose an iterative algorithm for the proposed neural KAA reconstruction problem using the theory of optimization transfer with quadratic surrogates \cite{Lange2000}. The resulting neural optimization transfer algorithm can decouple the non-linear neural-network learning from the tomographic reconstruction step without introducing hyper-parameters as compared to ADMM. The proposed algorithm is easy and efficient to implement in practice by using existing deep learning libraries. 
Specifically, it essentially deals with a $\gamma$-ray CT transmission reconstruction problem, resulting in the derivation of a quadratic surrogate function for optimization transfer and consequently a unique least-square formulation for neural network learning (see Fig. \ref{KAA vs Proposed} and Eq. (\ref{wMSE})). The proposed neural optimization transfer algorithm is expected to solve different DIP-type reconstruction problems for X-ray CT \cite{Baguer2020, Barutcu2021Limited}, magnetic resonance imaging \cite{Yoo2021, Zhao2024J}, and other imaging modalities \cite{Zhou2020Diffraction, Vu2021Deep} that employ a least-square reconstruction formula. Thus the algorithmic contribution from this work may have a broad impact in single-subject deep-learning reconstruction for tomographic imaging.

Part of this work was presented at the 2021 SPIE medical imaging conference \cite{Li2021a}. Here we gave the detailed derivation of the proposed \emph{neural optimization transfer} algorithm and theoretically proved its convergence. Besides, we tested the proposed method on more rigorous computer simulation study and implemented PET-enabled DECT imaging on uEXPLORER PET/CT scanner using a real phantom scan and \txtr{a patient scan}. The remaining of this paper is organized as follows. Section II introduces the background materials regarding PET-enabled dual-energy CT. Section III describes the proposed neural KAA method for gCT image reconstruction from time-of-flight PET data. The proposed neural optimization transfer algorithm is elaborated in Section IV. We then present a computer simulation study in Section V, a real phantom study in Section VI and \txtr{a real patient study in Section VII} to demonstrate the improvement of the proposed method. Finally, discussions and conclusions are drawn in Sections VIII and IX, respectively. 

\begin{figure}[h]
	\centering
	\includegraphics[trim=0cm 0cm 0cm 0cm, clip,width=2.5in]{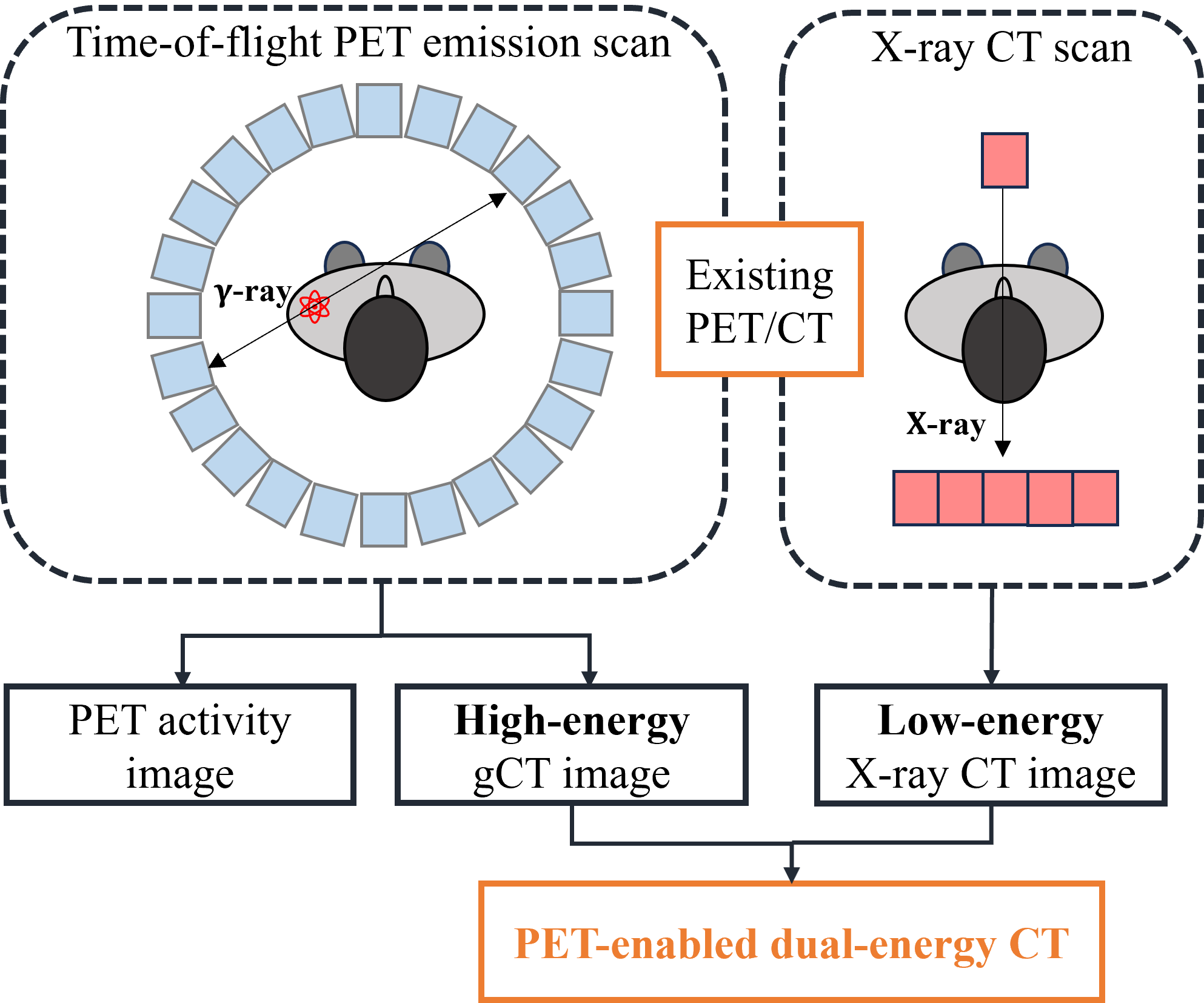}
	\caption{Illustrated concept of PET-enabled dual-energy CT imaging.}
	\label{PDECT}	
\end{figure}

\section{PET-Enabled Dual-Energy CT}
The PET-enabled DECT method merges the high-energy (511 keV) gCT image, reconstructed from time-of-flight PET emission data, with a low-energy X-ray CT image for dual-energy imaging \cite{Wang2020}, as illustrated in Fig. \ref{PDECT}. The gCT image is not acquired using an external radiation source but the internal $\gamma$-rays generated by annihilation radiation of PET radiotracer decays in a subject. In this work, we aim to propose a single-subject deep learning framework to improve gCT image reconstruction without the need of population-based pre-training. The unique advantage is that the proposed framework is more adaptive for this new imaging modality that is difficult for collecting a large training dataset.
\subsection{Statistical Model of PET Emission Data}
Commonly the PET measurement $\y$ is well modeled as independent Poisson random variables which follows the log-likelihood function:
\beq
L(\y|\lam, \muv) = \sum_{i=1}^{N_d}\sum_{m=1}^{N_t}y_{i,m}\log\overline{y}_{i,m}(\lam,\muv)-\overline{y}_{i,m}(\lam,\muv),
\label{log}
\eeq
where $i$ denotes the index of PET detector pair and $N_d$ is the total number of detector pairs. $m$
denotes the $m$th time-of-flight bin and $N_t$ is the number of time-of-flight bins. The expectation of the PET projection data $\overline{\y}$ is related to the radiotracer activity image $\lam$ and object attenuation image $\muv$ at 511 keV via
\beq
\overline{\y}_{m}(\lam, \muv) = \diag\{\n_m(\muv)\}\G_m\lam + \rv_m,
\label{ex}
\eeq
where $\G_m$ is the PET detection probability matrix for time-of-flight bin $m$. $\rv_m$ accounts for the expectation of random and scattered events. $\n_m(\muv)$ is the normalization factor with the $i$th element being
\beq
n_{i,m}(\muv) = c_{i,m} \cdot \exp(-[\A\muv]_i),
\eeq
where $c_{i,m}$ represents the multiplicative factor excluding the attenuation correction factor and $\A$ is the system matrix for transmission imaging.

\subsection{Standard MLAA Reconstruction}
The MLAA reconstruction algorithm \cite{Rezaei2012} simultaneously estimates the attenuation image $\muv$ and the activity image $\lam$ from the PET projection data $\y$ by maximizing the Poisson log-likelihood,
\beq
\hat{\lam}, \hat{\muv} = {\arg\max}_{\lam \geq 0, \muv \geq 0}L(\y|\lam,\muv).
\label{eq-mlaa}
\eeq
An iterative interleaved updating strategy is commonly used to seek the solution. 
At each iteration of the algorithm, $\lam$ is first obtained based on the attenuation image $\muv^n$ from the previous iteration $n$:
\beq
\lam^{n+1} = \arg\max_{\lam \geq 0} L(\y|\lam,\muv^n),
\eeq
which can be updated by the maximum-likelihood expectation maximization (MLEM) algorithm \cite{Shepp1982} with one subiteration,
\beq
\lam^{n+1} = \frac{\lam^n}{\pv^n} \cdot \left(\sum_{m}\G_{m}^{T} \left[\n_m(\muv^n) \cdot \frac{\y_m}{\bar{\y}_m(\lam^n, \muv^n)}\right]\right),
\label{eq-mlem}
\eeq
where $\pv^n$ denotes the updated sensitivity image from the iteration $n$,
$
\pv^n = \sum_{m}\G_{m}^{T}\n_{m}(\muv^n).
$

$\muv$ is then updated with the estimated $\lam$ following the maximum-likelihood transmission reconstruction formulation,
\bea
\muv^{n+1} &=& \arg\max_{\muv \geq 0} L(\y|\lam^{n+1},\muv)\nonumber,\\
&=&\arg\max_{\muv \geq 0} \sum_{i,m}h_{i,m}\big(\left[\A \muv\right]_{i}\big),
\label{MLTR}
\eea
where
\beq
h_{i,m}(l) \triangleq y_{i,m} \log(\hat{b}_{i,m}e^{-l} + r_{i,m}) - (\hat{b}_{i,m}e^{-l} + r_{i,m})
\label{h}
\eeq
with $\bm{l}=\A\muv$ and 
\beq
\hat{b}_{i,m} = c_{i,m} \cdot [\G_m \lam^{n+1}]_i.
\eeq
The sub-optimization problem Eq. (\ref{MLTR}) can be solved using the separable paraboloidal surrogate algorithm \cite{Erdogan1999}.

Note that conventional applications of MLAA mainly focused on improving PET attenuation correction (e.g., \cite{Panin2013,Bousse2016, Mehranian2015,Ahn2018,Defrise2014,Berker2016,Heuer2017MLAA}). A new application of MLAA demonstrated the potential of gCT reconstruction for improving the estimation of proton stopping-power in proton therapy \cite{Baumer2021Can}. Differently in our PET-enabled DECT method, the gCT image $\muv$ is combined with X-ray CT \txtr{image $\x$} to form a DECT image pair for multi-material decomposition \cite{Wang2020}.

\subsection{Kernel MLAA (KAA) for gCT Reconstruction}
The gCT image estimate by the MLAA method \cite{Rezaei2012} is commonly noisy due to the limited counting statistics of PET emission data. To suppress noise, the kernel MLAA or KAA integrates the X-ray CT prior image into the PET forward model by describing the gCT image intensity $\mu_j$ at pixel $j$ using a kernel representation \cite{Wang2020, Li2021},
\beq
\mu_j = \sum_{l\in\Nc_j}\alpha_l \kappa(\f_j, \f_l),
\eeq
where $\kappa (\cdot, \cdot)$ is the kernel function (e.g., radial Gaussian) with $\f_j$ and $\f_l$ denoting the feature vectors of pixel $j$ and $l$ that are extracted from the X-ray CT image \txtr{$\x$}. $\Nc_j$ defines the neighborhood  of pixel $j$, for example, selected by a k-nearest neighbor algorithm. $\alpha_l $ denotes the corresponding kernel coefficient of each neighboring pixel $l$ in $\Nc_j$. 

The equivalent matrix-vector form for the gCT image representation is 
\beq
\muv = \K\alp,
\label{eq-ker}
\eeq
where $\K$ is the kernel matrix and $\alp$ is the corresponding kernel coefficient image. 
Substituting Eq. (\ref{eq-ker}) into the MLAA formulation in Eq. (\ref{eq-mlaa}) gives the following KAA optimization formulation,
\beq
\hat{\lam},\hat{\alp} = {\arg\max}_{\lam \geq 0, \alp \geq 0}L\big(\y |\lam, \K\alp\big).
\eeq
Once $\hat{\alp}$ is determined, the gCT image is obtained as $\hat{\muv} = \K\hat{\alp}$. Note that the conventional MLAA can be considered as a special case of the KAA with $\K$ equal to an identity matrix.

The KAA problem is also solved using an interleaving optimization strategy between the activity image $\lam$  update and the kernel coefficient image $\alp$ update \cite{Wang2020}. In each iteration of KAA, $\lam^{n+1}$ is first obtained using the MLEM updating formula Eq. (\ref{eq-mlem})  and then $\alp^{n+1}$ is obtained using the following kernelized transmission reconstruction (KTR) optimization, 
\beq
\alp^{n+1}={\arg\max}_{\alp \geq 0}\sum_{i,m}h_{i,m}\big(\left[\A \K\alp\right]_{i}\big),\
\label{alp step}
\eeq
as illustrated in Fig. \ref{KAA vs Proposed}a.

\subsection{Material Decomposition Using PET-enabled DECT}
For each image pixel $j$, the gCT attenuation value $\mu_j$ and X-ray CT attenuation value $x_j$ jointly form a pair of dual-energy measurements $\uv_j\triangleq [x_j, \mu_j]^T$, which can be modeled by a set of material bases, such as air (A), soft tissue (S) or equivalently water, and bone (B):
\beq
\uv_j  = \U\ro_j,  \quad\U\triangleq\left( \begin{array}{ccc}
	x_A & x_S & x_B \\
	\mu_A & \mu_S & \mu_B
\end{array} \right), \ro_j\triangleq\left( \begin{array}{c}
	\rho_{j,A}\\
	\rho_{j,S}\\
	\rho_{j,B}
\end{array} \right),
\label{MMD-Eq}
\eeq
subject to
$
\sum_{k}\rho_{j,k} = 1.
$
The coefficients $\rho_{j,k}$ with $k = {A, S, B}$ are the fraction of each basis material in pixel $j$. The material basis matrix $\U$ consists of the linear attenuation coefficients of each basis material measured at the low and high energies. Finally, $\ro_j$ is estimated using the following least-square optimization for each image pixel,
\beq
\hat{\ro}_j =\arg\min_{\ro_j\geq\0}\left \|\uv_j - \U\ro_j \right\|^2.
\eeq

\section{Proposed Neural KAA for gCT Reconstruction}
\subsection{Kernel Model with Deep Coefficient Prior for gCT}
While demonstrating a substantially better performance than MLAA (e.g., in \cite{Wang2020}), KAA may still suffer from noise or artifacts in low-count cases. In this work, we exploit neural networks as a \txtr{conditional deep coefficient prior} for improving KAA for gCT image reconstruction. 

The kernel coefficient image $\alp$ for $\muv$ in Eq. (\ref{eq-ker}) is described as a function of the \txtr{conditional} neural networks,
\beq
\alp = \boldsymbol{\psi}(\tht|\z),
\label{CNN_a}
\eeq
where $\z$ is the available image prior from the corresponding X-ray CT in this work and $\boldsymbol{\psi}$ denotes the neural network mapping from the known input image $\z$ to the $\alp$ image with $\tht$ the parameters of the neural network. 

The gCT image is then modeled using the following kernel representation with the \txtr{conditional deep coefficient prior},
\beq
\muv = \K \boldsymbol{\psi}(\tht|\z).
\label{KDCP}
\eeq
Fig. \ref{fig: KDCP} shows a graphical illustration of the proposed model for representing a gCT image, of which the last layer is a linear kernel representation with pre-determined weights $\{\kappa_{j,l}\}$ that are also calculated from $\z$.

By substituting the gCT representation Eq. (\ref{KDCP}) into the forward model in Eq. \ref{ex}, we have the following forward projection model that is parameterized by the neural network parameters $\tht$:
\beq
\overline{\y}_{m} = \diag\{\bm{c}_m\cdot\e^{-\A\K \boldsymbol{\psi}(\tht|\z)}\}\G_m\lam + \rv_m.
\label{eq-kaamodel}
\eeq

The model is equivalent to the standard KAA image model in Eq. (\ref{ex}) if the neural network $\boldsymbol{\psi}$ is an identity mapping. It is also equal to a \txtr{conditional} DIP (CDIP) model directly in the gCT image domain if the kernel matrix $\K$ is an identity matrix, which leads to $\muv = \boldsymbol{\psi}(\tht|\z)$. 

\begin{figure}[t]
\centering
\subfloat[Existing KAA without deep learning]{\includegraphics[trim=0cm 0cm 0cm 0cm, clip,width=8cm]{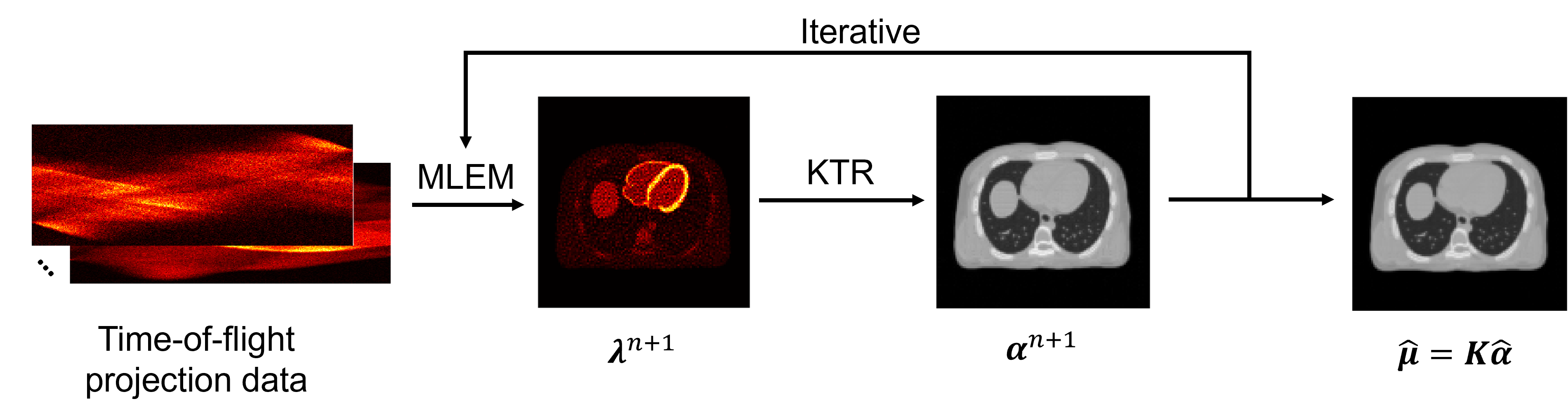}}\\
\subfloat[Proposed Neural KAA]{\includegraphics[trim=0cm 0cm 0cm 0cm, clip, width=8cm]{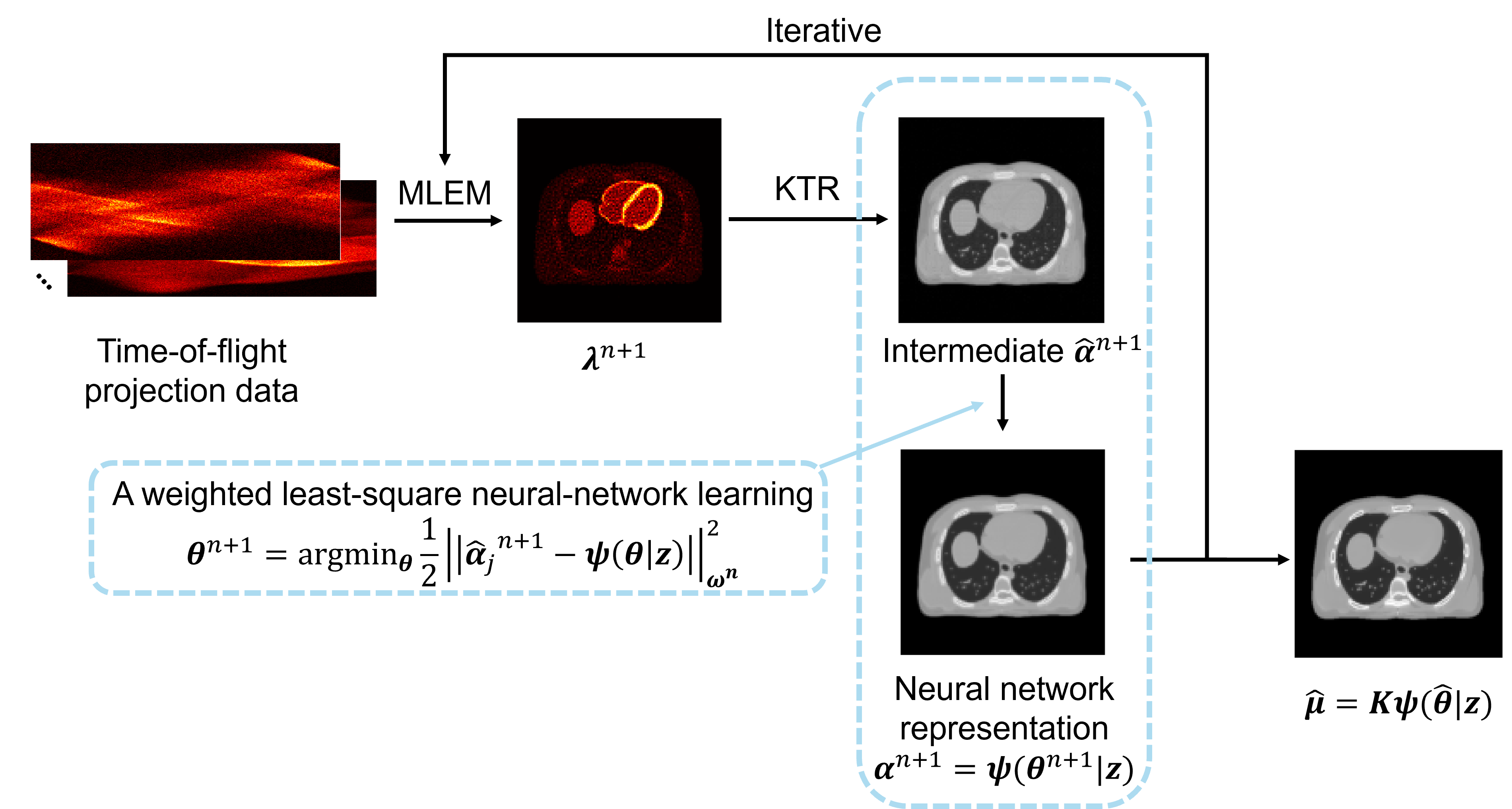}}
\caption{Graphical illustration of the standard KAA and proposed neural KAA. The blue dashed box indicates the proposed {\em neural optimization transfer} algorithm for which the weighted least-square loss function for network learning is theoretically derived, rather than arbitrarily defined.}
\label{KAA vs Proposed}
\end{figure}

\begin{figure}[t]
\centering
\includegraphics[trim=0.3cm 0cm 0cm 0.3cm, clip,width=2.5in]{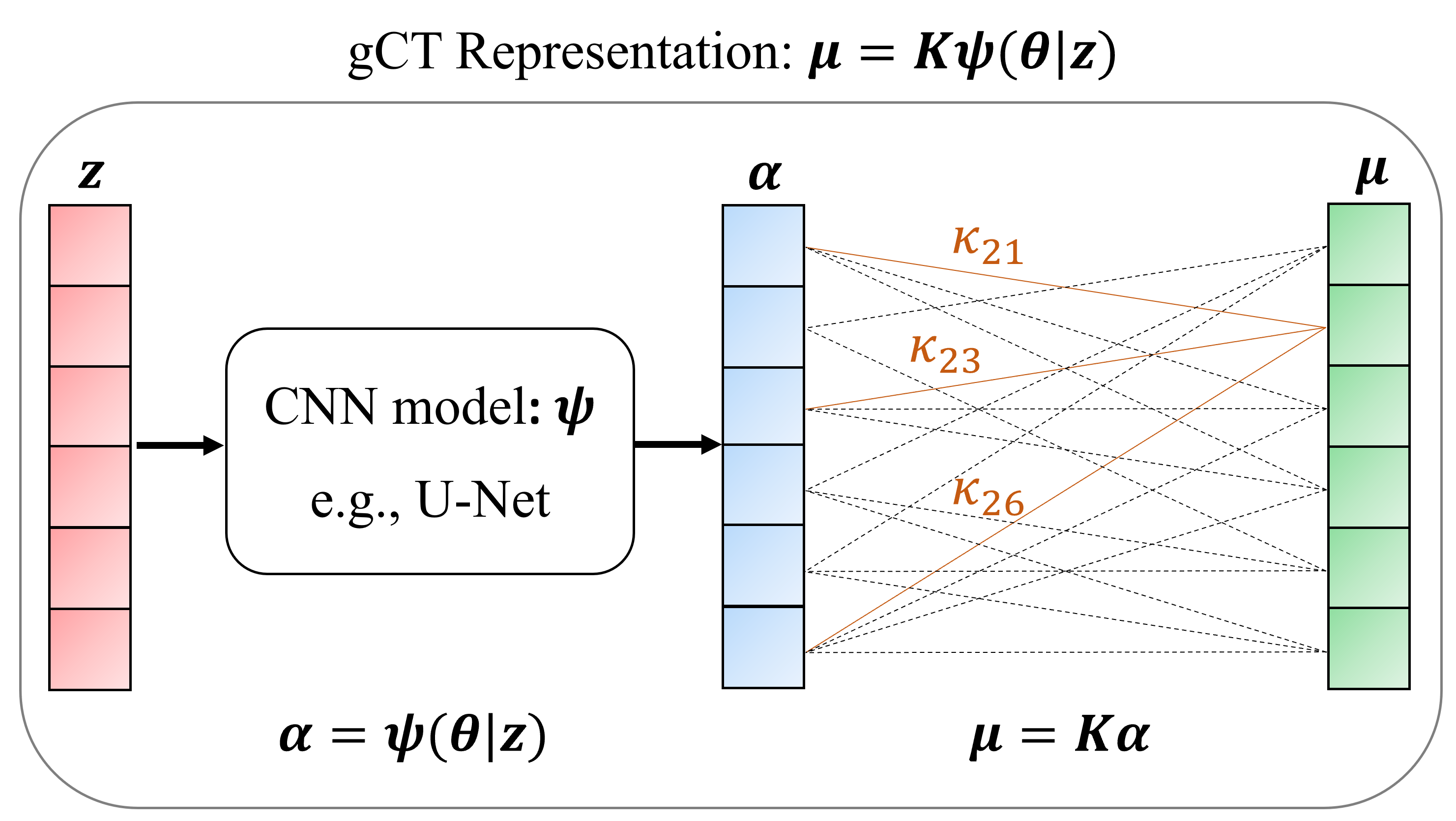}
\caption{Graphical illustration of the kernel model with a \txtr{conditional} deep coefficient prior for gCT $\muv$ representation.} 
\label{fig: KDCP}	
\end{figure}

\subsection{Optimization Formulation}
By combining the forward projection model with the MLAA formulation Eq. (\ref{eq-mlaa}), we have the proposed neural KAA optimization formulation,
\beq
\hat{\lam},\hat{\tht} = {\arg\max}_{\lam \geq 0, \tht}L\big(\y |\lam, \K\boldsymbol{\psi}(\tht|\z)\big).
\eeq
Similar to the optimization approach for MLAA and KAA, an interleaving updating strategy can be used here to estimate $\lam$ and $\tht$ iteratively,
\beq
\lam^{n+1} = \arg\max_{\lam\geq\0} L\big(\y |\lam, \K\boldsymbol{\psi}(\tht^n|\z)\big),
\label{nkml-lam}
\eeq
\beq
\tht^{n+1} = \arg\max_{\tht} L\big(\y |\lam^{n+1}, \K\boldsymbol{\psi}(\tht|\z)\big).
\label{nkml-tht}
\eeq
Once $\tht$ is estimated, the gCT image is obtained by
\beq
\hat{\muv} = \K\boldsymbol{\psi}(\hat{\tht}|\z).
\eeq

Compared to the previous KAA approach, the neural KAA approach combines the kernel representation with a neural network-based deep coefficient prior, which introduces an implicit regularization for KAA to improve the gCT image estimate $\hat{\muv}$ through Eq. (\ref{nkml-tht}).

\subsection{The Optimization Challenge}
In each iteration of the neural KAA, the $\lam$ estimation step in Eq. (\ref{nkml-lam}) can be directly implemented by using the MLEM algorithm Eq. (\ref{eq-mlem}). The $\tht$-estimation step Eq. (\ref{nkml-tht}) follows a KTR formulation but with using neural networks as \txtr{a conditional deep coefficient prior}. For simplicity, we rewrite this {\em neural} KTR problem in Eq. (\ref{nkml-tht}) as
\beq
\tht^{n+1} = \arg\max_{\tht}  J_n(\tht),
\label{Neural MLTR}
\eeq
where $J_n(\tht)$ is the {\em transmission} likelihood function at iteration $n$,
\bea
J_n(\tht) &\triangleq& L\big(\y |\lam^{n+1}, \K\boldsymbol{\psi}(\tht|\z)\big) \nonumber\\
&=&  \sum_{i,m}h_{i,m}\big(\left[\A \K\boldsymbol{\psi}(\tht|\z)\right]_{i}\big),
\label{J}
\eea
with $h_{i,m}$ defined in Eq. (\ref{h}) but the line integral $l$ following a kernelized neural-network model,
\beq
\bm{l}=\A\K\boldsymbol{\psi}(\tht|\z).
\label{nktr-proj}
\eeq
The optimization problem Eq. (\ref{Neural MLTR}) is challenging to solve because the unknown neural network parameters $\tht$ is nonlinearly involved in the projection domain for transmission imaging due to Eq. (\ref{nktr-proj}), resulting in a complex optimization problem. 

One commonly used solution would be a type of gradient descent algorithm that uses the chain rule to calculate the gradient of $J_n(\tht)$ with respect to $\tht$ \cite{Akcakaya2022Unsupervised, Baguer2020, Shu2022,Yoo2021},
\beq
\frac{\partial J_n(\tht)}{\partial\tht}=\frac{\partial J_n(\tht)}{\partial\psiv}\cdot\frac{\partial\psiv}{\partial\tht},
\eeq
which is then fed into an existing deep learning package to estimate the network parameters from the PET projection data $\y$. However, such an approach ties each calculation of $\frac{\partial\psiv}{\partial\tht}$, which relates to the neural network component, with a calculation of $\frac{\partial J_n(\tht)}{\partial\psiv}$, which relates to the PET tomographic reconstruction. While the former operation is usually efficient by using a deep learning library, the latter operation requires both forward and back projections of PET data and is computationally intensive due to the large size of the transmission system matrix $\A$. As a result, the whole algorithm can be slow due to the natural need for many iterations for training a neural network.

Another solution would be the ADMM algorithm as used for the DIP reconstruction \cite{Gong2019, Barutcu2021Limited, Zhao2024J}. This type of algorithms may separate the neural network learning step (that uses $\frac{\partial\psiv}{\partial\tht}$) from the tomographic reconstruction step (that uses $\frac{\partial J_n(\tht)}{\partial\psiv}$). However, a major weakness of ADMM is that it is difficult to tune the induced hyper-parameters in practice as demonstrated by many studies \cite{Zhang2020Review, Ben2020Deep}. 

In this work, we develop an approach to decoupling the reconstruction step and neural network learning step without adding extra hyper-parameters towards a more practical and efficient implementation.
\begin{figure}[h]
\centering
\includegraphics[trim=0cm 0cm 0cm 0cm, clip,width=2.5in]{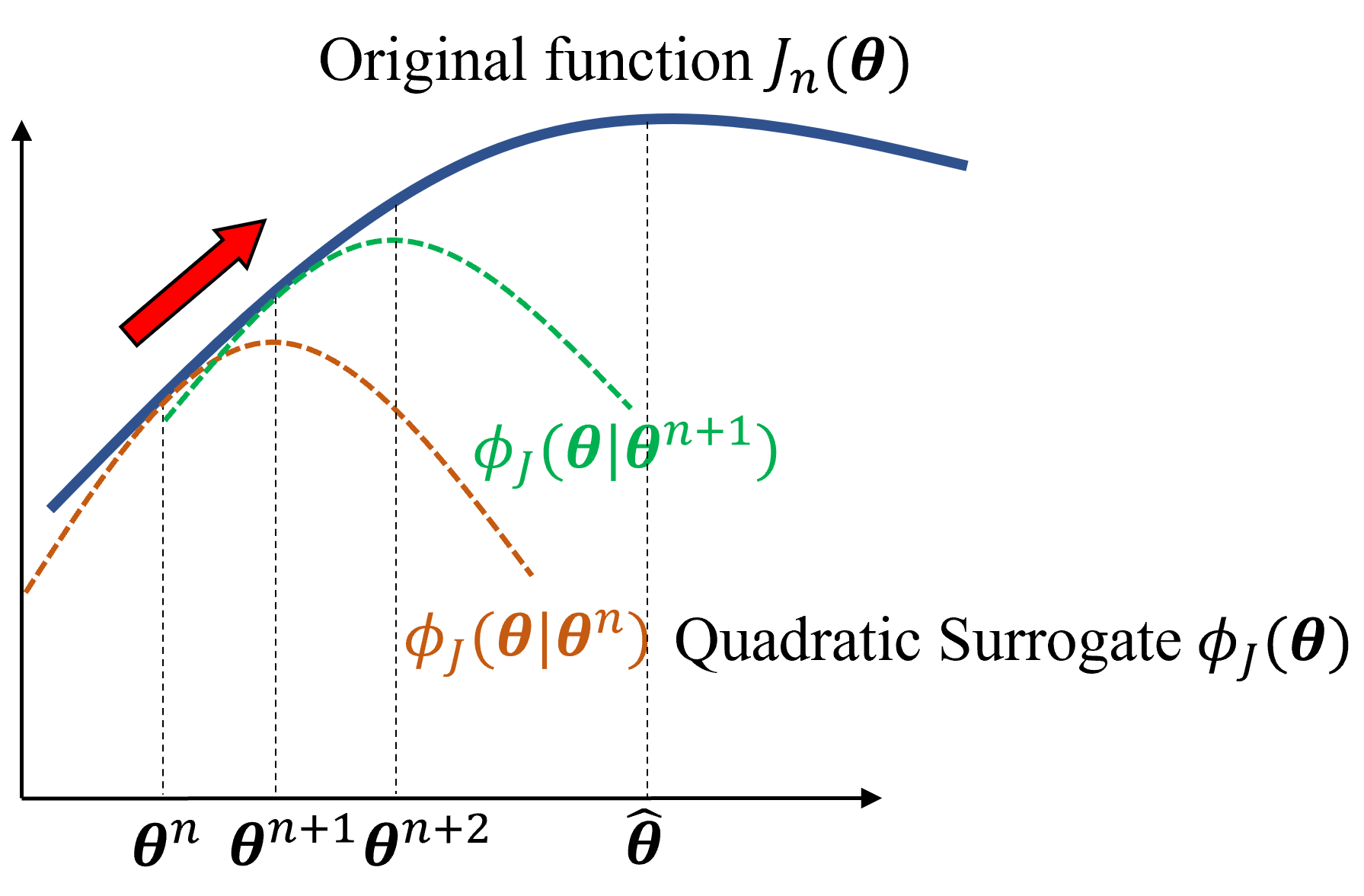}
\caption{Graphical illustration of the basic idea of optimization transfer with quadratic surrogates. The surrogate function $\phi_J(\tht)$ minorizes the original objective function $J_n(\tht)$. The new update $\tht^{n+1}$, which maximizes $\phi_J(\tht|\tht^n)$, will guarantee a monotonic increase in $J_n(\tht)$ and converge to the local solution $\hat{\tht}$.}
\label{OT}	
\end{figure}

\section{A Neural Optimization Transfer Algorithm for gCT}
\label{sec4}
\subsection{Optimization Transfer with Quadratic Surrogates}
The neural network reconstruction problem shares the same complication as the tomographic reconstruction of nonlinear kinetic parameters in dynamic PET \cite{Wang2013} which, however, can be solved by a quadratic surrogate-based optimization transfer algorithm \cite{Wang2009}. Thus we apply the principle of optimization transfer \cite{Lange2000} to convert the original difficult problem for $J_n(\tht)$ into a quadratic surrogate optimization problem, as illustrated in Fig. \ref{OT}. The quadratic surrogate function $\phi_J(\tht|\tht^n)$ minorizes the transmission likelihood function $J_n(\tht)$,
\beq
\phi_J(\tht|\tht^n) \leq J_n(\tht),
\eeq
\beq
\phi_J(\tht^n|\tht^n) = J_n(\tht^n).
\eeq
Then the maximization of $J_n(\tht)$ is transferred into maximizing the surrogate function,
\beq
\tht^{n+1} = \arg\max_{\tht}\phi_J(\tht|\tht^n).
\eeq
The quadratic surrogate $\phi_J(\tht|\tht^n)$ is designed to be easy to solve. The new update $\tht^{n+1}$ is guaranteed to monotonically increase the original likelihood $J_n(\tht)$, i.e.,
\beq
J_n(\tht^{n+1}) \geq J_n(\tht^{n}),
\eeq
as also demonstrated by Fig. \ref{OT}.

Note that the concept of optimization transfer has been also applied for the KAA reconstruction in which the gCT projection is linear with respect to the unknown attenuation image $\muv$ \cite{Wang2020}. However, in the neural KAA, the unknown becomes the neural network parameters $\tht$ which is non-linearly involved in the gCT projection domain. Our derivations will show how the quadratic surrogate functions are built to decouple the linear tomographic reconstruction step and the nonlinear neural network learning step from each other, as also shown in Fig. \ref{KAA vs Proposed}.

\subsection{Paraboloidal Surrogates in the Projection Domain}

One difficulty with dealing the objective function $J_n(\tht)$ directly is the Poisson log-likelihood function follows a non-quadratic form. Following a derivation similar to \cite{Fessler1998, Erdogan1999}, a paraboloidal surrogate function can be constructed for $J_n(\tht)$,
\beq
\begin{aligned}
J_n(\tht)&\geq S(\tht|\tht^n)\\
&= \sum_{m,i}h_{i,m}(l_i^n)+ \dot{h}_{i,m}(l_i^n)\Delta l_i - \frac{\eta_{i,m}(l_i^n)}{2}\Delta l_i ^2,
\end{aligned}
\label{ps}
\eeq
where $l_i^n$ is $l_i$ calculated with $\tht^n$ and $\Delta l_i =l_i-l_i^n$. $\eta_{i,m}(l)$ is chosen by design as the optimal curvature of the Poisson log-likelihood \cite{Fessler1998},
\beq
\eta_{i,m}(l)=\left\{
\begin{aligned}
&\frac{2}{l^2}[h_{i,m}(l)-h_{i,m}(0)-l\dot{h}_{i,m}(l)], &l>0 \\
&-\ddot{h}_{i,m}(l),&l=0
\end{aligned}
\right.
\label{eta}
\eeq
where $\dot{h}_{i,m}(l)$ and $\ddot{h}_{i,m}(l)$ are the first and second derivatives of $h_{i,m}(l)$, respectively \cite{Fessler1998}.

With several algebraic operations similar to the way in \cite{Wang2009}, we then can derive the following equivalent quadratic form for $S(\tht|\tht^n)$ in Eq. (\ref{ps}),
\beq
S(\tht|\tht^n) = -\frac{1}{2}\big|\big|\hat{\boldsymbol{l}}^{n+1}-\A \K\boldsymbol{\psi}(\tht|\z)\big|\big|_{\hat{\etav}^n}^2 + C_S^n,
\label{OT-1}
\eeq
where $C_S^n$ is the remainder that is independent of the unknown parameter $\tht$. $\hat{\boldsymbol{l}}^{n+1}$ is an intermediate gCT projection data \cite{Wang2020},
\beq
\hat{l}_i^{n+1} = l_i^n + \frac{\sum_m\dot{h}_{i,m}(l_i^n)}{\sum_m\eta_{i,m}(l_i^n)},
\label{l}
\eeq
and $\hat{\etav}^n$ is an intermediate weight also in the projection domain, 
\beq
\hat{\eta}_i^n = \sum_m\eta_{i,m}(l_i^n).
\label{eta_i}
\eeq

Based on this quadratic surrogate $S(\tht|\tht^n)$, the maximization of $J_n(\tht)$ for the neural KTR is transferred to an equivalent weighted least-square reconstruction problem,
\beq
\hat{\tht}^{n+1} = \arg\min_{\tht} \frac{1}{2}\big|\big|\hat{\boldsymbol{l}}^{n+1}-\A \K\boldsymbol{\psi}(\tht|\z)\big|\big|_{\hat{\etav}^n}^2.
\label{WLS}
\eeq
This quadratic form is simpler than the original non-quadratic likelihood function $J_n(\tht)$. However, the nonlinear neural network model $\boldsymbol{\psi}(\tht|\z)$ is still coupled in the projection domain.

\subsection{Separable Quadratic Surrogates in the Image Domain}
By considering $\A\K$ as a single matrix and using the convexity of Eq. (\ref{OT-1}) to build a separable quadratic surrogate \cite{Erdogan1999b}, we can construct the following surrogate $s(\tht|\tht^n)$ for $S(\tht|\tht^n)$,
\beq
\begin{aligned}
S(\tht|\tht^n)& \geq  s(\tht|\tht^n) \\ 
&=S(\tht^n|\tht^n)+(\g^n)^T\Delta\boldsymbol{\psi}(\tht|\z)  - \frac{1}{2}\big|\big|\Delta\boldsymbol{\psi}(\tht|\z)\big|\big|^2_{\ome^n}
\end{aligned}
\label{SQS}
\eeq
where $\Delta\boldsymbol{\psi}(\tht|\z) \triangleq \boldsymbol{\psi}(\tht|\z) - \boldsymbol{\psi}(\tht^n|\z)$. $\g^n$ is the gradient of $S(\tht|\tht^n)$ with respect to $\psiv$ at iteration $n$,
\beq
\g^n = \K^T\A^T \diag(\hat{\etav}^n)\Big(\hat{\boldsymbol{l}}^{n+1}-\A\K\boldsymbol{\psi}(\tht^n|\z)\Big),
\label{g}
\eeq
and $\ome^{n}$ is an intermediate weight image,
\beq
\ome^n = \K^T\A^T \diag(\hat{\etav}^n)\A\K \mathbf{1},
\label{ome}
\eeq
where $\mathbf{1}$ is the all-one vector. Note that $\g^n$ is also equal to the gradient of $J_n(\tht)$ with respect to $\psiv$ at iteration $n$.

After matching a quadratic function, we can have the following equivalent form for $s(\tht|\tht^n)$,
\beq
s(\tht|\tht^n) = \frac{1}{2}\big|\big|\hat{\alp}_j^{n+1}-\psiv(\tht|\z)\big|\big|_{\ome^n}^2+ C_s^n
\eeq
where $\hat{\alp}^{n+1}$ is an intermediate kernel coefficient image calculated from
\beq
\hat{\alp}^{n+1}=\alp^{n} + \frac{\g^{n}}{\ome^{n}}, 
\label{kernel MLTR}
\eeq
with $\alp^n \triangleq \boldsymbol{\psi}(\tht^n|\z)$. $C_s^n$ denotes the corresponding remainder term that is independent of $\tht$. 
Note that $\hat{\alp}^{n+1}$ is equivalent to one iteration of the KTR reconstruction algorithm in \cite{Wang2020}. 

Thus,  the weighted least-square reconstruction problem Eq. (\ref{WLS}) defined in the projection domain is transferred to the following neural-network learning problem that is fully defined in the image domain,
\beq
\boxed{
\tht^{n+1}  = \arg\min_{\tht} \frac{1}{2}\big|\big|\hat{\alp}_j^{n+1}-\psiv(\tht|\z)\big|\big|_{\ome^n}^2.
}
\label{wMSE}
\eeq
In particular, the cost function here follows a weighted $l_2$-norm loss that can be easily used with an existing deep learning package (e.g., PyTorch). 

It is straightforward to prove that the surrogate function $s(\tht|\tht^n)$ minorizes the original likelihood function $J_n(\tht)$,
\beq
s(\tht|\tht^n) \leq S(\tht|\tht^n) \leq J_n(\tht),
\label{c1}
\eeq
\beq
s(\tht^n|\tht^n) = S(\tht^n|\tht^n) = J_n(\tht^n).
\eeq

Note that the optimization transfer algorithm in \cite{Wang2020} for standard KAA (Fig. \ref{KAA vs Proposed}a) is only a special case of the derived algorithm for the neural KAA (Fig. \ref{KAA vs Proposed}b) here without having the least-square neural network learning step. Therefore we call the new algorithm the {\em neural optimization transfer} algorithm to highlight its estimation of the neural network model parameters. The associated least-square form in Eq. (\ref{wMSE}) for network training is not arbitrarily defined but analytically derived from the theory of optimization transfer for a non-linear model. Furthermore, no any new hyper-parameters are introduced from the optimization transfer process, which represents an important advantage as compared to a possible ADMM algorithm.

\begin{table}[t]
\normalsize
\begin{tabularx}{3.4in}{p{0.1cm}X}
\toprule
&{\bf Algorithm 1} The neural optimization transfer algorithm for gCT reconstruction\\
\midrule
1: & Input parameters: Maximum iteration number {\ttfamily MaxIt}, initial $\lam^1$, and initial $\tht^1$ to provide $\alp^1=\psiv(\tht^1|\z)$.\\
2: & {\bf for} $n = 1$ to {\ttfamily MaxIt} {\bf do}\\
3: & Obtain the activity image update $\lam^{n+1}$ using Eq. (\ref{eq-mlem}); \\
4: & Get the intermediate kernel coefficient image $\hat{\alp}^{n+1}$:\\
&$\hat{\alp}^{n+1}=\alp^{n} + \frac{\g^{n}}{\ome^{n}}$,\\
& \txtc{where $\g^{n}$ and $\ome^{n}$ are calculated based on $\hat{\etav}^n$ and $\hat{\lv}^{n+1}$ in Eq. (\ref{g}) and Eq. (\ref{ome});}\\
5: & \txtc{Perform the least-square neural-network learning in Eq. (\ref{wMSE}) to update $\tht^{n+1}$ and $\alp^{n+1}=\psiv(\tht^{n+1}|\z)$:}\\
& $\tht^{n+1}  = \arg\min_{\tht} \frac{1}{2}\big|\big|\hat{\alp}_j^{n+1}-\psiv(\tht|\z)\big|\big|_{\ome^n}^2$;\\
6: & {\bf end for}\\
7: & {\bf return} $\hat{\muv} = \K \boldsymbol{\psi}(\hat{\tht}|\z)$ \\
\bottomrule
\end{tabularx}
\end{table}
\begin{figure}[h]
\centering
\includegraphics[trim=-0.3cm 2.0cm 0.5cm 0cm, clip,width=3.0in]{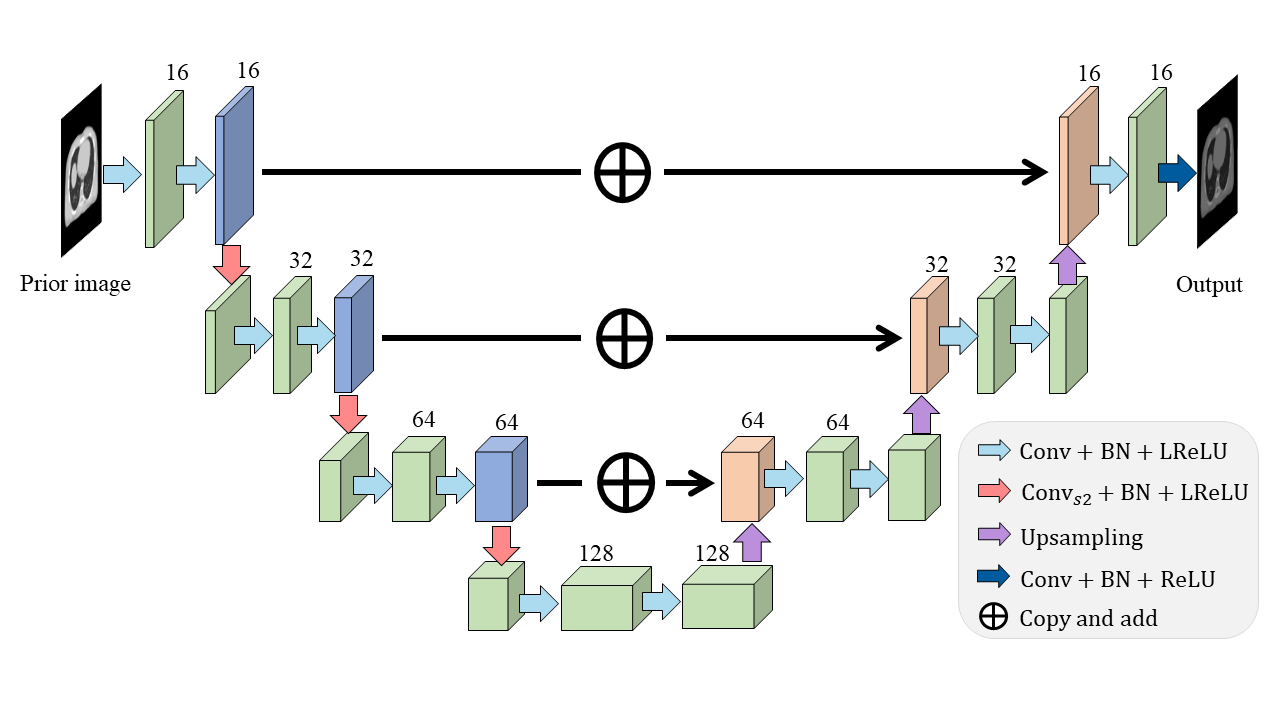}
\caption{The residual U-Net $\boldsymbol{\psi}(\tht|\z)$ used for kernel coefficient image representation in our work.}
\label{fig: Unet}	
\end{figure}

\begin{figure*}[t]
\centering
\subfloat[]{\includegraphics[trim=2cm 1cm 1cm 0.0cm, clip, width=3.93cm]{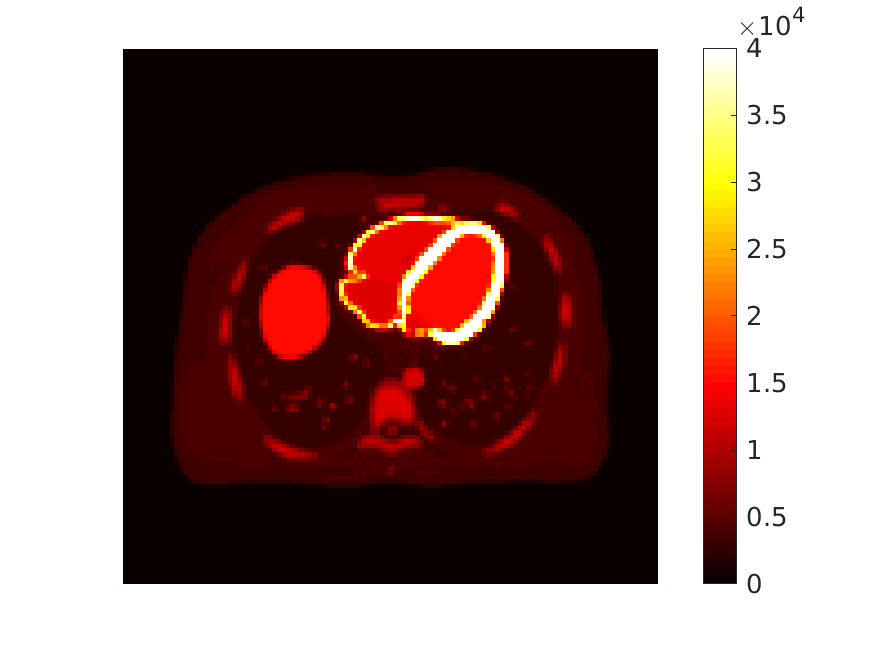}}
\subfloat[]{\includegraphics[trim=2cm 1cm 1cm 0.5cm, clip, width=4cm]{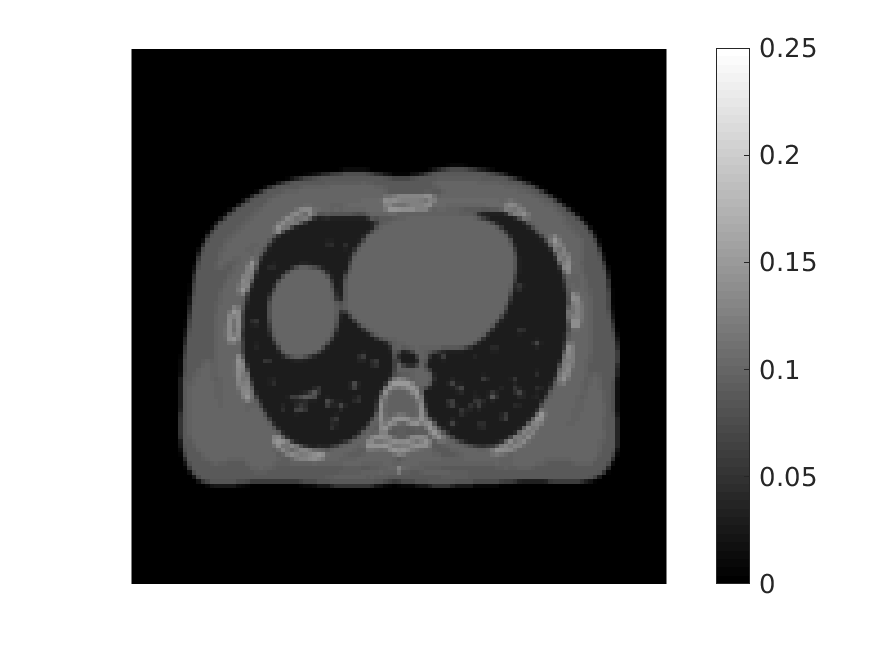}}
\subfloat[]{\includegraphics[trim=2cm 1cm 1cm 0.5cm, clip, width=4cm]{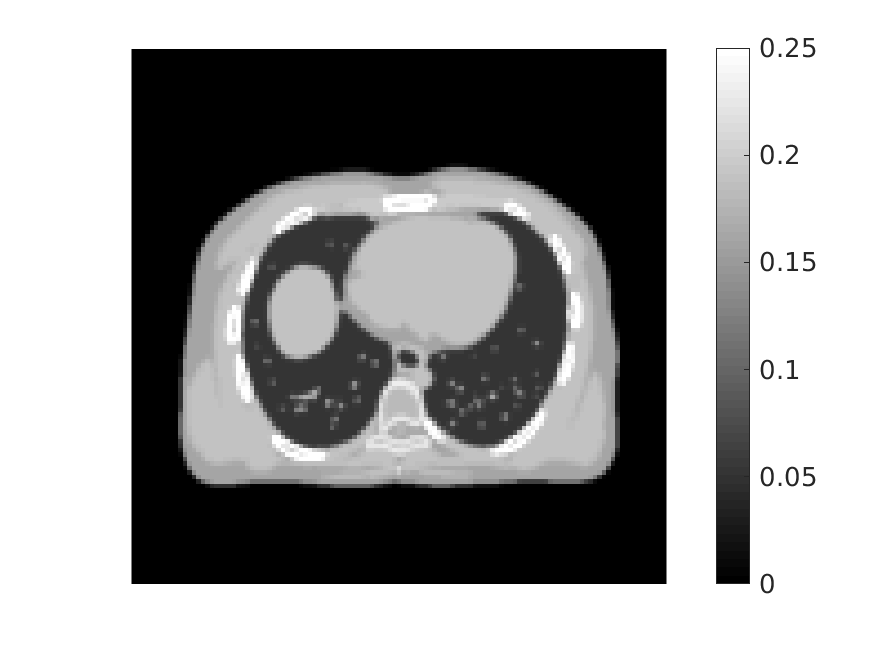}}
\subfloat[]{\includegraphics[trim=2cm 1.6cm 1cm 0.5cm, clip, width=4cm]{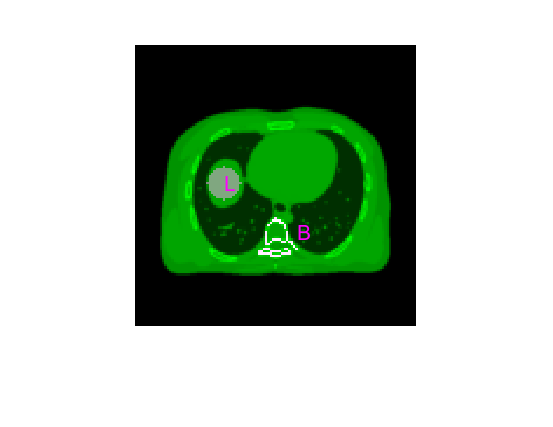}}
\caption{The digital phantom used in the PET/CT computer simulation. (a) PET activity image in Bq/cc; (b) PET attenuation image at 511 keV in cm$^{-1}$; (c) X-ray CT image at 80 keV; (d) illustration of a liver ROI `L' and a spine bone ROI `B'.}
\label{fig-pht-xcat}
\end{figure*}

\subsection{Summary of the Algorithm}
A pseudo-code of the proposed neural optimization transfer algorithm for gCT reconstruction is summarized in Algorithm 1. The algorithm consists of three separate steps in each iteration:
\begin{enumerate}
\item {PET activity image reconstruction}: Obtain the PET activity image update $\lam^{n+1}$ using \txtr {one iteration of} MLEM algorithm in Eq. (\ref{eq-mlem}).
\item {Kernelized gCT image reconstruction}: Obtain an intermediate kernel coefficient image update $\hat{\alp}^{n+1}$ using one iteration of the KTR algorithm in Eq. (\ref{kernel MLTR});
\item \textit{Least-square} neural-network learning for gCT image approximation: Update the network model parameters $\tht^{n+1}$ using Eq. (\ref{wMSE}) to approximate $\hat{\alp}^{n+1}$;
\end{enumerate}
Step 1 and step 2 are updated analytically. The tomographic reconstruction of projection data are only involved in these two steps. Step 3 can be implemented efficiently using existing deep learning packages without involving any projection data directly. Thus, the neural network learning step is decoupled from the image reconstruction steps and is easy to implement in practice. \txtr{The initialized weights of the neural network learning at the reconstruction iteration $n+1$ are inherited from the previous iteration $n$ rather than a random initialization. Note that random initialization was also tested but demonstrated a less stable performance.} The algorithm is guaranteed to increase the likelihood function monotonically due to the nature of optimization transfer \cite{Lange2000}.

Note that many other imaging modalities such as X-ray CT \cite{Baguer2020, Barutcu2021Limited}, MRI \cite{Yoo2021, Zhao2024J}, and optical tomography \cite{Zhou2020Diffraction, Vu2021Deep} may often use the same weighted least-square reconstruction formula in Eq. (\ref{WLS}) (but with constant $\hat{\boldsymbol{l}}^{n+1}$ and $\hat{\etav}^n$). Thus the neural optimization transfer algorithm in Algorithm 1 may be also applicable to these reconstruction problems, though without the need for Step 1 which is specific for the neural KAA for gCT reconstruction. 

\subsection{Neural Network Structure and Settings}
Any neural-network model that is suitable for image representation would be compatible with the proposed algorithm. As an example, here we use a popular residual U-Net architecture to represent the kernel coefficient image, as illustrated in Fig. \ref{fig: Unet}, following its previous successful use for PET activity image reconstruction \cite{Gong2019, Li2023}. The model comprises the following key components: (1) A convolution layer (kernel size: 3) + batch normalization (BN) + leaky rectified linear unit (LReLU) for feature extraction; (2) The strided convolution layer (strided size: 2, kernel size: 3) for down-sampling; (3) Bilinear/trilinear interpolation for up-sampling; (4) The additive operation used for skip connection between encoder and decoder paths; (5) A ReLU operation prior to the output layer used for generating the non-negative kernel coefficient image. The number of feature maps for each layer is also included in Fig. \ref{fig: Unet}. 

In this work, the input image of the neural network was set to an X-ray CT image, that can be explained as the conditional deep image prior (CDIP) \cite{Cui2022}. We used the Adam algorithm with a learning rate of $10^{-3}$ and 150 sub-iterations for the least-square neural network learning, which were the same setting as previous work \cite{Li2023}. The neural-network learning step was implemented with PyTorch on a PC with an Intel i9-9920X CPU with 64GB RAM and a NVIDIA GeForce RTX 2080Ti GPU.

\begin{figure}[h]
\centering
{\includegraphics[trim=0.1cm 0cm 1cm 0.2cm, clip, width=2in]{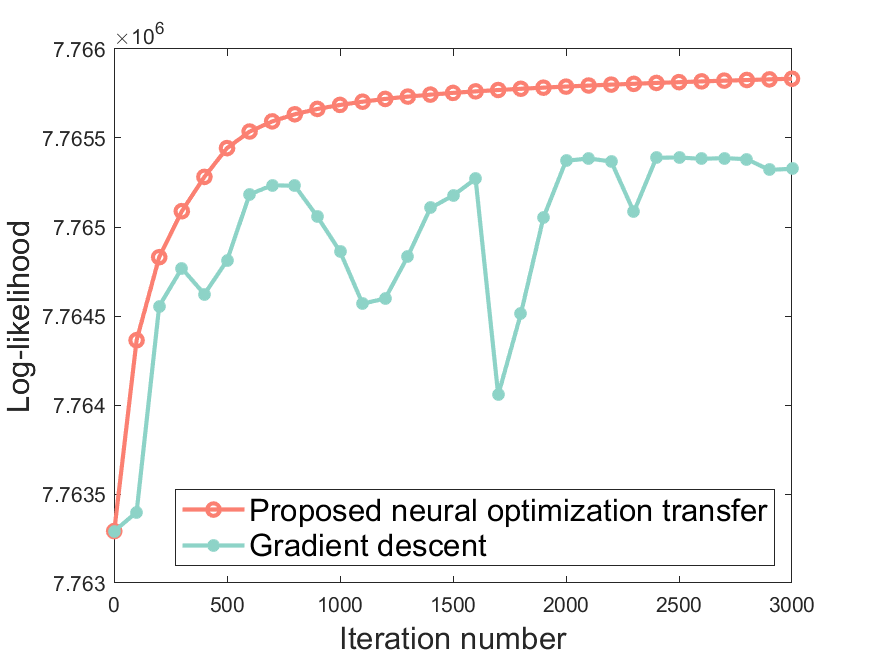}}
\caption{Comparison of the plots of the log-likelihood function for the gradient descent and proposed optimization transfer algorithms.}
\label{Neural OT}
\end{figure}


\begin{figure*}[t]
\centering

\subfloat[]{
\begin{minipage}[t]{0.22\linewidth}
	\centering
	\includegraphics[trim=2.5cm 1cm 1cm 0.0cm, clip, height=3.5cm]{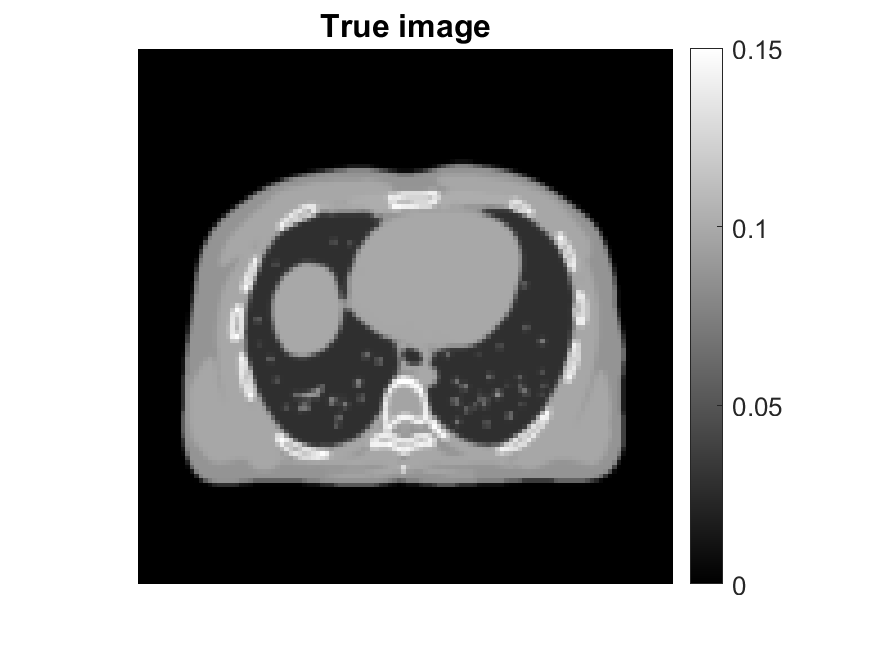}\\
	\includegraphics[trim=2.5cm 1cm 1cm 0.0cm, clip, height=3.5cm]{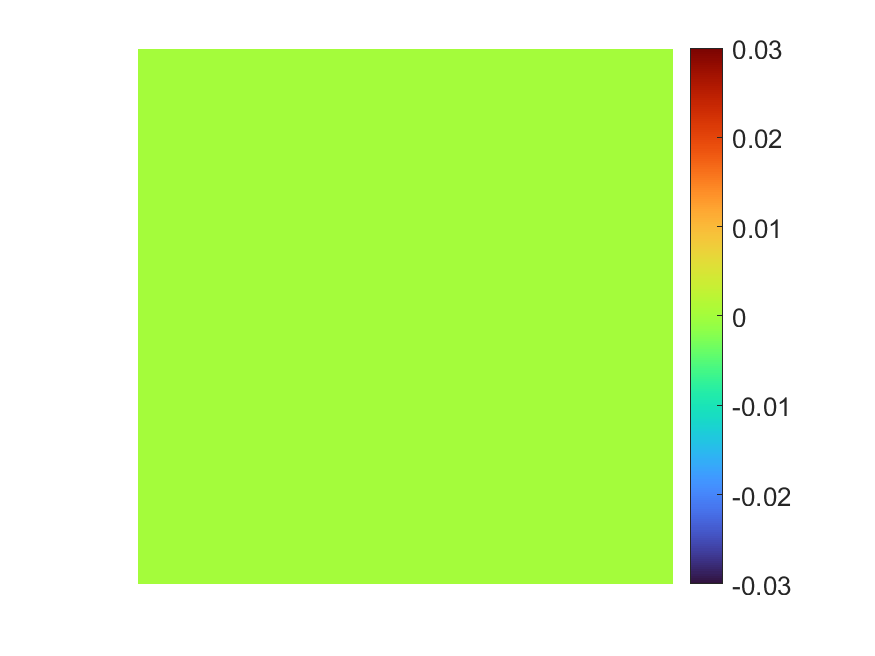}\\
\end{minipage}%
}%
\subfloat[]{
\begin{minipage}[t]{0.18\linewidth}
	\centering
	\includegraphics[trim=2.5cm 1cm 3.2cm 0.0cm, clip, height=3.5cm]{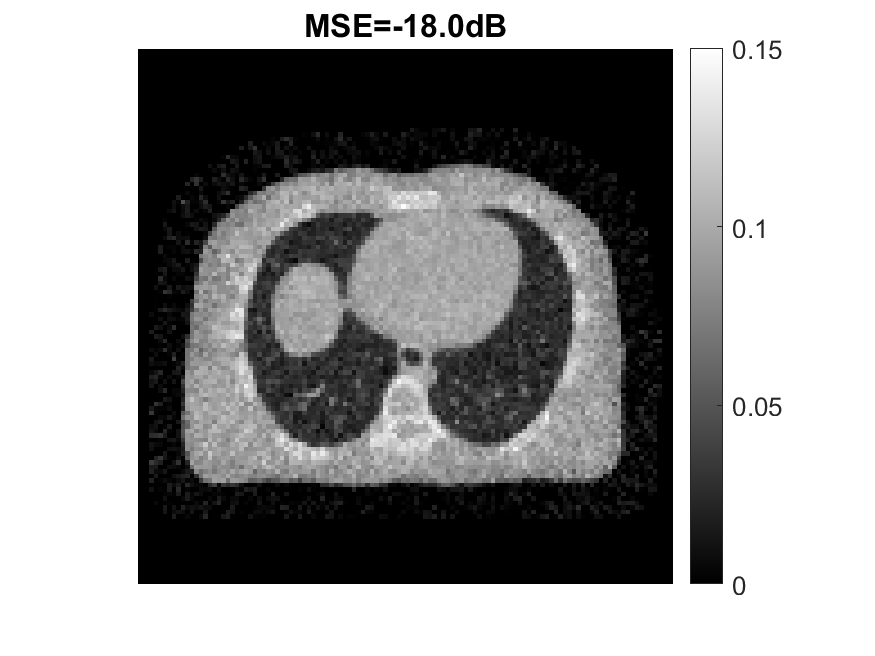}\\
	\includegraphics[trim=2.5cm 1cm 3.2cm 0.0cm, clip, height=3.5cm]{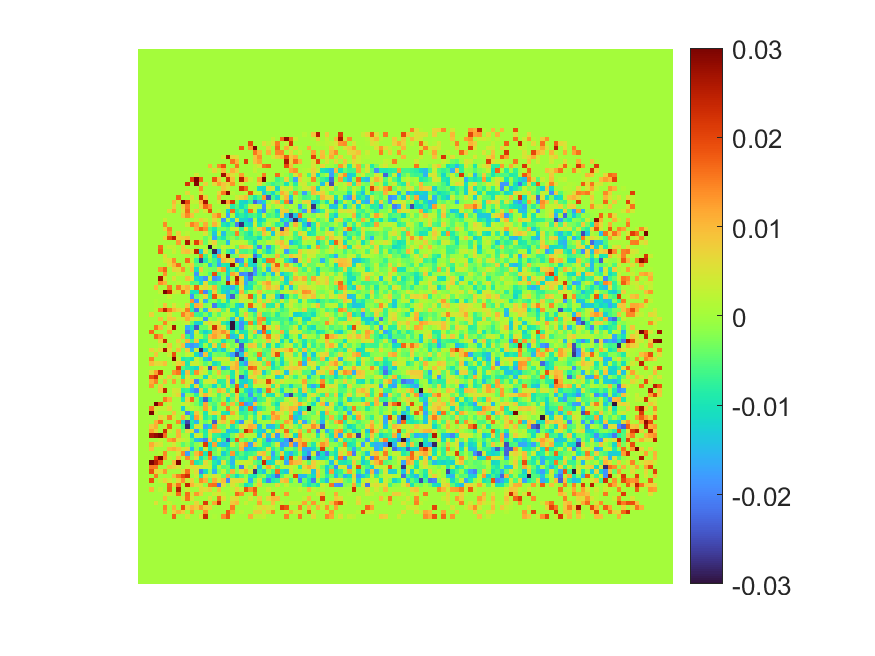}\\
\end{minipage}%
}%
\subfloat[]{
\begin{minipage}[t]{0.18\linewidth}
	\centering
	\includegraphics[trim=2.5cm 1cm 3.2cm 0.0cm, clip, height=3.5cm]{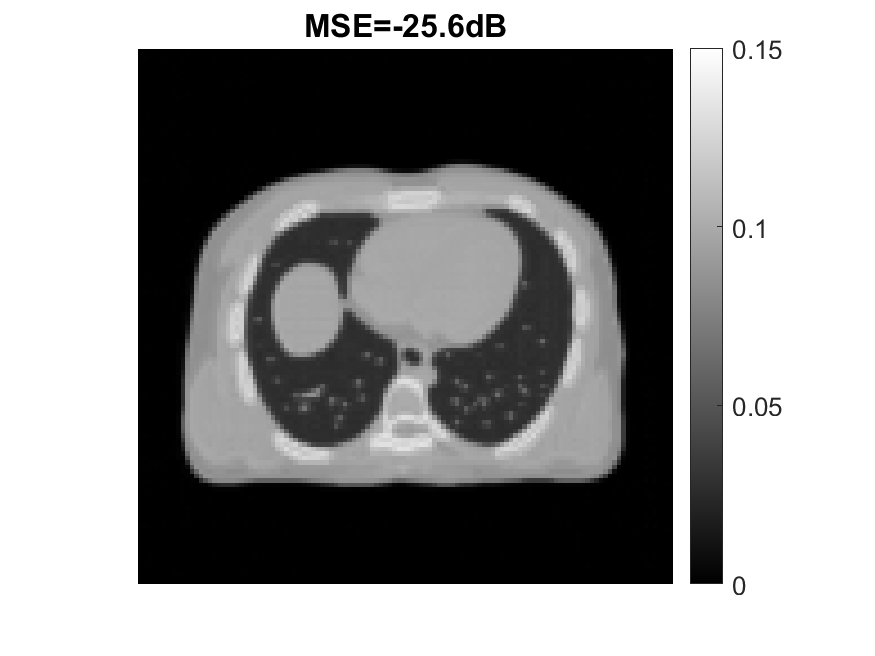}\\
	\includegraphics[trim=2.5cm 1cm 3.2cm 0.0cm, clip, height=3.5cm]{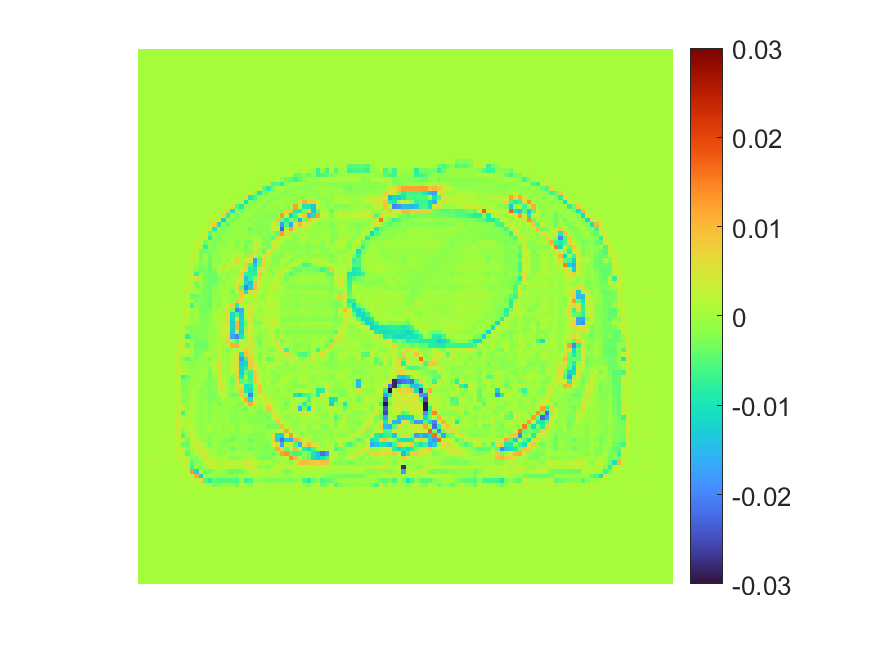}\\
\end{minipage}%
}%
\subfloat[]{
\begin{minipage}[t]{0.18\linewidth}
	\centering
	\includegraphics[trim=2.5cm 1cm 3.2cm 0.0cm, clip, height=3.5cm]{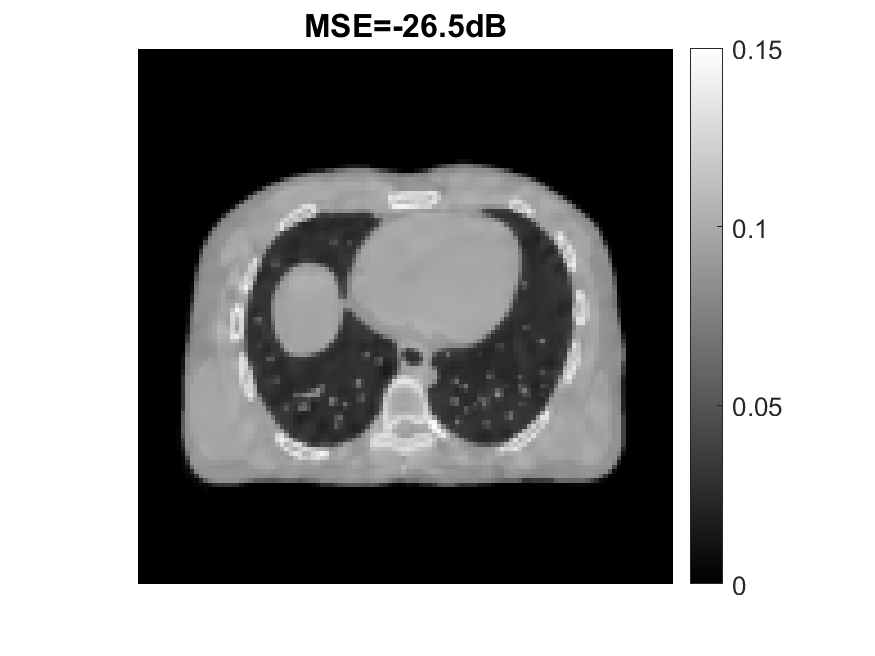}\\
	\includegraphics[trim=2.5cm 1cm 3.2cm 0.0cm, clip, height=3.5cm]{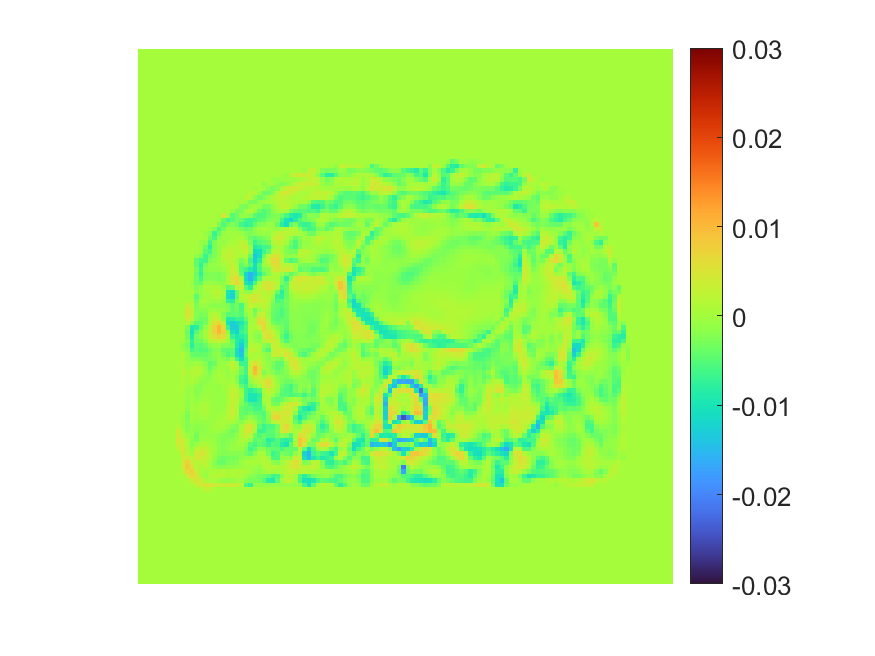}\\
\end{minipage}%
}%
\subfloat[]{
\begin{minipage}[t]{0.18\linewidth}
	\centering
	\includegraphics[trim=2.5cm 1cm 3.2cm 0.0cm, clip, height=3.5cm]{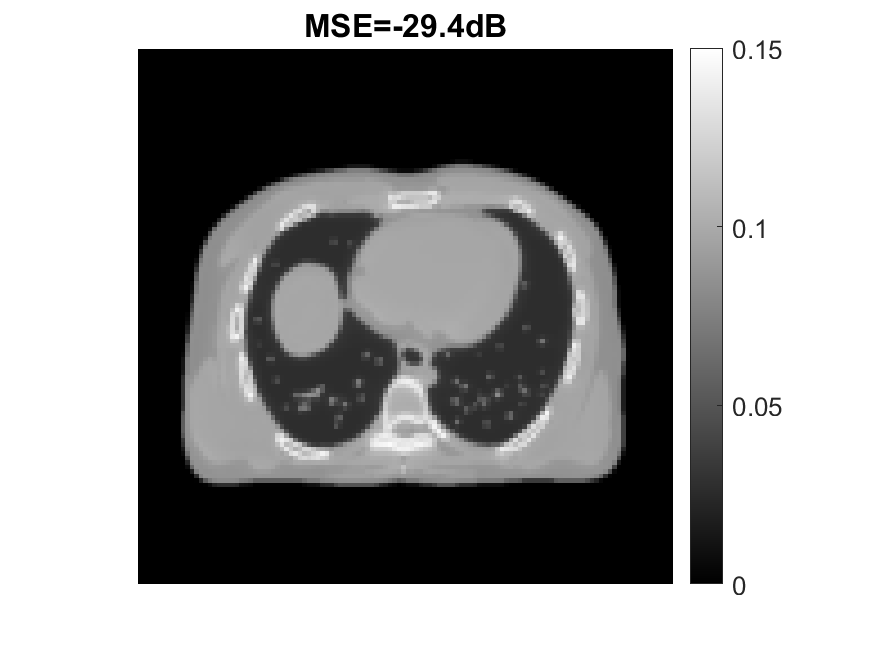}\\
	\includegraphics[trim=2.5cm 1cm 3.2cm 0.0cm, clip, height=3.5cm]{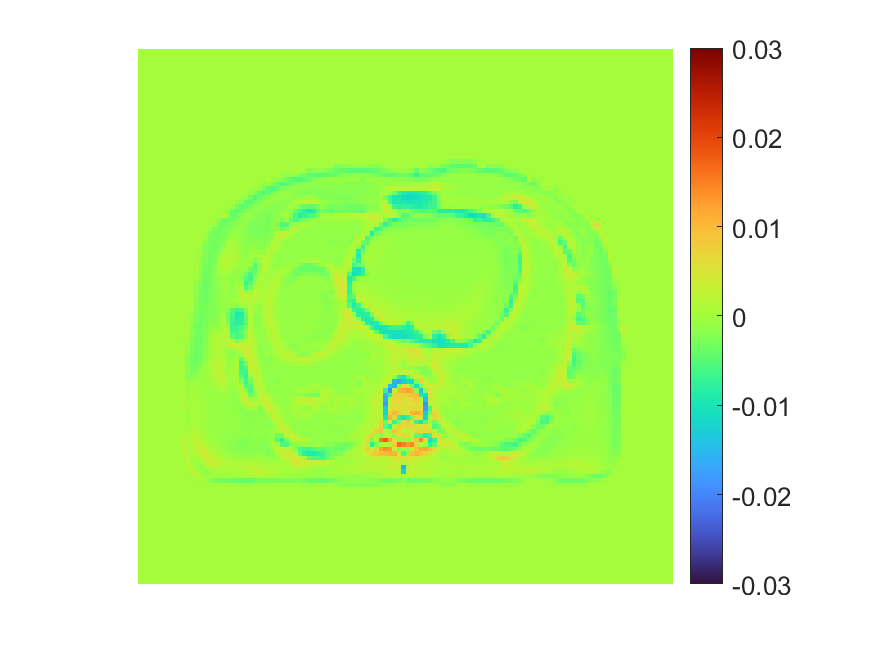}\\
\end{minipage}%
}%
\caption{gCT images (top) by different reconstruction algorithms and their corresponding error images (bottom). (a) Ground truth, (b) MLAA, (c) KAA, (d) \txtr{CDIP}, and (e) proposed neural KAA.}
\label{gCT image}
\end{figure*}
\begin{figure*}[h]
\centering
\subfloat[]{\includegraphics[trim=0.5cm -0.4cm 0.3cm 0.6cm, clip,height=3.85cm]{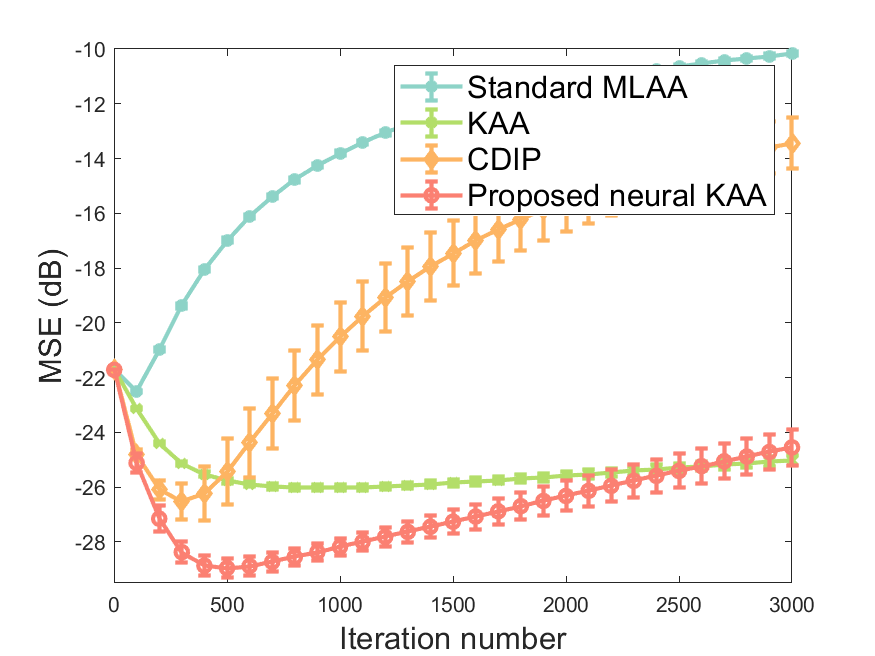}}
\subfloat[]{\includegraphics[trim=0.3cm 0cm 0.3cm 0.6cm, clip, height=3.8cm]{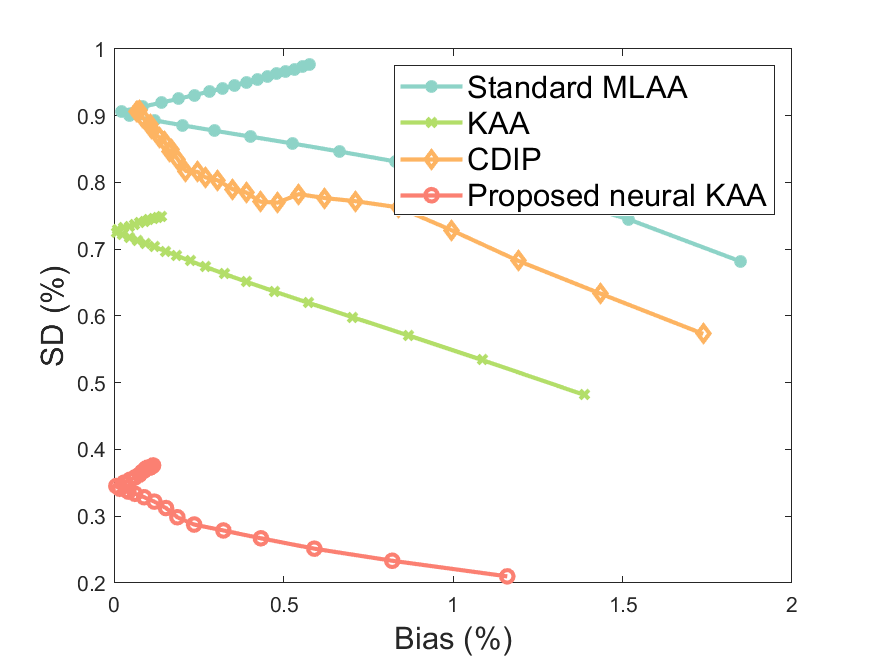}}
\subfloat[]{\includegraphics[trim=0.3cm 0cm 0.3cm 0.6cm, clip, height=3.8cm]{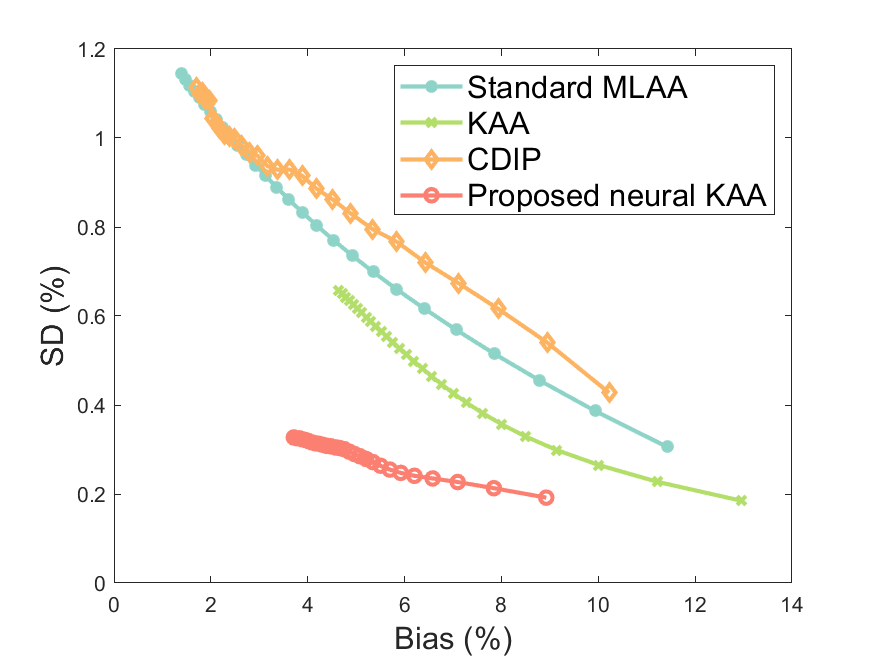}}
\caption{Quantitative comparison of different reconstruction algorithms for gCT image quality. (a) Plot of gCT image MSE as a function of iteration number; (b-c): Plot of bias versus SD trade-off for gCT image quantification in (b) a liver ROI and (c) a bone ROI.}
\label{gCT-plots}
\end{figure*}

\begin{figure*}[h]
\centering
\subfloat[]{
\begin{minipage}[t]{0.22\linewidth}
	\centering
	\includegraphics[trim=2.5cm 0.5cm 1.5cm 0cm, clip, height=3.5cm]{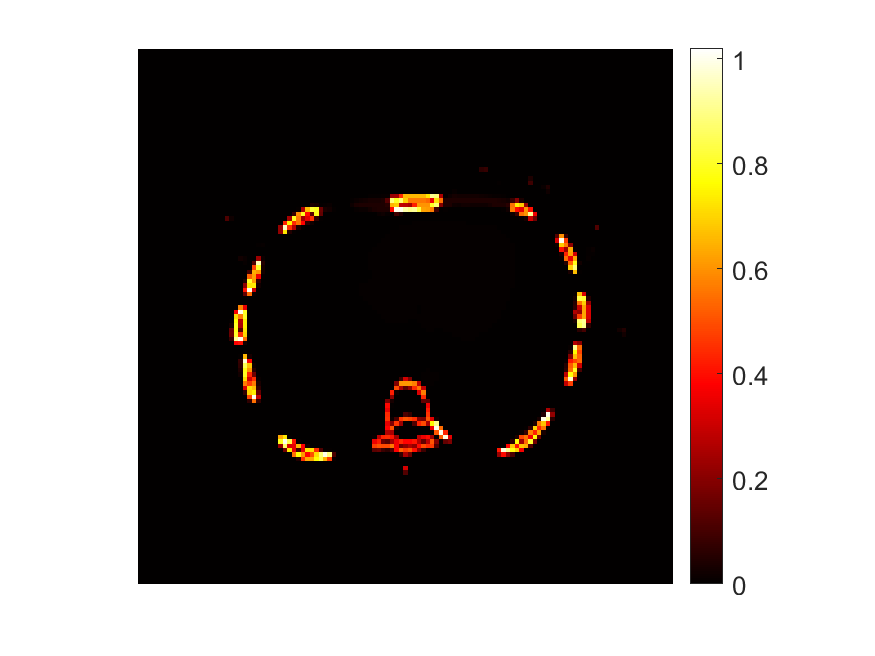}\\
	\includegraphics[trim=2.5cm 0.5cm 1.5cm 0cm, clip, height=3.5cm]{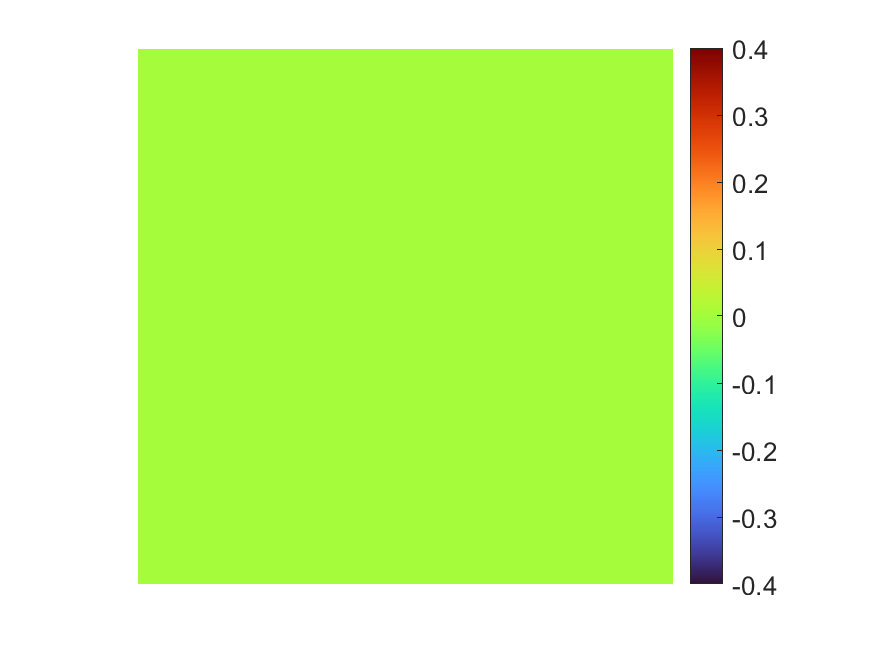}\\
	\includegraphics[trim=2.5cm 0.5cm 1.5cm 0cm, clip, height=3.5cm]{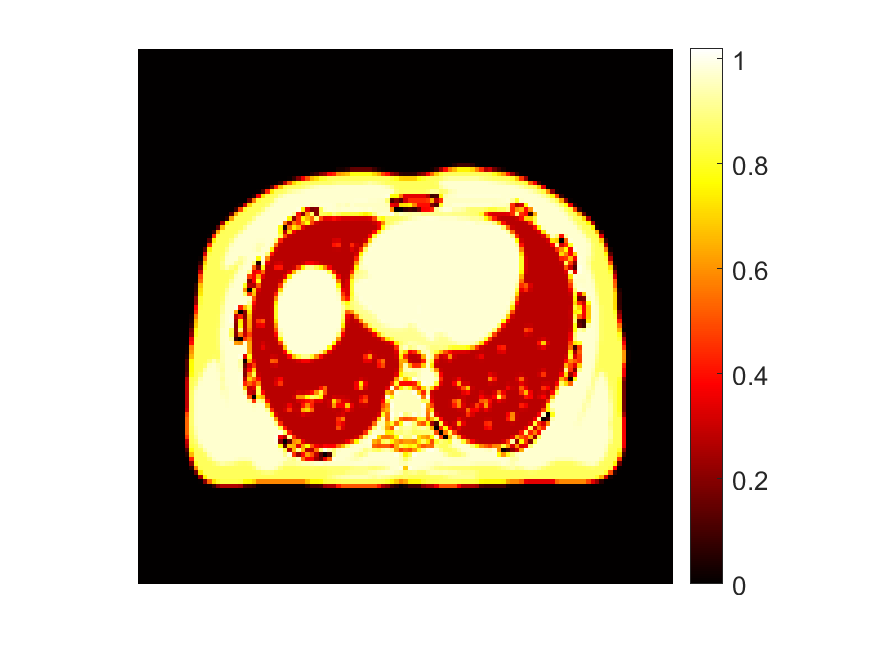}\\
	\includegraphics[trim=2.5cm 0.5cm 1.5cm 0cm, clip, height=3.5cm]{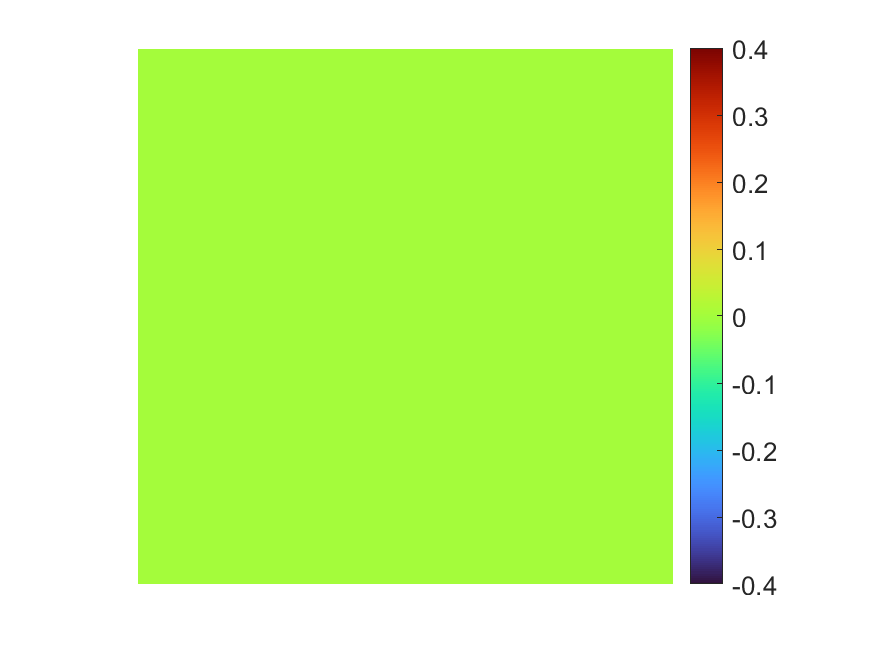}\\
\end{minipage}%
}%
\subfloat[]{
\begin{minipage}[t]{0.18\linewidth}
	\centering
	\includegraphics[trim=2.5cm 0.5cm 3.2cm 0cm, clip, height=3.5cm]{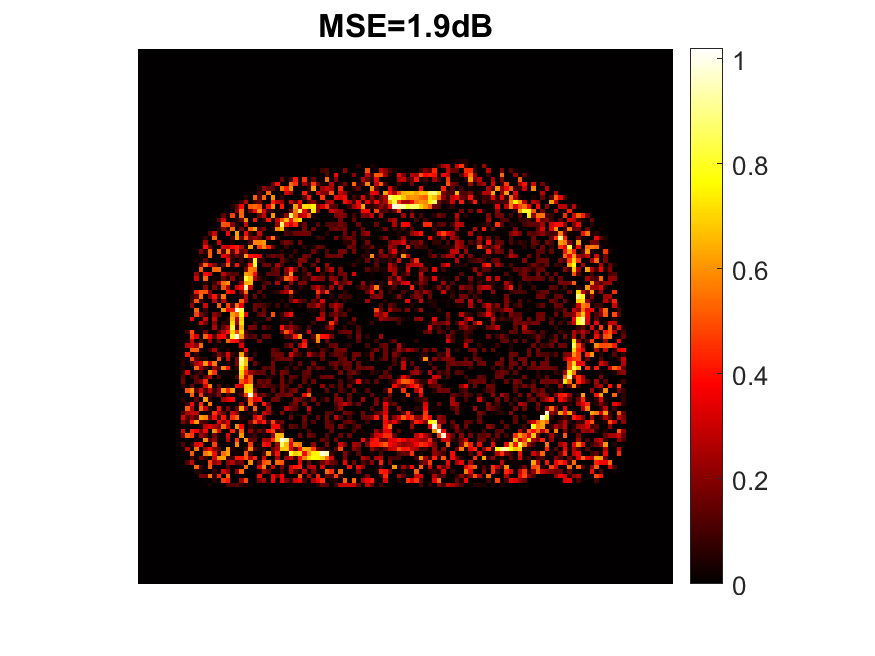}\\
	\includegraphics[trim=2.5cm 0.5cm 3.2cm 0cm, clip, height=3.5cm]{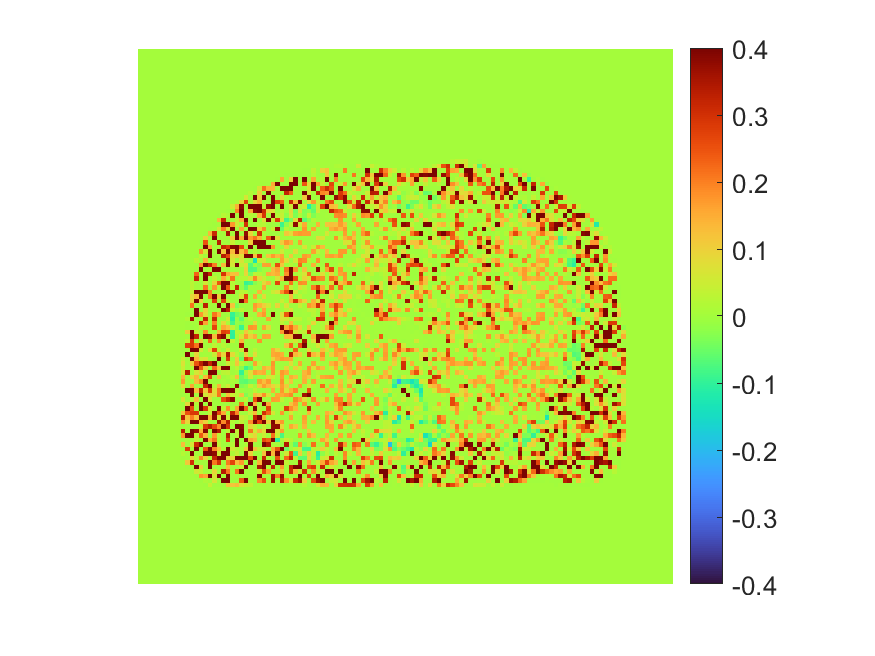}\\
	\includegraphics[trim=2.5cm 0.5cm 3.2cm 0cm, clip, height=3.5cm]{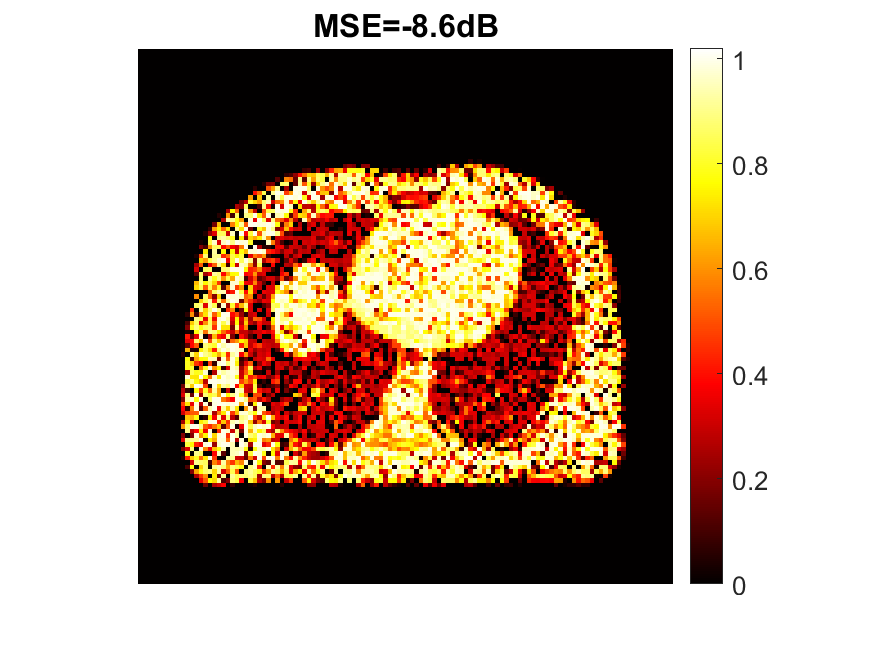}\\
	\includegraphics[trim=2.5cm 0.5cm 3.2cm 0cm, clip, height=3.5cm]{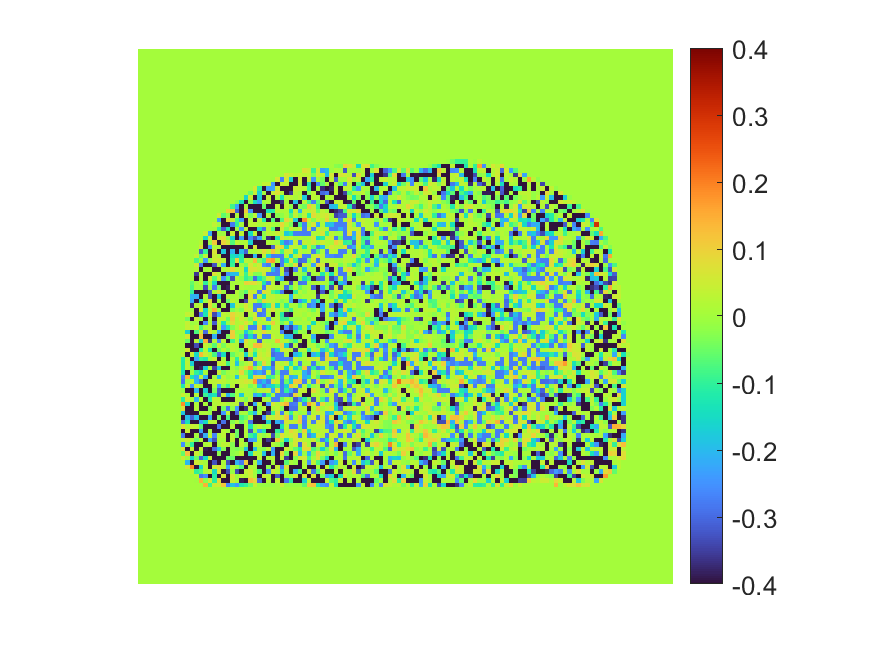}\\
\end{minipage}%
}%
\subfloat[]{
\begin{minipage}[t]{0.18\linewidth}
	\centering
	\includegraphics[trim=2.5cm 0.5cm 3.2cm 0cm, clip, height=3.5cm]{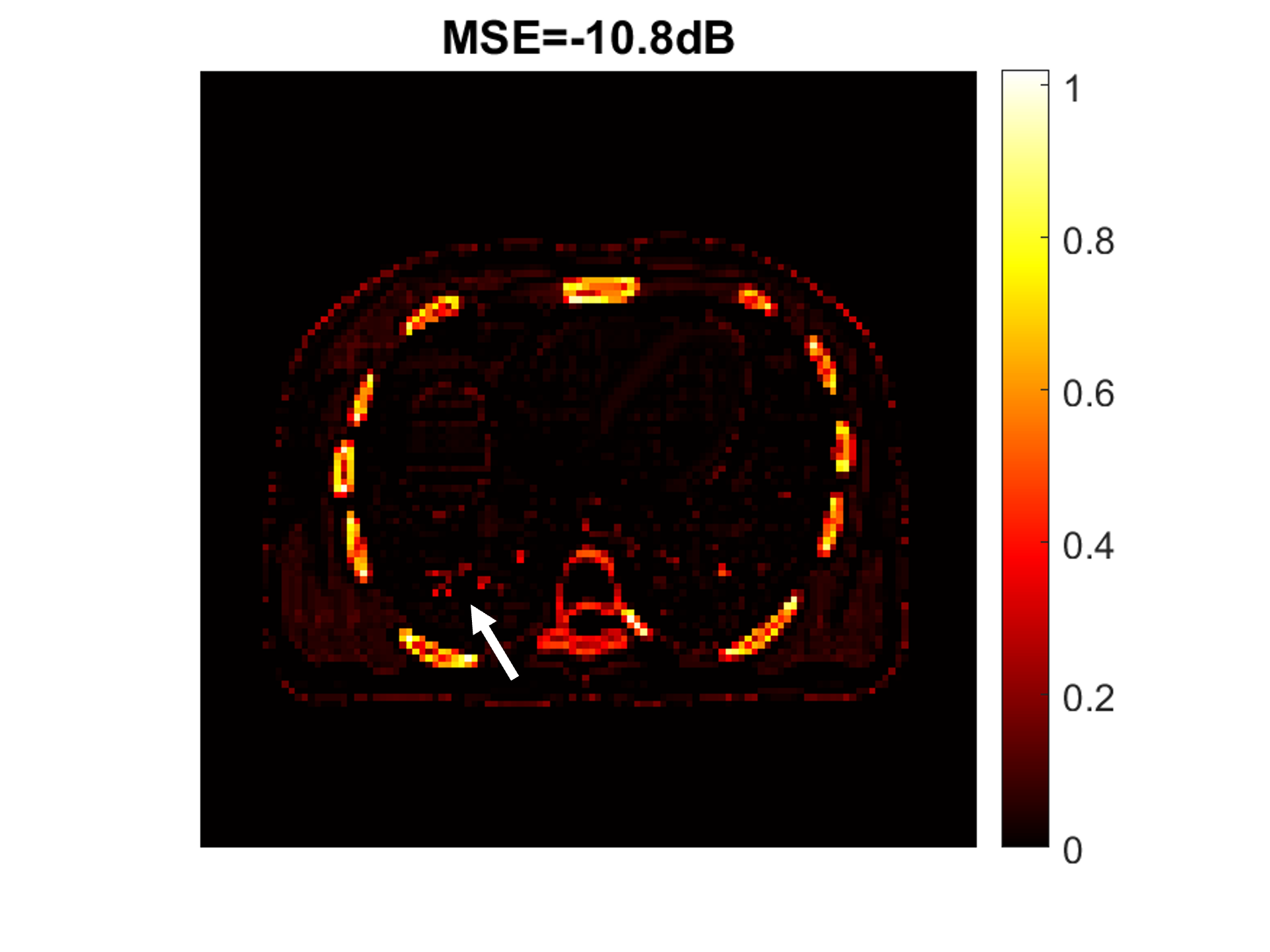}\\
	\includegraphics[trim=2.5cm 0.5cm 3.2cm 0cm, clip, height=3.5cm]{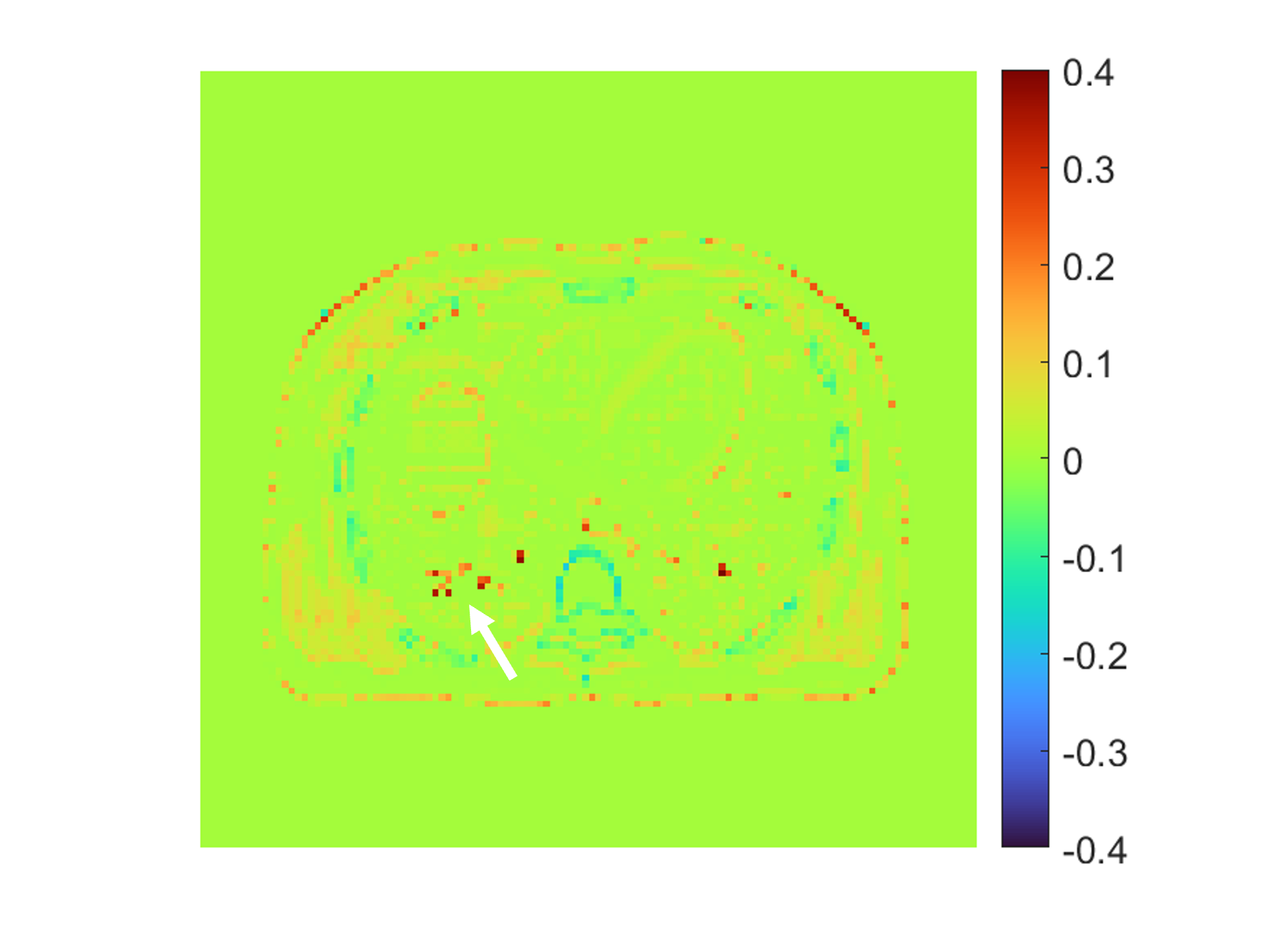}\\
	\includegraphics[trim=2.5cm 0.5cm 3.2cm 0cm, clip, height=3.5cm]{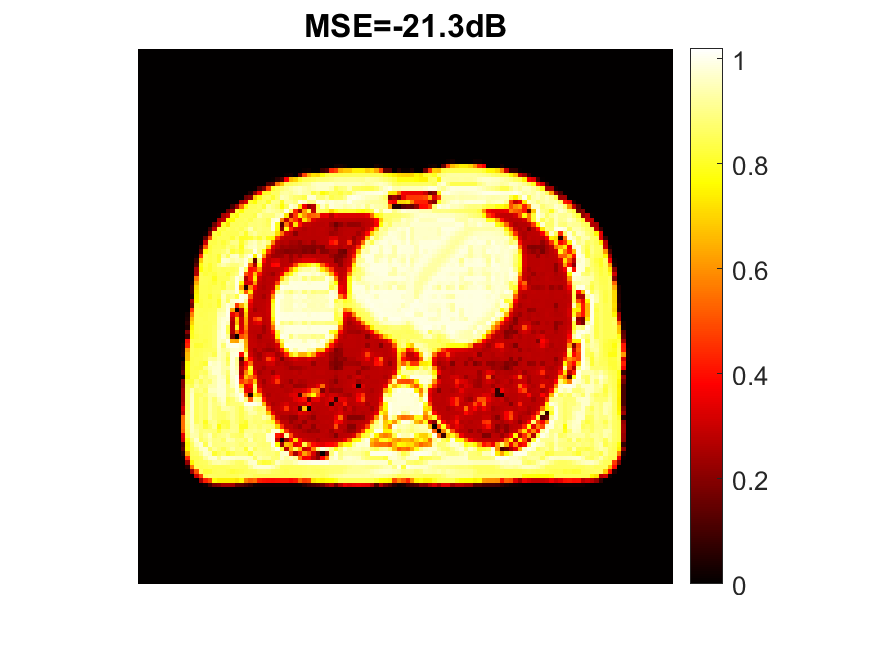}\\
	\includegraphics[trim=2.5cm 0.5cm 3.2cm 0cm, clip, height=3.5cm]{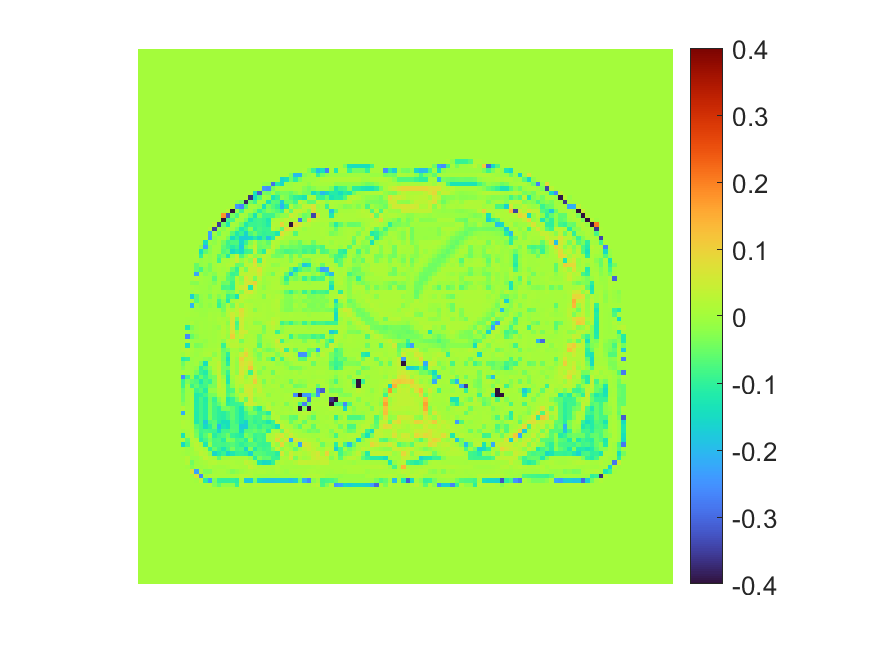}\\
\end{minipage}%
}%
\subfloat[]{
\begin{minipage}[t]{0.18\linewidth}
	\centering
	\includegraphics[trim=2.5cm 0.5cm 3.2cm 0cm, clip, height=3.5cm]{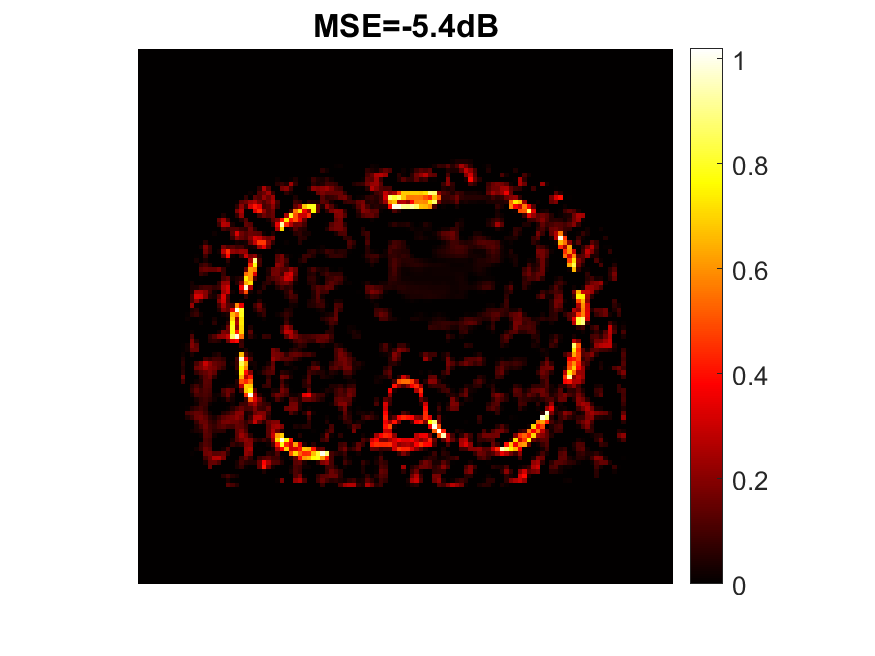}\\
	\includegraphics[trim=2.5cm 0.5cm 3.2cm 0cm, clip, height=3.5cm]{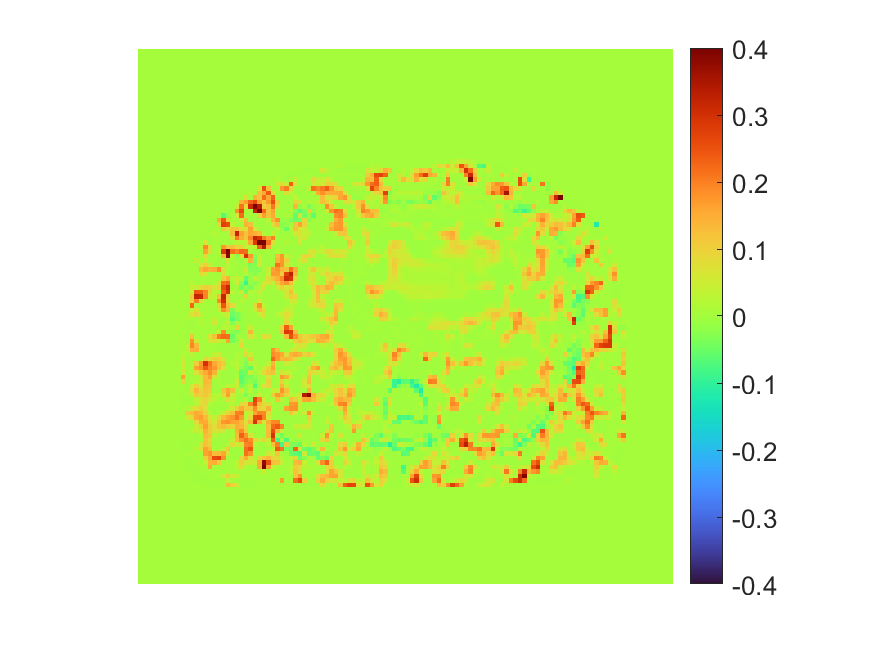}\\
	\includegraphics[trim=2.5cm 0.5cm 3.2cm 0cm, clip, height=3.5cm]{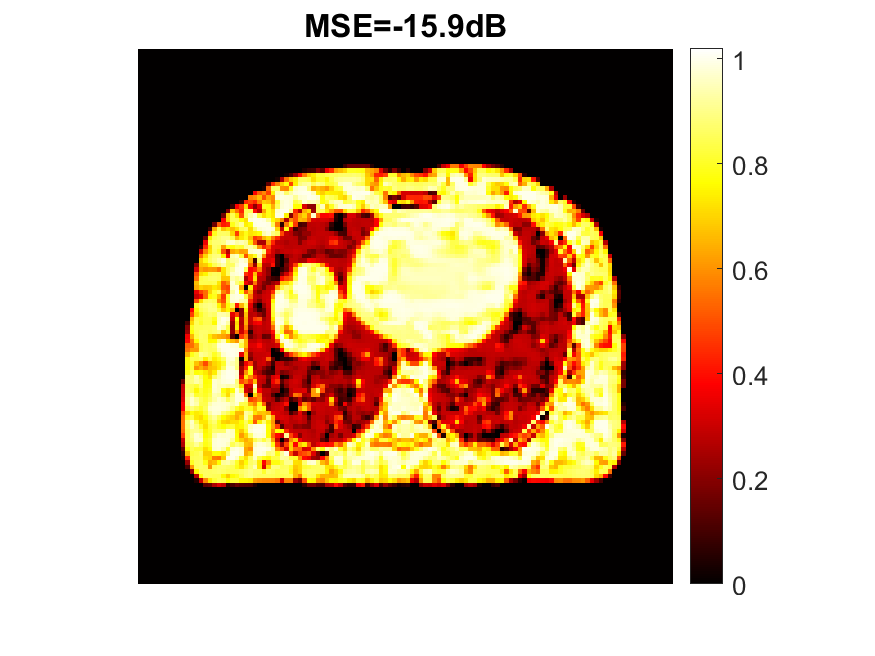}\\
	\includegraphics[trim=2.5cm 0.5cm 3.2cm 0cm, clip, height=3.5cm]{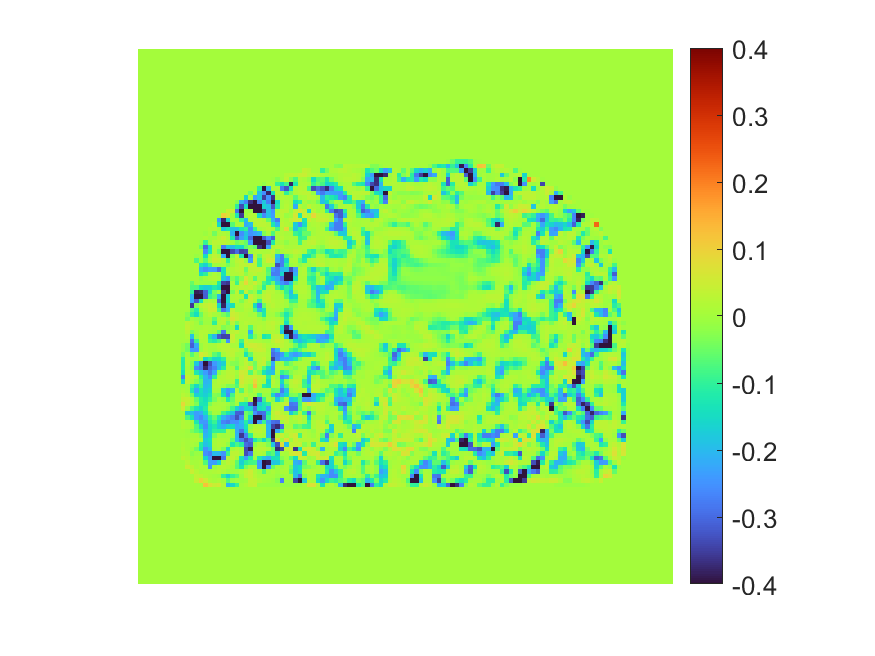}\\
\end{minipage}%
}%
\subfloat[]{
\begin{minipage}[t]{0.18\linewidth}
	\centering
	\includegraphics[trim=2.5cm 0.5cm 3.2cm 0cm, clip, height=3.5cm]{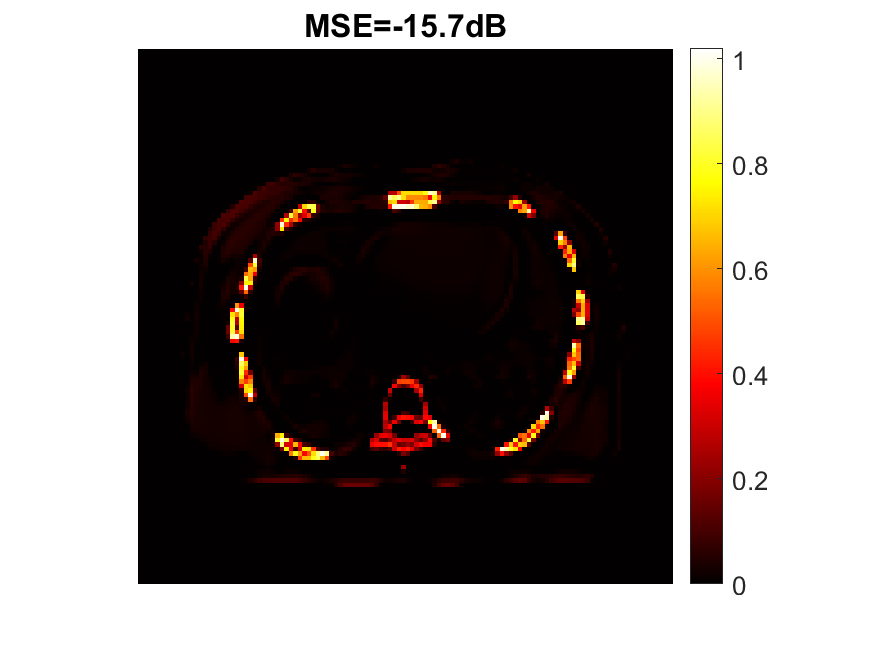}\\
	\includegraphics[trim=2.5cm 0.5cm 3.2cm 0cm, clip, height=3.5cm]{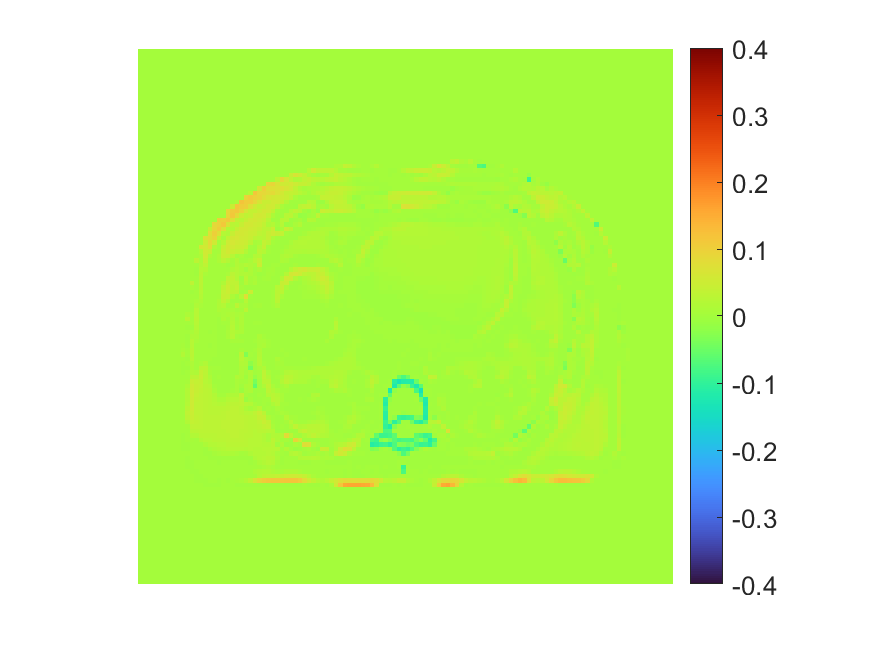}\\
	\includegraphics[trim=2.5cm 0.5cm 3.2cm 0cm, clip, height=3.5cm]{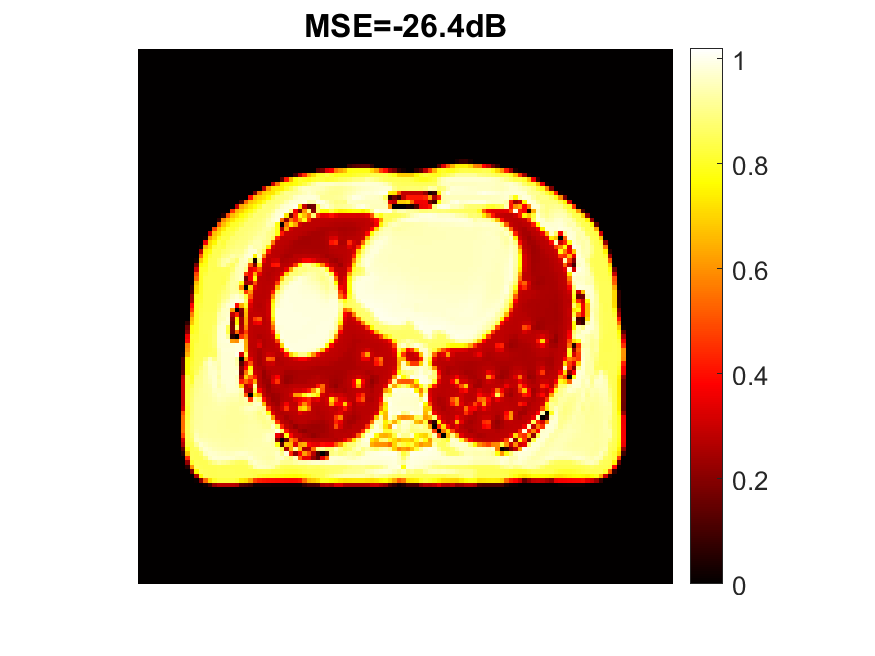}\\
	\includegraphics[trim=2.5cm 0.5cm 3.2cm 0cm, clip, height=3.5cm]{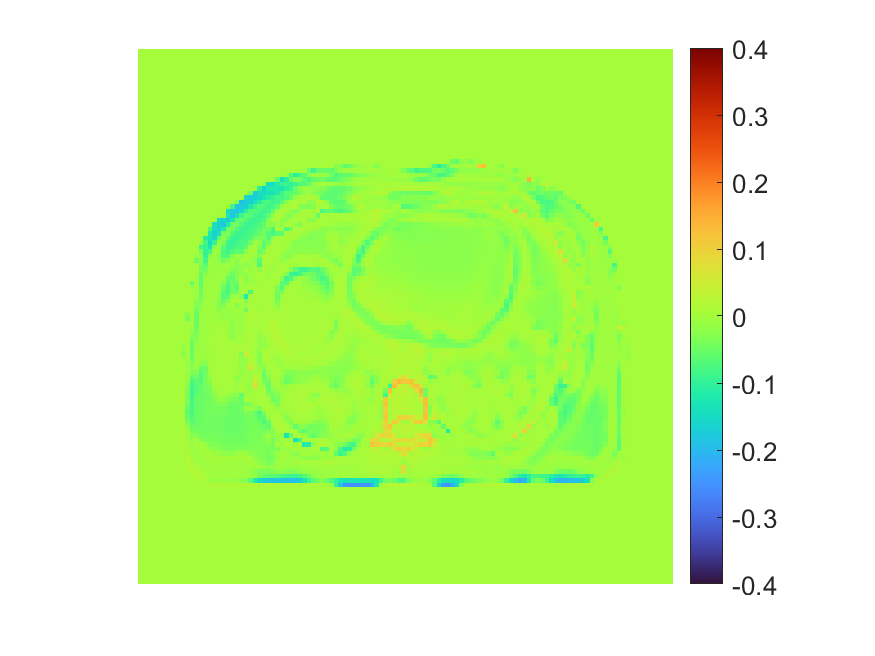}\\
\end{minipage}%
}%
\caption{True and estimated fractional images of two basis materials using different reconstruction algorithms, as well as their corresponding error images: bone (top two rows) and soft tissue (bottom two rows). (a) Ground truth, (b) standard MLAA, (c) KAA, (d) \txtr{CDIP}, and (e) proposed neural KAA.}
\label{MMD}
\end{figure*}

\section{Computer Simulation Studies}
\label{sec5}
\subsection{Simulation Setup}
We first conducted a two-dimensional computer simulation study following the GE Discovery 690 PET/CT scanner geometry. This PET scanner has a time-of-flight (TOF) timing resolution of approximately 550 ps. The simulation was conducted using one chest slice of the XCAT phantom \cite{Segars2008}. Fig. \ref{fig-pht-xcat}a and  Fig. \ref{fig-pht-xcat}b show the simulated ground truth of PET activity image and 511 keV gCT attenuation image, respectively. The low-energy X-ray CT image was simulated from XCAT at 80 keV and is shown in Fig. \ref{fig-pht-xcat}c. The activity and gCT images were first forward projected to generate noise-free emission sinograms with the size of \txtr{281 (radial bins) $\times$ 288 (angular bins) $\times$ 11 (TOF bins)}. A 40\% uniform background was included to simulate random and scattered events. Poisson noise was then generated using 5 million expected events. The projection data was reconstructed into PET activity and gCT images of 180$\times$180 with a pixel size of 3.9$\times$3.9 mm$^2$. Ten noisy realizations were simulated and reconstructed for comparing reconstruction methods.

\subsection{Compared Methods and Implementation Details}
In this work, four types of reconstruction methods were compared, including (1) the standard MLAA \cite{Rezaei2012}, (2) existing KAA \cite{Wang2020}, (3) a \txtr{CDIP} reconstruction method, which is equal to the proposed neural KAA with $\K = \I$, and (4) proposed neural KAA that uses the neural optimization transfer algorithm.

For constructing kernels, the feature vector $\f_j$ was chosen as the pixel intensities of X-ray CT image in a $3\times3$ image patch centered at pixel $j$. The radial Gaussian kernel function, $\kappa(\f_j,\f_l) = \exp (-||\f_j - \f_l||^2/2\sigma^2)$, was used to build the kernel matrix $\K$ using $\sigma = 1$ and $k$NN search with $k = 50$, in the same way as \cite{Wang2020}.


The initial estimate of the PET activity image was set to a uniform image. Following \cite{Wang2020}, we used the X-ray CT-converted 511 keV attenuation map as the initial estimate of gCT image for accelerated convergence. All reconstructions were run for 3000 iterations for investigating the convergence behaviors of different algorithms.

\subsection{Evaluation Metrics}
As the focus of this work is on gCT for DECT, the evlauation of PET activity image quality is not concerned. Different reconstruction algorithms were first compared for gCT image quality using the mean squared error (MSE),
\beq
\rm{MSE}(\hat{\muv}) = 10\log_{10}\left(|| \hat{\muv} - \muv^{\rm{true}}||^2 / ||\muv^{\rm{true}} ||^2\right)~~\rm(dB),
\eeq
where $\hat{\muv}$ represents the reconstructed gCT image by each method and $\muv^{\rm{true}}$ denotes the ground truth. The error image, defined as $\hat{\muv} - \muv^{\rm{true}}$, was used for highlighting the differences in key parts.
The ensemble bias and standard deviation (SD) of the mean intensity in a regions of interest (ROI) were also calculated to evaluate ROI quantification in a liver region and a bone region (Fig. \ref{fig-pht-xcat}(d)),
\beq
\textup{Bias} = \frac{\left|\overline{c} - c^{\rm{true}} \right|}{c^{\rm{true}}},\quad\textup{SD} = \frac{1}{c^{\rm{true}}}\sqrt{\frac{\sum_{i=1}^{N_r}|c_i - \overline{c}|^2}{N_r- 1}},
\eeq
where $c^{\rm{true}}$ is the true average intensity in a ROI and $\overline{c} = \frac{1}{N_r}\sum_{i=1}^{N_r}c_i$ denotes the mean of $N_r$ realizations ($N_r = 10$). 

Different reconstruction algorithms were further compared for DECT multi-material decomposition. Similarly, the image MSE, error image, and ROI-based bias and SD were calculated for each of the material basis fractional images.

\subsection{Neural Optimization Transfer versus Gradient descent}
Fig. \ref{Neural OT} shows the plots of Poisson log-likelihood as a function of iteration number for one simulated data for the proposed neural optimization transfer and gradient descent algorithms. Here the implementation of gradient descent algorithm followed \cite{Baguer2020} with an optimized step size using grid search. While the gradient descent algorithm not surprisingly demonstrates an oscillating behavior \txtr{because of the non-convex nature of the original likelihood function (Eq. \ref{Neural MLTR})}, the proposed neural optimization transfer algorithm shows a monotonic increase in the likelihood function as the iteration increases, which is guaranteed by the theory of optimization transfer.  

\subsection{Comparison for gCT Image Quality}
Fig. \ref{gCT image} shows the true and reconstructed 511-keV gCT images using different algorithms with 400 iterations, as well as corresponding error images. The image MSE results were included for quantitative comparison. \txtr{The selection of 400 iterations was a balance for different methods (KAA, CDIP and the proposed method) to show good image quality according to the MSE performance}. The MLAA reconstruction was noisy. The KAA method substantially improved the gCT image quality, but with a lower contrast in the bone region as compared to the ground truth (as evidenced in its error image). The \txtr{CDIP} method had a slightly better MSE than KAA, but it induced artifacts, which was in turn propagated into the material decomposition results as shown later in Fig. \ref{MMD}.  In comparison, the proposed neural KAA demonstrated the least level of noise with good visual quality, and achieved the lowest MSE among different algorithms.

Fig. \ref{gCT-plots}a shows the MSE plots as a function of iteration number for different algorithms. The iteration number varied from 0 to 3000 with a step of 100 iterations and error bars were calculated over the 10 noisy realizations. Compared to KAA, the proposed neural KAA showed a lower MSE among different methods. The curve also shows that similar to other algorithms, early stopping of the iterations is beneficial for the neural KAA to obtain good image quality while keep computational efficiency.

Fig. \ref{gCT-plots}b and Fig. \ref{gCT-plots}c further show the comparison of ensemble bias versus SD for gCT quantification in a liver region and a bone region. The curves were obtained by varying the iteration number from 300 to 3000 iterations with an interval of 100 iterations. As iteration number increases, the bias of ROI quantification is reduced while the SD increases. At a comparable bias level, the proposed neural KAA had a lower noise SD than the other three methods.

\begin{figure}[t]
\centering
\subfloat[]{\includegraphics[trim=0.3cm 0cm 1cm 0.6cm, clip, width=4.5cm]{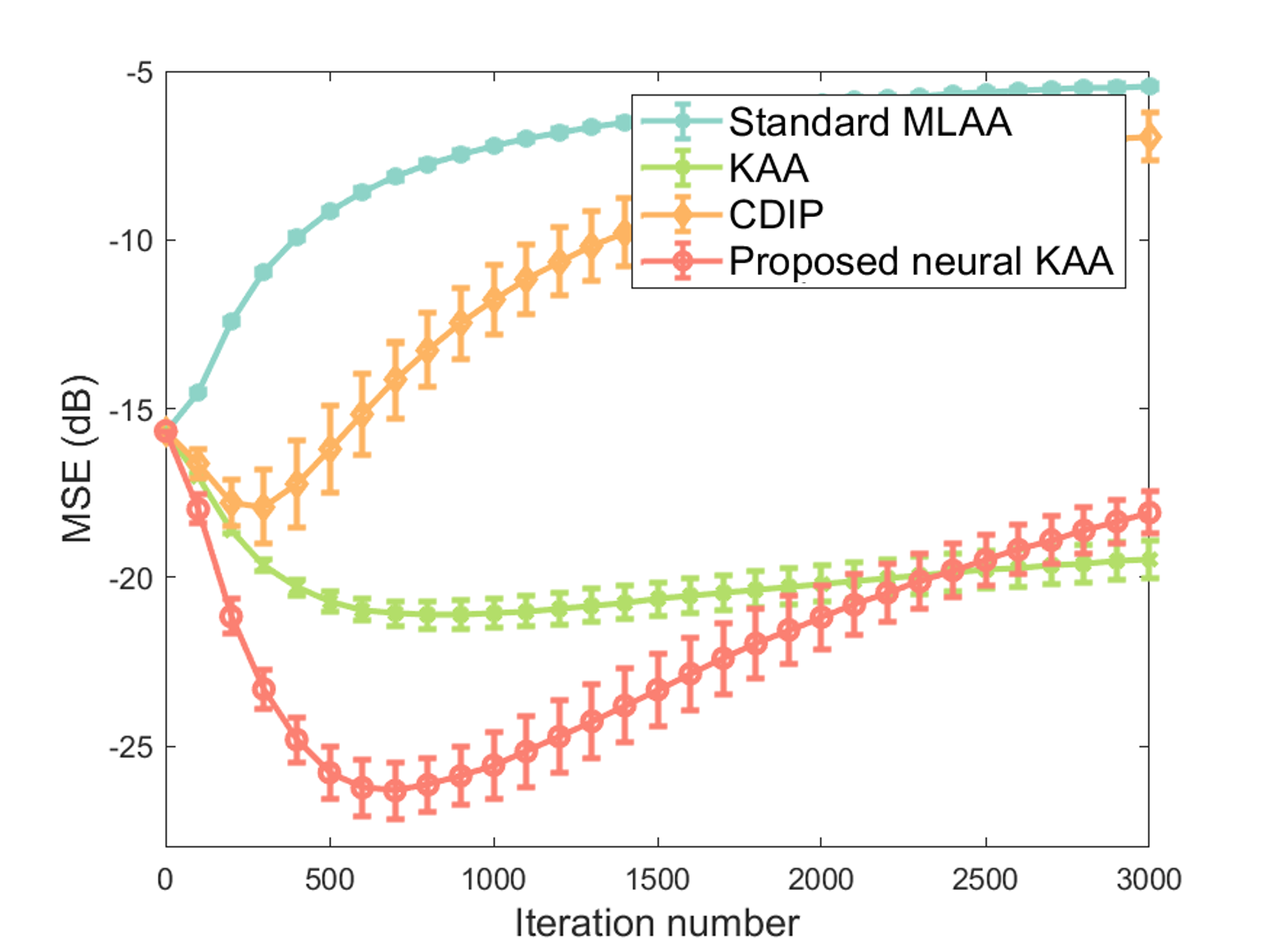}}
\subfloat[]{\includegraphics[trim=0.3cm 0cm 1cm 0.6cm, clip, width=4.5cm]{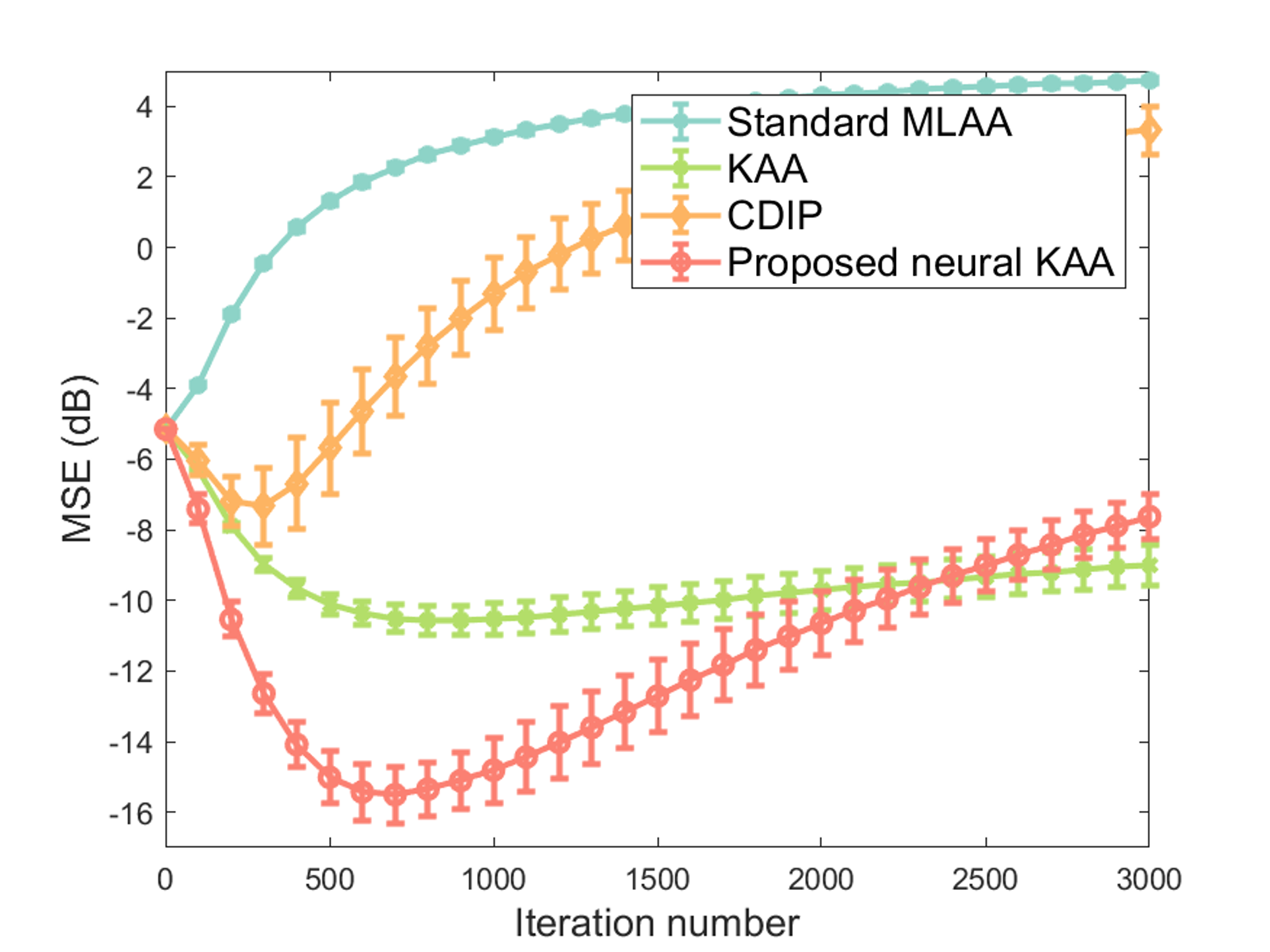}}
\caption{Plot of image MSE as a function of iteration number for (a) soft-tissue fractional image and (b) bone fractional image.}
\label{MMD-MSE}
\end{figure}

\begin{figure}[t]
\centering
\subfloat[]{\includegraphics[trim=0.3cm 0cm 1cm 0.6cm, clip, width=4.5cm]{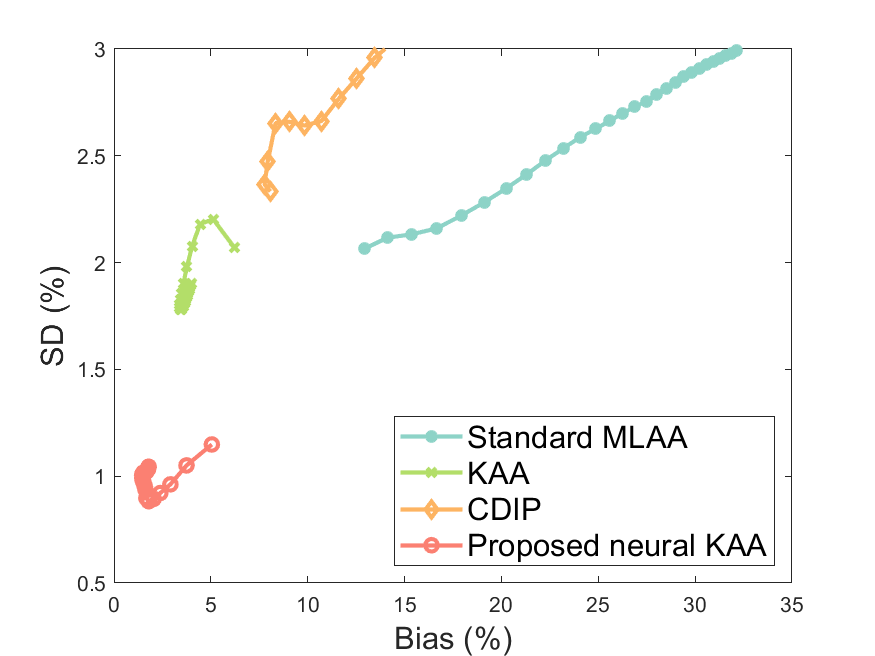}}
\subfloat[]{\includegraphics[trim=0.3cm 0cm 1cm 0.6cm, clip, width=4.5cm]{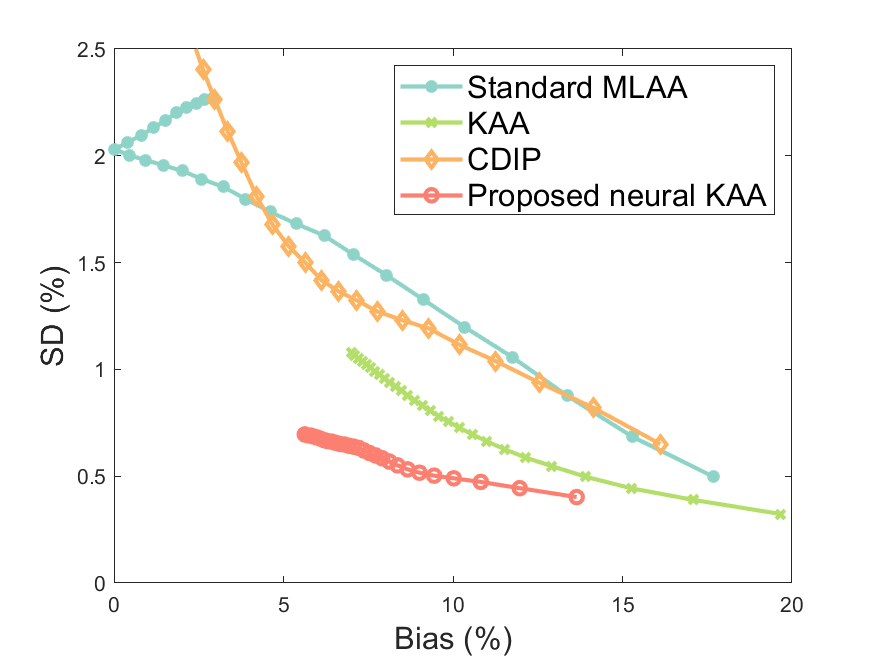}}
\caption{Bias versus SD trade-off for ROI quantification on the fractional image of (a) soft tissue and (b) bone basis materials.}
\label{MMD-ROI}
\end{figure}
\vspace{-5pt}
\subsection{Comparison for Material Decomposition}
We also conducted a comparison of different reconstruction methods for multi-material decomposition (MMD). Fig. \ref{MMD} shows the fractional basis images of bone and soft tissue obtained from MMD of the PET-enabled DECT images with 400 iterations, as well as their corresponding error images. The ground truth of the soft tissue and bone bases was generated using the noise-free pair of low-energy X-ray CT image \txtr{($\x$ in Eq. (\ref{MMD-Eq}))} and the 511 keV gCT image. The conventional KAA method outperformed MLAA but were still with noise artifacts, as pointed by arrows. Interestingly, even though the \txtr{CDIP} reconstruction had a better MSE for gCT images than KAA, the benefit did not propagate into the MMD basis images. The proposed neural KAA achieved less noise and artifacts than KAA thanks to the CDIP-based regularization on the kernel coefficient image $\alp$.

Fig. \ref{MMD-MSE} further shows image MSE as a function of iteration number for each basis fractional image. Again, the proposed neural KAA demonstrated the lower minimum MSE result in each basis image among different reconstruction methods.

Fig. \ref{MMD-ROI} shows the quantitative comparisons of ensemble bias versus SD for ROI quantification on the soft tissue (a) and bone (b) fractional images by varying the reconstruction iteration number. Similar to the results of gCT ROI quantification, the neural KAA achieved the lowest noise level at a comparable bias level. It is noticeable that for the bone ROI quantification, both the standard KAA and neural KAA showed a bias when compared to MLAA. The bias was propagated from the gCT reconstruction (as shown in Fig. \ref{gCT-plots}c). 
\begin{figure}[t]
\centering
{\includegraphics[trim=0.4cm 0cm 1cm 0.5cm, clip,width=2in]{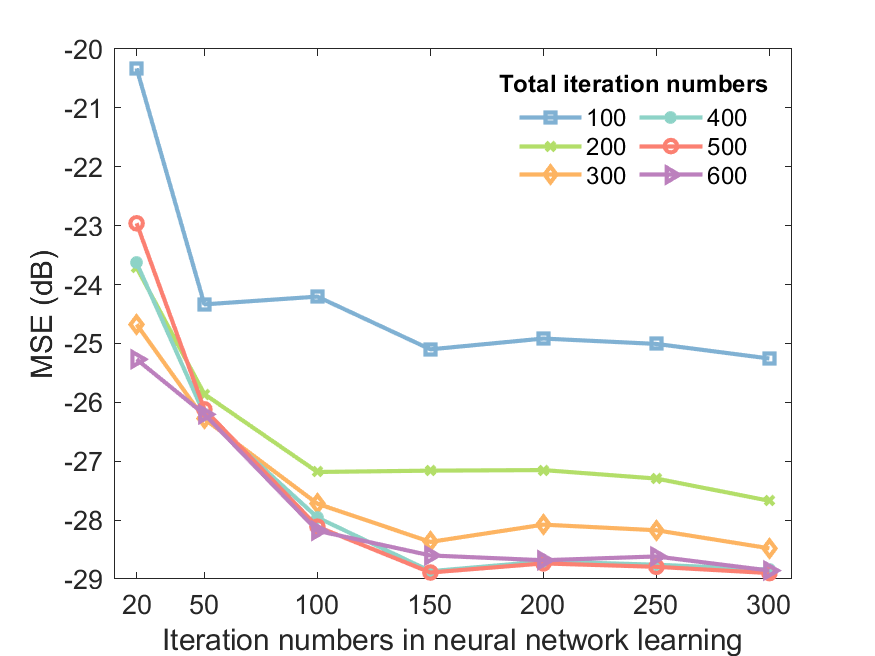}}
\caption{\txtr{Effect of sub-iteration numbers of neural network learning on the gCT image MSE under different total outer-iteration numbers of the neural KAA.}}
\label{MSE-sub}
\end{figure}
\vspace{-5pt}
\subsection{Investigation of Neural Network Learning Settings}
Our experiments indicated that neural network learning is stable when the learning rate in the Adam optimizer ranges from $10^{-4}$ to $10^{-2}$. A larger rate may make the learning difficult to converge, while a smaller rate may reduce the convergence rate. The sub-iteration number used in the least-square neural network learning may also have influenced the results. \txtr{Fig. \ref{MSE-sub} shows the effect of this sub-iteration number of neural network learning and total outer-iteration number of the neural KAA on the gCT image MSE. The results suggest the algorithm becomes relatively stable after 150 sub-iterations under different total outer-iterations of the neural KAA.}

\begin{figure}[h]
\centering
\includegraphics[trim=0cm 0cm 0cm 0cm, clip,width=1.7in]{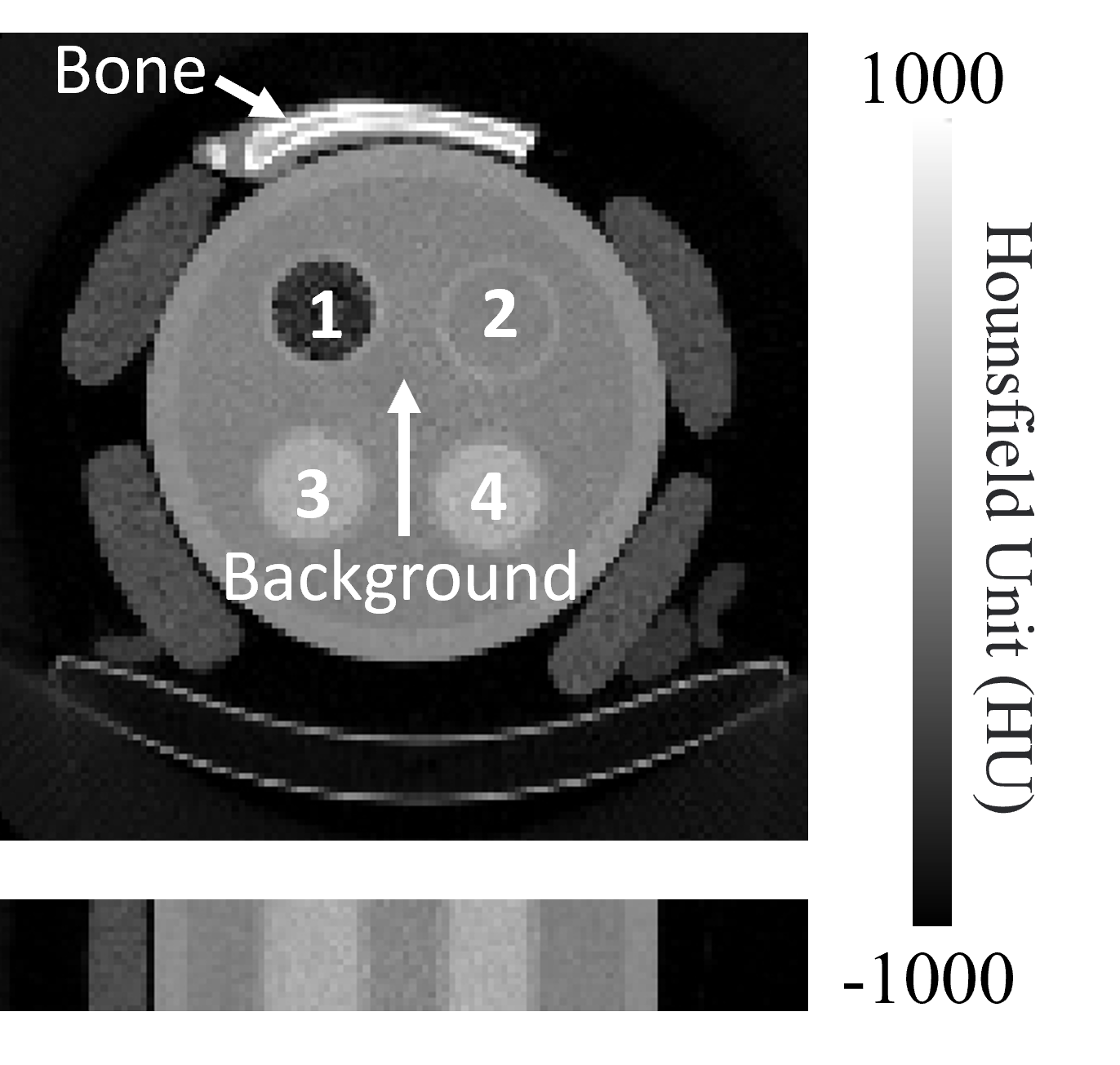}
\caption{Transverse (top) and coronal (bottom) slices of the 80 kVp X-ray CT image.}
\label{CT-80}
\vspace{-15pt}
\end{figure}
\begin{figure*}[t]
\centering
\includegraphics[trim=0cm 0cm 0cm 0cm, clip,width=6in]{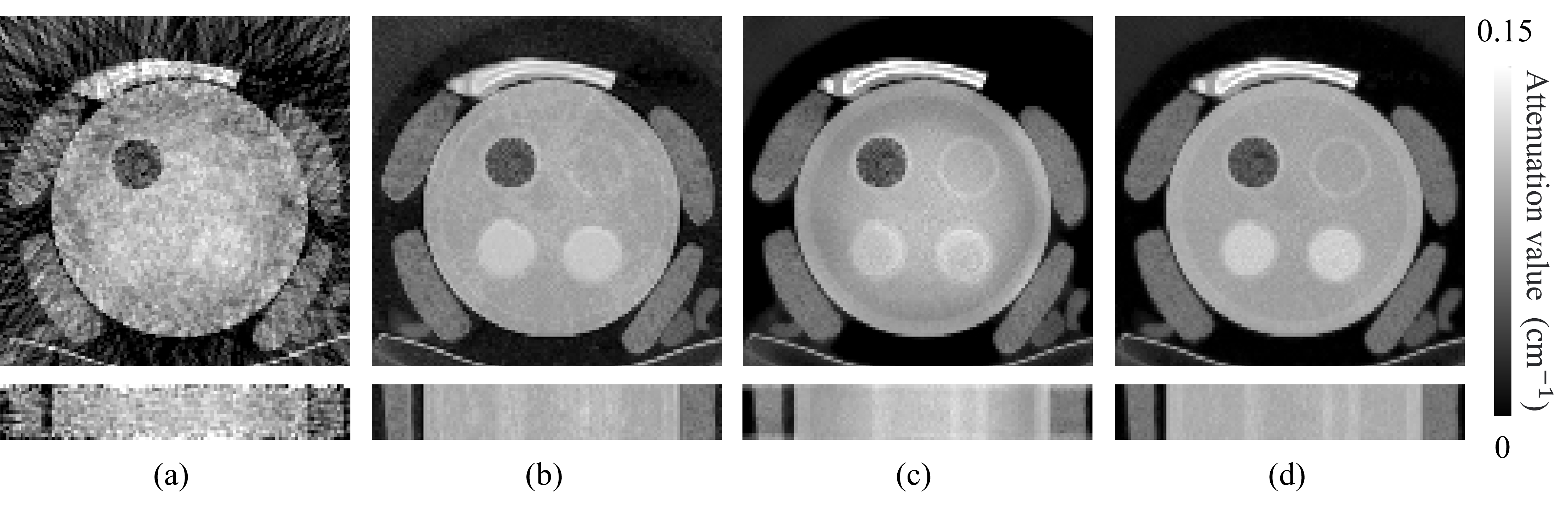}
\caption{Reconstructed 3D gCT images for the real phantom data using (a) MLAA, (b) KAA, (c) \txtr{CDIP}, and (d) proposed neural KAA. Each reconstruction is shown in the transverse (top) and coronal (bottom) views.}
\label{gCT-phantom}	
\end{figure*}
\begin{figure*}[h]
\centering
\includegraphics[trim=0cm 0cm 0cm 0cm, clip,width=6.5in]{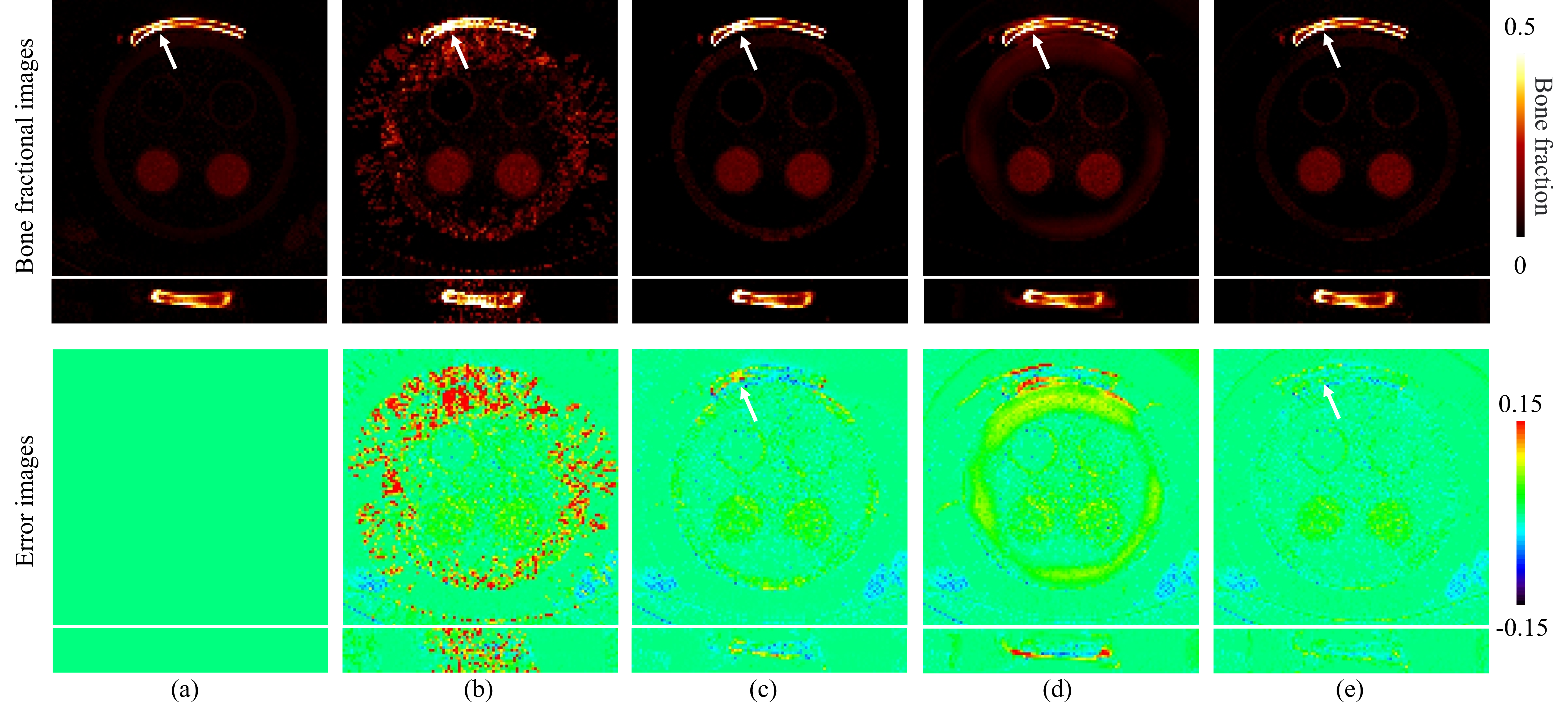}
\caption{Estimated bone fractional images and error images from (a) X-ray DECT images and (b-e) PET-enabled DECT reconstructed with (b) MLAA, (c) KAA, (d) \txtr{CDIP}, and (e) proposed neural KAA. Each 3D image is shown in the transverse and coronal views.}
\label{MMD-phantom}	
\end{figure*}

\begin{figure}[h]
\centering
\includegraphics[trim=0.0cm 0cm 1cm 0cm, clip,width=2in]{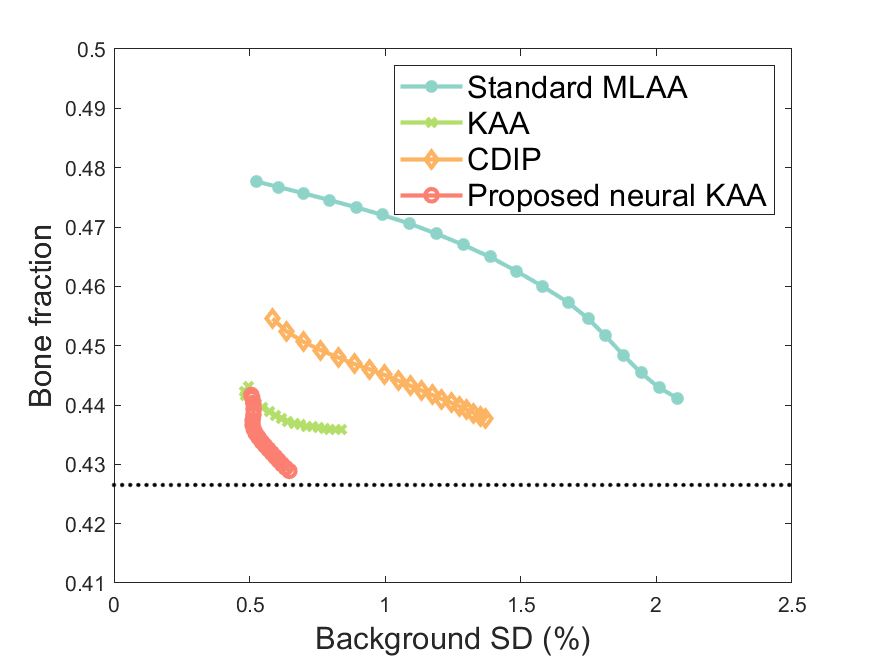}
\caption{Plot of bone fraction versus background noise for a bone ROI in the bone fractional image by varying iteration number from 40 to 400. Dash line denotes the bone fraction obtained from X-ray DECT-based bone basis image.}
\label{ROI_p}
\vspace{-10pt}
\end{figure}

\begin{figure*}[t]
\centering
\includegraphics[trim=0cm 0cm 0cm 0cm, clip,width=6in]{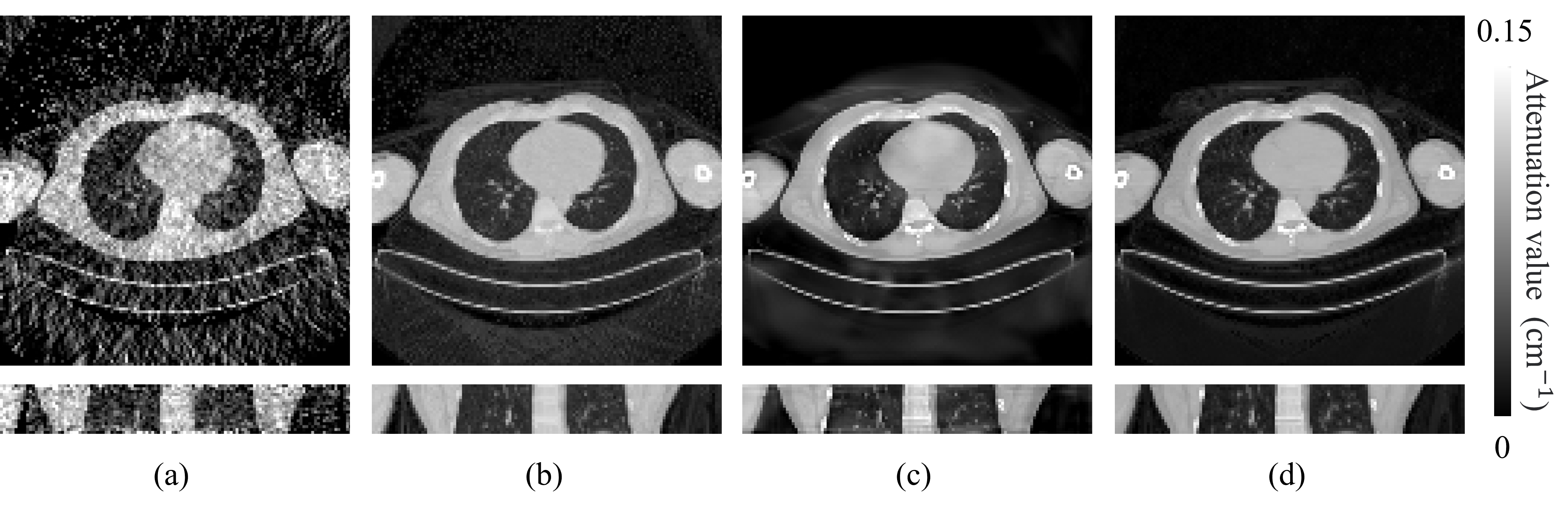}
\caption{\txtr{Reconstructed gCT images for the patient data using (a) MLAA, (b) KAA, (c) \txtr{CDIP}, and (d) proposed neural KAA. Each reconstruction is shown in the transverse (top) and coronal (bottom) views.}}
\label{gCT-real}	
\end{figure*}
\begin{figure}[h]
\centering
\includegraphics[trim=0.0cm 0cm 1cm 0cm, clip,width=2in]{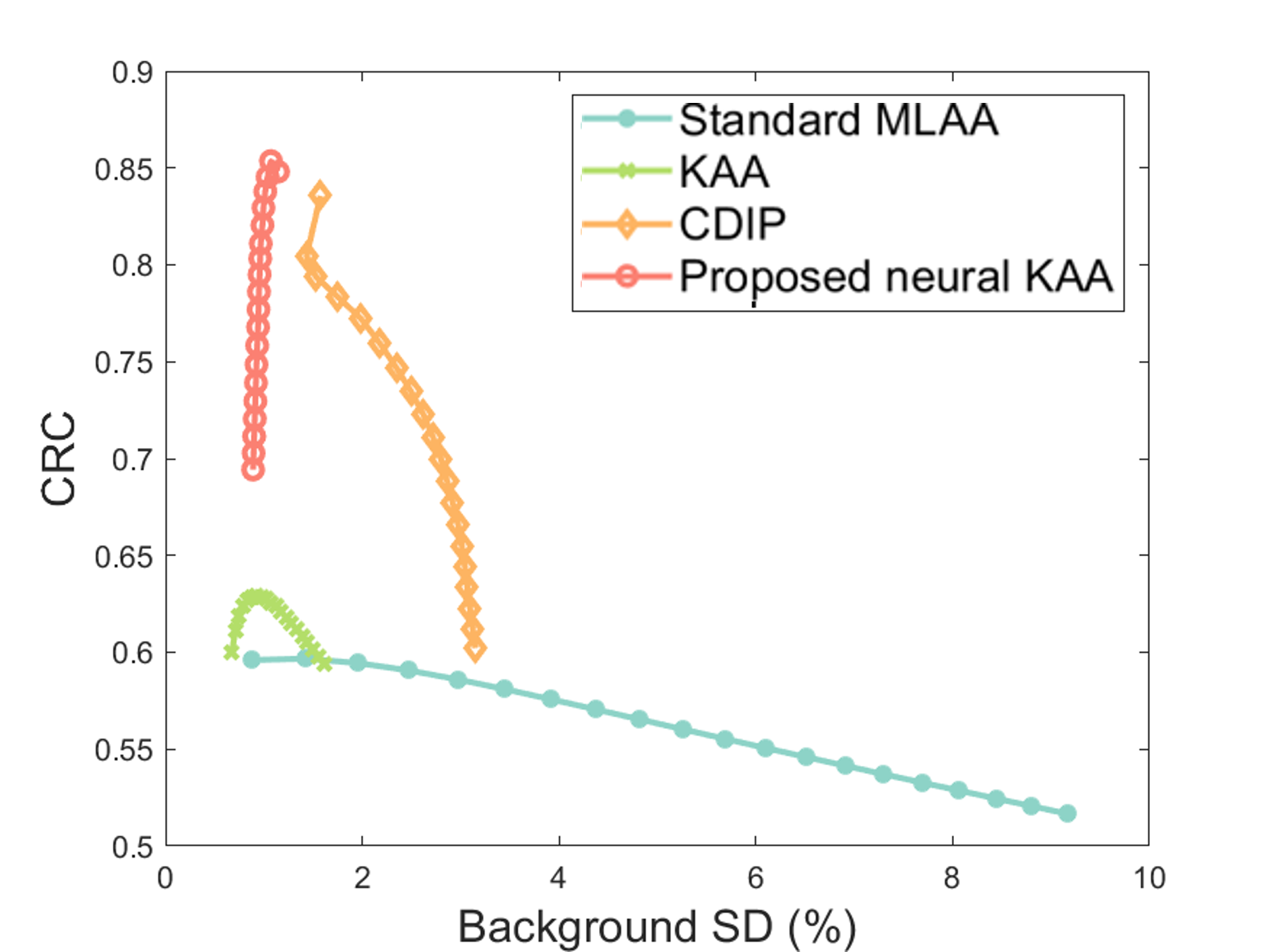}
\caption{\txtr{Plot of CRC versus background noise for a bone ROI in the gCT image by varying iteration number from 40 to 400.}}
\label{ROI_c}
\vspace{-10pt}
\end{figure}
\begin{figure*}[h]
\centering
\includegraphics[trim=0cm 0cm 0cm 0cm, clip,width=6in]{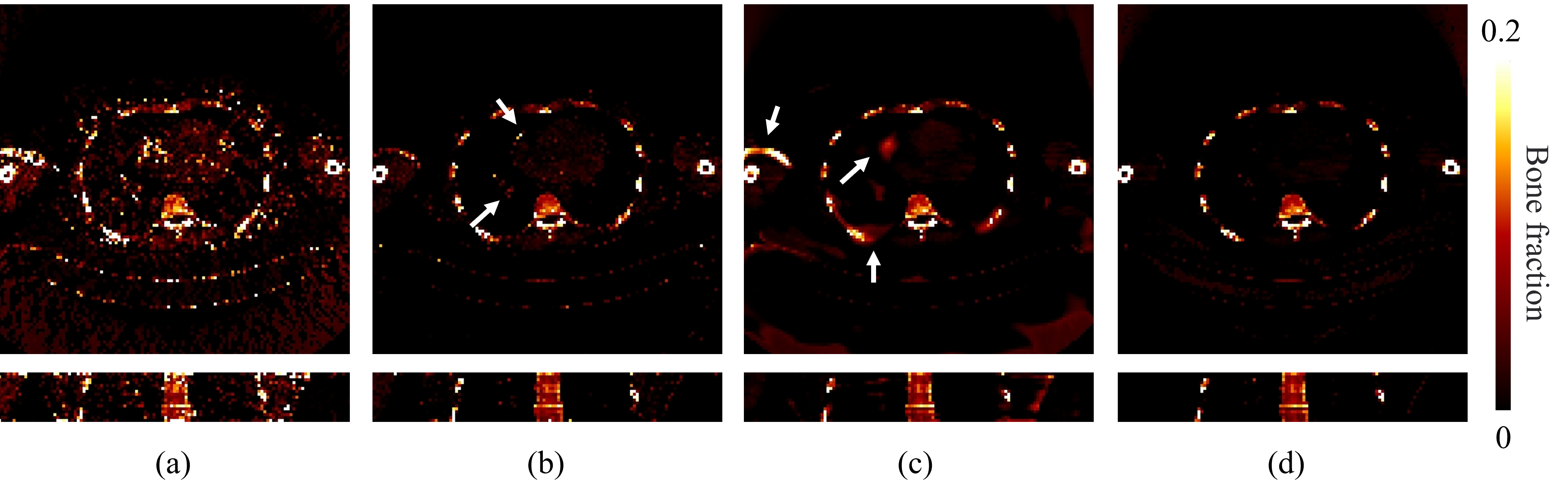}
\caption{\txtr{Estimated bone fractional images from PET-enabled DECT reconstructed with (a) MLAA, (b) KAA, (c) \txtr{CDIP}, and (d) proposed neural KAA. Each 3D image is shown in the transverse and coronal views.}}
\label{MMD-real}	
\end{figure*}

\section{Evaluation on Real Phantom Data}
\label{sec6}
\subsection{Phantom Data Acquisition}
We have further evaluated different reconstruction methods using a real three-dimensional (3D) phantom scan on the uEXPLORER PET/CT scanner \cite{Spencer2021} at UC Davis. This phantom \cite{Zhu2022} was filled with water in the background and four inserts were filled with (1) lung tissue equivalent material, (2) water, and (3\&4) salt water, as shown in Fig. \ref{CT-80}. Other attenuation materials composed of fat tissue-equivalent materials and bovine rib bones were wrapped around the phantom. A $^{18}$F-FDG solution was uniformly filled in all five compartments. Two X-ray CT scans, one at 80 kVp and the other at 140 kVp, were acquired to provide the reference for MMD analysis. In this study, the reconstructions were performed on a truncated dataset of 2-min scan duration and 7 cm axial length to reduce computational time. \txtr{The projection data were acquired with 27 TOF bins for a TOF resolution of 505ps \cite{Spencer2021}. The 3D sinogram for each TOF bin was of 533 radial bins, 420 angular bins, and 576 direct and oblique detector planes.} The reconstructed gCT image size was $150\times150\times17$ with a voxel size of $4\times4\times4$ mm$^3$. \txtr{For MMD using a PET-enabled DECT image pair, the X-ray CT image was downsampled to match with the gCT voxel size using linear interpolation.}

\subsection{Reconstruction Methods}
We used the 80 kVp CT image as the image prior to generate the kernel matrix $\K$ and the input of the neural network. The settings of the neural network in the \txtr{CDIP} method and proposed neural KAA were the same as in the simulation study except using a 3D version of U-Net to match with the data. All reconstructions were implemented using the CASToR package \cite{Merlin2018} as described in \cite{Zhu2022} and run for 400 iterations. The 80 kVp X-ray CT-converted attenuation map was used as the initial for gCT. 

\subsection{Results}
Fig. \ref{gCT-phantom} shows the gCT images reconstructed using different algorithms with 400 iterations. Similar to the results shown in the simulation, the MLAA was extremely noisy. Both the standard KAA and \txtr{CDIP} methods reduced noise significantly but still contained significant artifacts. In contrast, the proposed neural KAA demonstrated the best visual quality.

Fig. \ref{MMD-phantom} shows the bone fractional images from MMD and corresponding error images using different approaches. Compared to the reference from X-ray DECT, the image by MLAA demonstrated heavy noise. Both the KAA and \txtr{CDIP} methods suppressed the noise, but not without additional noise or artifacts. In comparison, the proposed neural KAA demonstrated the least artifacts and noise in the uniform regions and achieved the most similar bone fraction pattern with the reference image, as pointed by the arrows.

Fig. \ref{ROI_p} further shows a quantitative comparison for ROI quantification of bone fraction. Here the ROI quantification is plotted versus the background noise SD measured in the water region by varying the iteration number from 40 to 400 with an interval of 20 iterations. As the iteration number increases, the estimated bone fraction becomes lower while the SD increases. Compared to the other three methods, the proposed neural KAA was the closest to the X-ray DECT reference (dashed line) and also achieved the lowest background noise level.
\txtr{\section{Evaluation on Real Patient Data}
\label{sec7}
We have also evaluated different reconstruction methods for PET-enabled DECT imaging using a cancer patient scan on the uEXPLORER scanner. The injection dose of $^{18}$F-FDG was around 10 mCi. Similar to the phantom study, we extracted the last 2-min emission scan data for gCT reconstruction. All the other settings were the same with the real phantom experiment.}

\txtr{Fig. \ref{gCT-real} shows the reconstructed gCT images of the patient data using different algorithms with 400 iterations. Again, the MLAA was extremely noisy. Both the standard KAA and \txtr{CDIP} methods yielded the similar noise behavior to the phantom results in Fig. \ref{gCT-phantom}. By contrast, the proposed neural KAA largely overcame the issues and demonstrated a better visual quality, though no ground truth was available.}

\txtr{Fig. \ref{ROI_c} further shows a plot of contrast recovery coefficient (CRC) versus background noise SD for a bone ROI quantification in the gCT image. The CRC was calculated by $CRC=|\overline{B}-\overline{S}|/\overline{S}$, where $\overline{B}$ is the mean intensity in the bone ROI and $\overline{S}$ is the ROI mean of the muscle background. The plot was generated by varying the iteration number from 40 to 400 with an interval of 20 iterations. Compared to other three methods, the proposed neural KAA achieved the best trade-off of CRC versus background noise.}

\txtr{Finally, Fig. \ref{MMD-real} shows the bone fractional images of MMD from PET-enabled DECT using different approaches. Even though there was no available X-ray DECT imaging as the reference, observations were similar to the physical phantom results. The bone fractional image derived from MLAA demonstrated heavy noise. Both KAA and \txtr{CDIP} results showed noticeable noise or artifacts, as pointed by the arrows. In comparison, the proposed neural KAA achieved a good visual quality with reduced noise and artifacts.}

\section{Discussion}
\label{sec8}
This work has developed a single-subject deep-learning framework to improve gCT reconstruction in PET-enabled DECT imaging. The proposed neural KAA approach estimates neural network parameters from PET projection data for gCT image reconstruction and results in a challenging optimization problem. The often used gradient descent algorithm for DIP reconstruction was slow and not able to provide stable results (Fig. \ref{Neural OT}). While ADMM is another possible option, our comparisons (results not shown) have indicated ADMM was also unstable and it was difficult to tune the hyper-parameters. We have solved the optimization problem by developing a neural optimization transfer algorithm, which decouples the optimization problem into modular steps that can be easily implemented using existing libraries for image-based neural network learning and projection-based tomographic reconstruction, respectively. Particularly, the network learning step follows a least-square form (Eq. (\ref{wMSE})) in the image domain that is widely used in deep learning, which is not empirically assigned in this work but theoretically derived from the theory of optimization transfer with quadratic surrogates. This least-square type of optimization transfer can be also applied to other imaging modalities that employ the least-square reconstruction with neural networks (e.g., X-ray CT \cite{Baguer2020, Barutcu2021Limited}, MRI \cite{Yoo2021, Zhao2024J}, and optical tomography \cite{Zhou2020Diffraction, Vu2021Deep}). 

\txtr{Both the standard KAA \cite{Wang2020} and CDIP demonstrated an improvement as compared to MLAA for gCT image reconstruction. However, the former still suffered from noise and the latter resulted in over-smoothing which in turn impacted materiel decomposition (Fig. \ref{MMD}). By combining them together, the proposed neural KAA achieved a much better performance by inheriting the advantages of each method to balance noise suppression and oversmoothing.}

There are other applications of deep neural networks for 511 keV attenuation image enhancement \cite{Hwang2018, Hwang2019, Toyonaga2022, Shi2023}, which typically target post-reconstruction image processing for a attenuation correction purpose. \txtr{Another promising direction is the unrolled model-based deep-learning reconstruction that considers PET physical process \cite{Corda-D'Incan2022,Lim2020,Mehranian2020}. For example, it is possible to extend the idea of \cite{Mehranian2020}, which is designed for PET activity image reconstruction, to develop an unrolled forward-backward splitting transmission-reconstruction (FBSTR-Net) algorithm for gCT imaging, as shown in the Supplemental Material. However, all these methods require a population-based training database of patient scans, which has not been available for the development of the novel PET-enabled DECT imaging method. Here as an alternative, we used simulated data to compare our proposed single-subject learning method with the FBSTR-Net. The results are provided in the supplemental Fig. 24 and Fig. 25. The FBSTR-Net was better than CDIP, comparable to KAA, but not as good as the neural KAA, indicating the benefit of the deep coefficient prior that is more adaptive to individual patients. Furthermore, the proposed neural KAA does not require pre-training and is directly applicable to single subjects, which is critical for PET-enabled DECT as a new imaging method for which no prior database is available, even though online learning is needed with an increased computational cost.} The use of deep neural networks in single-subject deep learning may be also achieved by another way that uses convolutional neural network to improve the kernel construction \cite{Li2021}. This direction is complementary to, not competing with, the neural KAA in this paper because the former improves $\K$ while the latter improves $\alp$ in the kernel model $\muv=\K\alp$. Our future work will combine them together.

Similar to many other deep learning approaches, the neural network learning module of the proposed algorithm involves parameter tuning, such as the sub-iteration number and learning rate. However, these parameters were all set to be the same as we used in our other works \cite{Li2023}. The stable performance that we have observed indicates the robustness of the neural optimization transfer algorithm, despite the different tomographic reconstruction tasks. Future studies may continue the evaluation of the stability using more datasets.

\txtr{The current implementation of the algorithm may need hundreds of iterations (e.g., 400) for gCT image reconstruction. However, ordered subsets can be used to accelerate the algorithm \cite{Presotto2015}. If 20 subsets are used, the needed number of iterations would be approximately 20, which is computationally feasible for clinical practice. While the neural KAA requires a neural network training in each reconstruction iteration, each training only took 10 seconds for the 3D real patient data without any code optimization. To further accelerate the online training, it may be possible to apply transfer learning by fixing the first few layers (for common feature extraction) but fine-tuning the later layers for adaption \cite{Gong2019PET}. In this way, the benefits of population-based deep learning and single-subject learning may be also jointly explored to improve the performance of deep reconstruction.}

There are limitations in this work. We observed a marginally increased bias in bone region quantification compared to the standard MLAA method, \txtr{meanwhile the intensity inside the spine was overestimated as compared to the ground truth in the simulation study. This might be due to suboptimal spatial correlations embedded in the kernel matrix, causing oversmoothing.} Further improvement of the kernel matrix $\K$ could be achieved by using trained kernels following the deep kernel concept \cite{Li2022}, which will be investigated in future work. \txtr{In addition, patient movement and physiological motion between a PET scan and an X-ray CT scan would affect the kernel construction and gCT image reconstruction. This effect may be mitigated by registering the X-ray CT image with the non-attenuation-corrected PET image and will be explored in our future work. While the current PET-enabled DECT imaging is implemented with the PET resolution, we are also developing a super-resolution reconstruction approach to improve the gCT resolution to the X-ray CT resolution \cite{Zhu2024}.}

\section{Conclusion}
\label{sec9}
In this paper, we have developed a neural KAA approach that combines the existing KAA with a neural network-based deep coefficient prior to improve gCT image reconstruction for PET-enabled DECT imaging. A neural optimization transfer algorithm has been further developed to address the optimization challenge for the tomographic estimation of neural network parameters. This leads to an efficient modular implementation that decouples the tomographic reconstruction steps from the neural network learning step with the latter following a unique least squares form in Eq. (\ref{wMSE}). Computer simulation, real phantom and patient results for gCT image reconstruction and multi-material decomposition have demonstrated the feasibility of the algorithm and shown noticeable improvements over the existing methods for PET-enabled DECT imaging.

\newpage

\section*{Supplemental Material: \\ Comparison With Model-Based Deep Reconstruction Method}
	Model-based deep-learning methods (e.g., \cite{Corda-D'Incan2022,Lim2020,Mehranian2020}) have been developed for PET but are mainly used for reconstruction of activity images. Here, we applied the Forward-Backward Splitting (FBS) idea of \cite{Mehranian2020} to develop an unrolled model for gCT image reconstruction and compare it to our proposed single-subject deep-learning reconstruction method. As for a new imaging modality for which no prior training database of patient scans is available, this comparison was performed on 2D simulation data in this preliminary study.

	\subsection{Unrolled FBS Model for gCT Reconstruction}
	Based on the maximum-a-posteriori (MAP) reconstruction of PET emission data, we can update the attenuation image $\muv$ following
	
	\begin{equation}
	\muv^{n+1} = \arg\max_{\muv \geq 0} L(\y|\lam^{n+1},\muv)\nonumber - \beta\R(\muv), \tag{47}
	\label{MAP-MLTR}
	\end{equation}
	where $\R(\muv)$ is a regularization term that imposes prior information on $\muv$, controlled by the regularization parameter $\beta$. By using the FBS algorithm as the same as used in \cite{Mehranian2020}, the optimization of Eq. (\ref{MAP-MLTR}) is performed in the following steps
	\begin{equation}
	\muv_{\rm{Reg}}^{n+1} = \muv^{n} - \gamma \beta \nabla \R(\muv^{n}), \tag{48}
	\label{Reg}
	\end{equation}
	\begin{equation}
	\muv^{n+1} =  \arg\max_{\muv \geq 0} L(\y|\lam^{n+1},\muv)\nonumber - \frac{1}{2\gamma}||\muv - \muv_{\rm{Reg}}^{n+1}||^2, \tag{49}
	\label{OT-Reg}
	\end{equation}
	where Eq. (\ref{Reg}) is a gradient descent step of $\R$ with the step size of $\gamma$, which can be solved by a deep-learning model, e.g., U-Net. Eq. (\ref{OT-Reg}) is an optimization problem regarding the log-likelihood function with  $1/\gamma$ as the regularization hyperparameter. Different from \cite{Mehranian2020} for PET activity image reconstruction, here our focus is for gCT image reconstruction and the separable paraboloidal surrogate (SPS) algorithm is thus used
	\begin{equation}
	\mu^{n+1}_{j} =  \arg\max_{\mu_j \geq 0} -\frac{1}{2} \omega^{n}_{j}(\mu_{j} - \mu^{n+1}_{j,\rm{SPS}})^2 -\frac{1}{2\gamma}(\mu_j - \mu_{j,\rm{Reg}}^{n+1})^2, \tag{50}
	\label{SPS}
	\end{equation}
	where $\muv_{\rm{SPS}}^{n+1}$ is updated by
	\begin{equation}
	\muv_{\rm{SPS}}^{n+1} = \muv^{n} + \frac{\g^n}{\ome^n},\tag{51}
	\label{SPS-update}
	\end{equation}
	where $\g^n$ and $\ome^n$ are the gradient image and intermediate weight image respectively as given by same formulas to Eq. (38) and Eq. (39) with an identity matrix $\K$ and an identity mapping $\psiv$.
	
	Finally, by setting the derivative of Eq. (\ref{SPS}) to zero, we get a closed-form solution
	\begin{equation}
	\mu^{n+1}_{j} =  \frac{\omega^{n}_{j}\mu^{n+1}_{j,\rm{SPS}} + \gamma \mu_{j,\rm{Reg}}^{n+1} }{\omega^{n}_{j} + \gamma}. \tag{52}
	\label{Fusion}
	\end{equation}
	We call this new algorithm a transmission-reconstruction or FBSTR to emphasize its nature for gCT attenuation image reconstruction.
	\begin{figure}[t]
		\centering
		\renewcommand{\thefigure}{21}
		\includegraphics[trim=0cm 0cm 0cm 0cm, clip,width=3.5in]{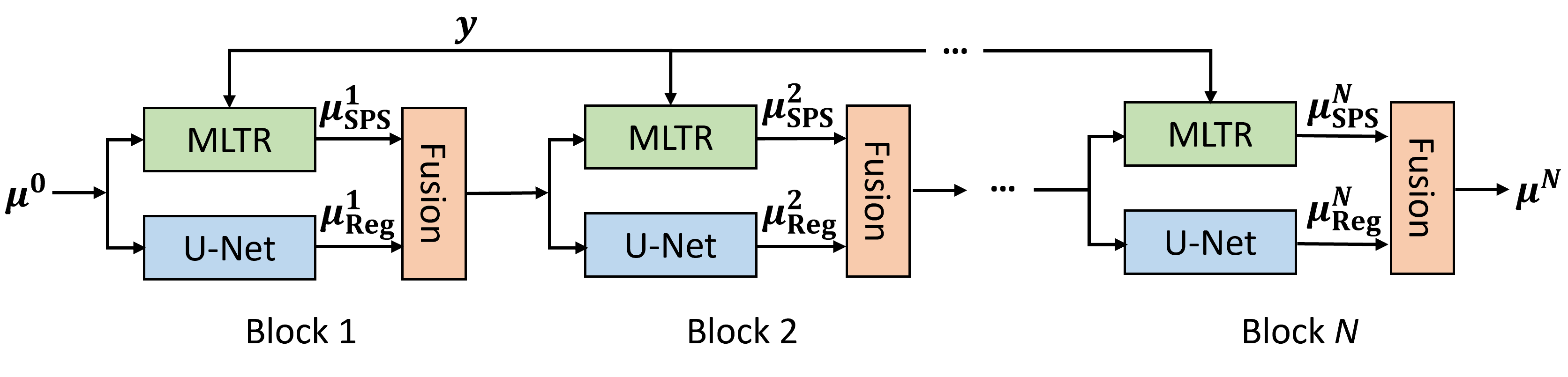}
		\caption{Graphical illustration of the FBSTR-Net with $N$ blocks.}
		\label{Fig 21}
	\end{figure}
	\begin{figure}[h]
		\centering
		\renewcommand{\thefigure}{22}
		\includegraphics[trim=1cm 0cm 1cm 0cm, clip,width=3.5in]{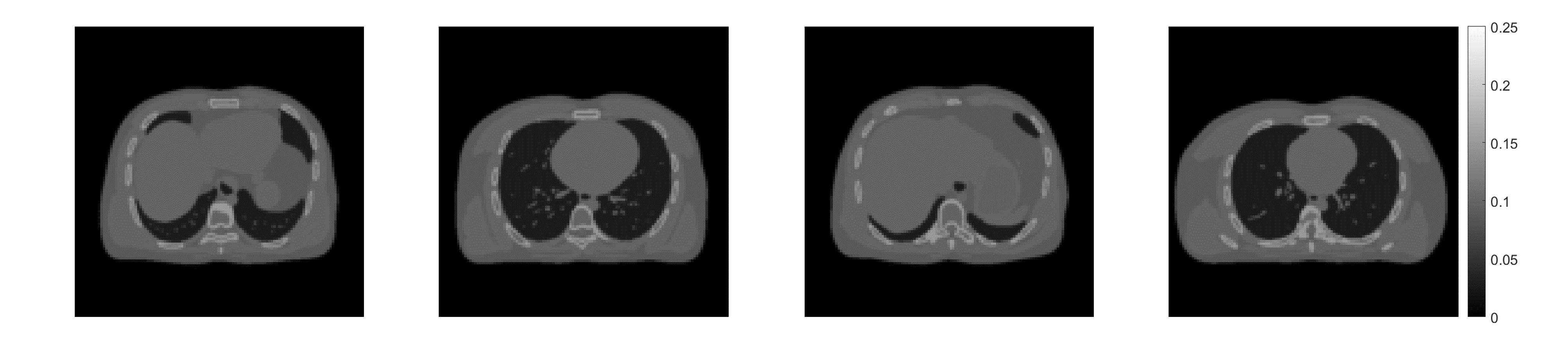}
		\caption{Examples of simulated gCT images for training.}
		\label{gCT}
		\vspace{-10pt}
	\end{figure}
	\begin{figure}[h]
		\centering
		\renewcommand{\thefigure}{23}
		\includegraphics[trim=0cm 0cm 1cm 0cm, clip,width=2in]{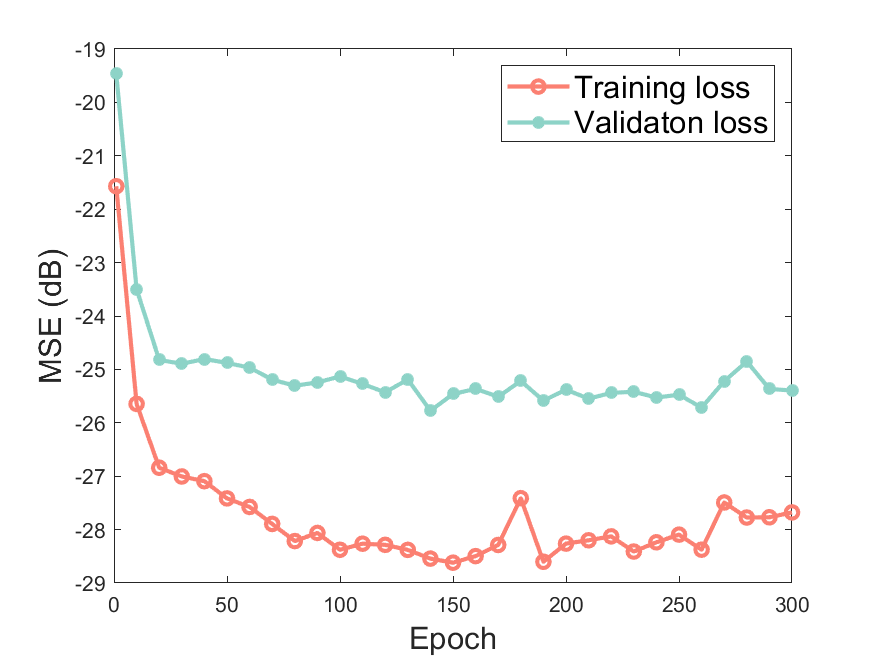}
		\caption{Plots of training and validation MSE losses.}
		\label{loss fig}
		\vspace{-10pt}
	\end{figure}
	\begin{figure*}[h]
		\centering
		\renewcommand{\thefigure}{24}
		\subfloat[]{
			\begin{minipage}[t]{0.18\linewidth}
				\centering
				\includegraphics[trim=2.5cm 1cm 1cm 0.0cm, clip, height=2.7cm]{True.png}\\
				\includegraphics[trim=2.5cm 1cm 1cm 0.0cm, clip, height=2.7cm]{Diff_Truth.png}\\
			\end{minipage}%
		}%
		\subfloat[]{
			\begin{minipage}[t]{0.15\linewidth}
				\centering
				\includegraphics[trim=2.5cm 1cm 3.2cm 0.0cm, clip, height=2.7cm]{MLAA_5.png}\\
				\includegraphics[trim=2.5cm 1cm 3.2cm 0.0cm, clip, height=2.7cm]{diff_MLAA.png}\\
			\end{minipage}%
		}%
		\subfloat[]{
			\begin{minipage}[t]{0.15\linewidth}
				\centering
				\includegraphics[trim=2.5cm 1cm 3.2cm 0.0cm, clip, height=2.7cm]{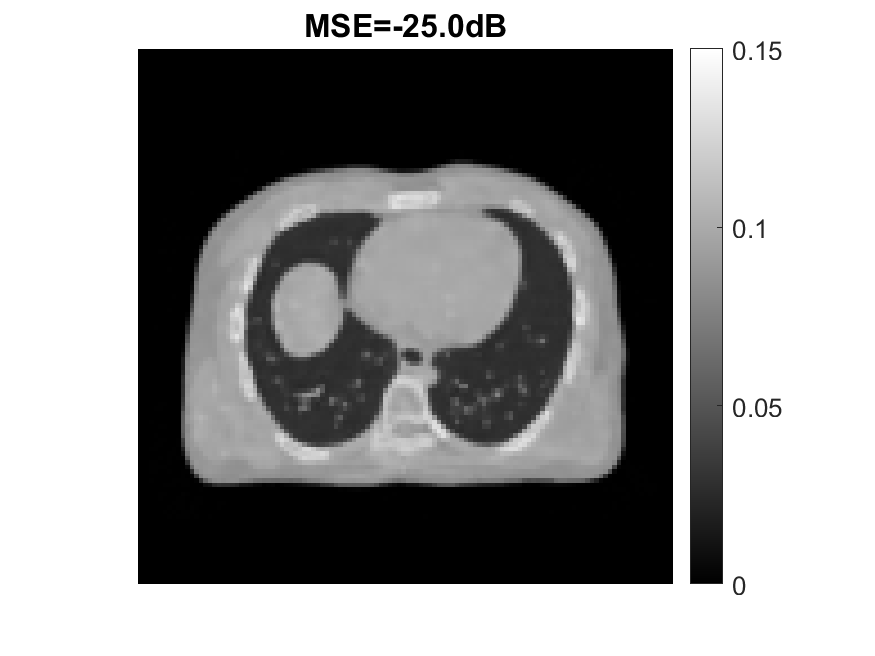}\\
				\includegraphics[trim=2.5cm 1cm 3.2cm 0.0cm, clip, height=2.7cm]{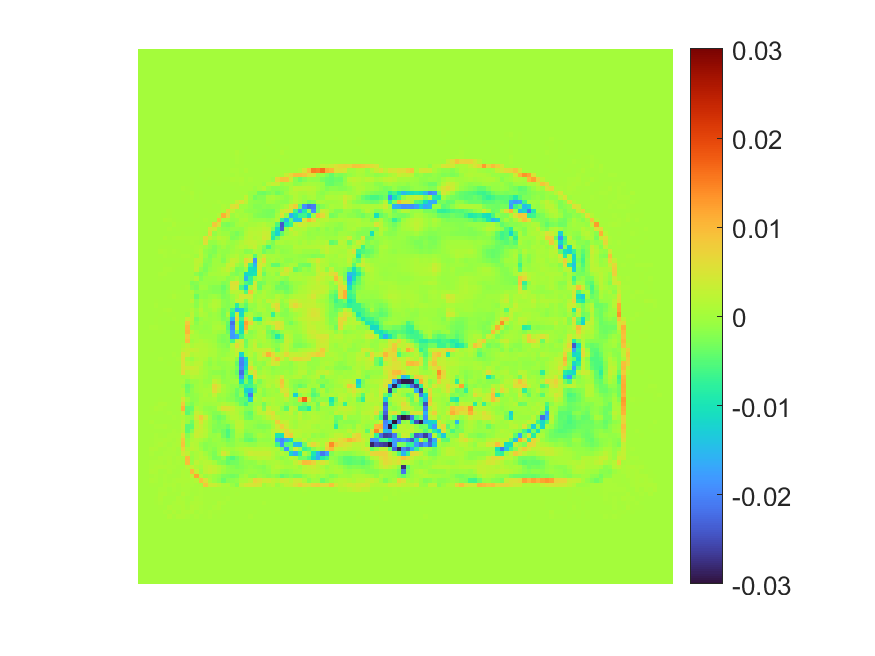}\\
			\end{minipage}%
		}%
		\subfloat[]{
			\begin{minipage}[t]{0.15\linewidth}
				\centering
				\includegraphics[trim=2.5cm 1cm 3.2cm 0.0cm, clip, height=2.7cm]{KMLAA_5.png}\\
				\includegraphics[trim=2.5cm 1cm 3.2cm 0.0cm, clip, height=2.7cm]{diff_KAA.png}\\
			\end{minipage}%
		}%
		\subfloat[]{
			\begin{minipage}[t]{0.15\linewidth}
				\centering
				\includegraphics[trim=2.5cm 1cm 3.2cm 0.0cm, clip, height=2.7cm]{NMLAA_5.png}\\
				\includegraphics[trim=2.5cm 1cm 3.2cm 0.0cm, clip, height=2.7cm]{diff_DIP.png}\\
			\end{minipage}%
		}%
		\subfloat[]{
			\begin{minipage}[t]{0.15\linewidth}
				\centering
				\includegraphics[trim=2.5cm 1cm 3.2cm 0.0cm, clip, height=2.7cm]{NKMLAA_5.png}\\
				\includegraphics[trim=2.5cm 1cm 3.2cm 0.0cm, clip, height=2.7cm]{diff_NKAA.png}\\
			\end{minipage}%
		}%
		\caption{gCT images (top) by different reconstruction algorithms and their corresponding error images (bottom). (a) Ground truth, (b) MLAA, (c) FBSTR-Net, (d) KAA, (e) \txtr{CDIP}, and (f) proposed neural KAA.}
		\label{gCT image}
	\end{figure*}
	\begin{figure*}[h]
		\centering
		\renewcommand{\thefigure}{25}
		\subfloat[]{
			\begin{minipage}[t]{0.18\linewidth}
				\centering
				\includegraphics[trim=2.5cm 0.5cm 1.5cm 0cm, clip, height=2.7cm]{xcat_proj5m_bone_True.png}\\
				\includegraphics[trim=2.5cm 0.5cm 1.5cm 0cm, clip, height=2.7cm]{diff_bone_Standard_MLAA.png}\\
				\includegraphics[trim=2.5cm 0.5cm 1.5cm 0cm, clip, height=2.7cm]{xcat_proj5m_soft_True.png}\\
				\includegraphics[trim=2.5cm 0.5cm 1.5cm 0cm, clip, height=2.7cm]{diff_water_Standard_MLAA.png}\\
			\end{minipage}%
		}%
		\subfloat[]{
			\begin{minipage}[t]{0.15\linewidth}
				\centering
				\includegraphics[trim=2.5cm 0.5cm 3.2cm 0cm, clip, height=2.7cm]{xcat_proj5m_bone_MLAA.png}\\
				\includegraphics[trim=2.5cm 0.5cm 3.2cm 0cm, clip, height=2.7cm]{diff_bone_Standard_kernel_MLAA.png}\\
				\includegraphics[trim=2.5cm 0.5cm 3.2cm 0cm, clip, height=2.7cm]{xcat_proj5m_soft_MLAA.png}\\
				\includegraphics[trim=2.5cm 0.5cm 3.2cm 0cm, clip, height=2.7cm]{diff_water_Standard_kernel_MLAA.png}\\
			\end{minipage}%
		}%
		\subfloat[]{
			\begin{minipage}[t]{0.15\linewidth}
				\centering
				\includegraphics[trim=2.5cm 0.5cm 3.2cm 0cm, clip, height=2.7cm]{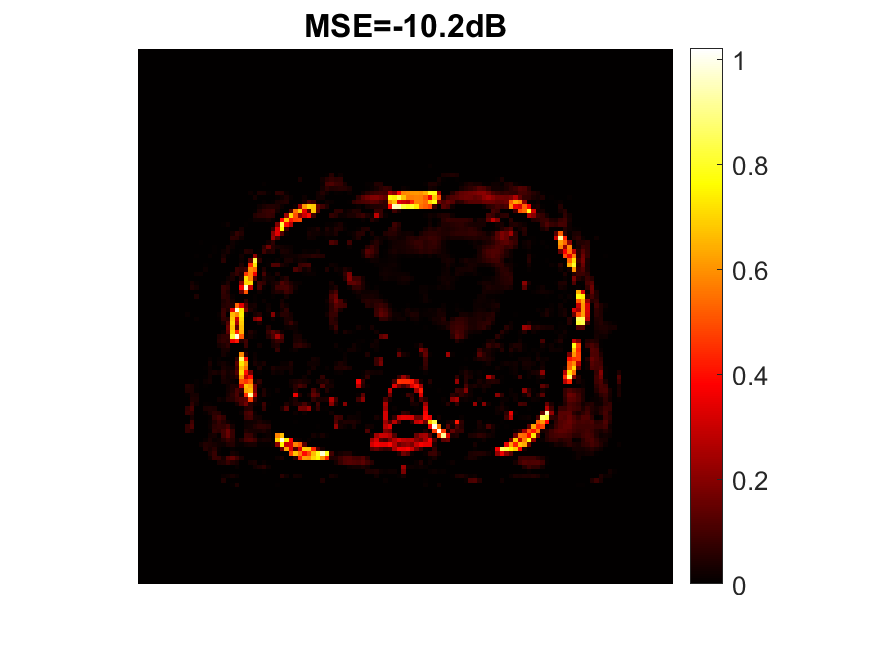}\\
				\includegraphics[trim=2.5cm 0.5cm 3.2cm 0cm, clip, height=2.7cm]{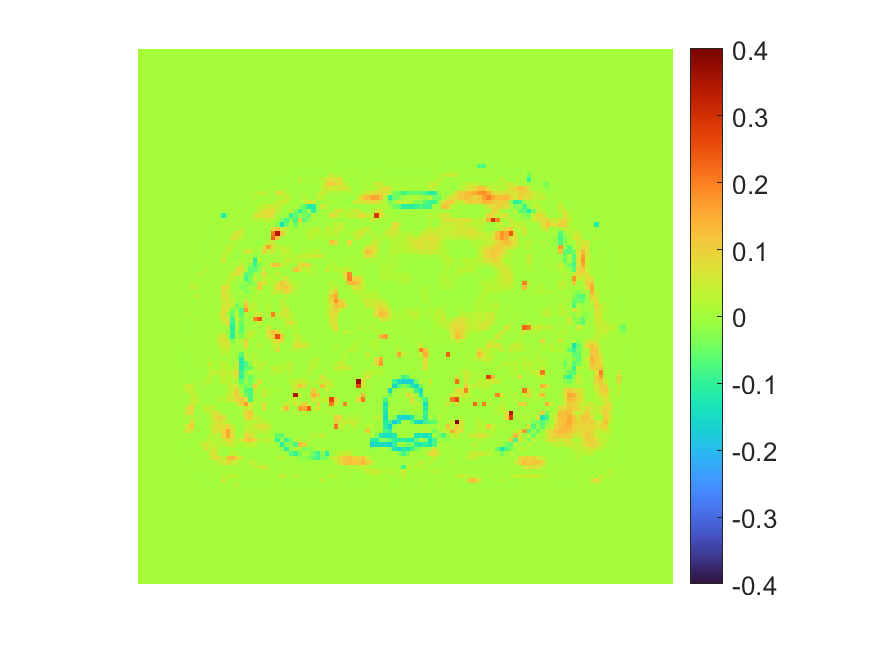}\\
				\includegraphics[trim=2.5cm 0.5cm 3.2cm 0cm, clip, height=2.7cm]{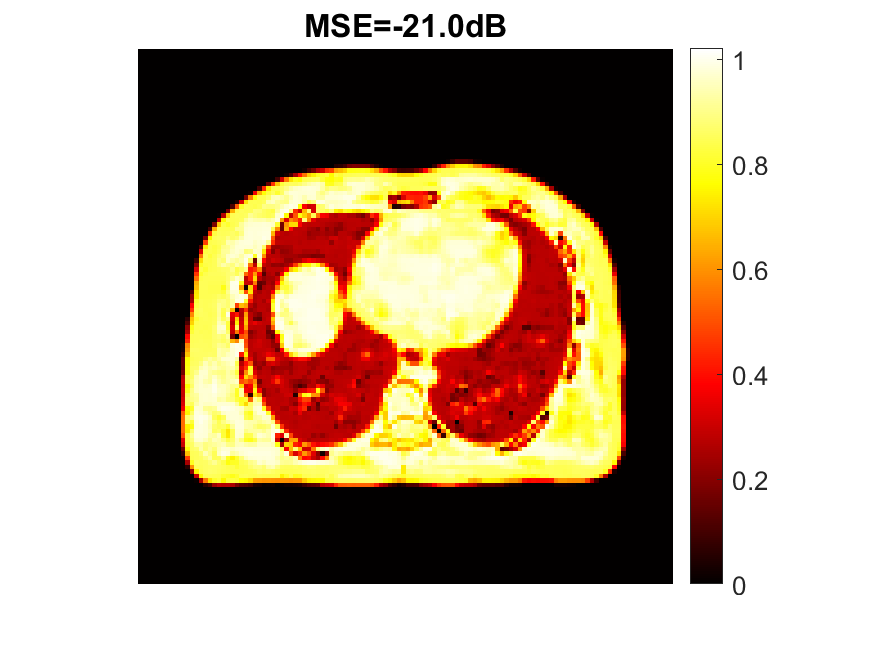}\\
				\includegraphics[trim=2.5cm 0.5cm 3.2cm 0cm, clip, height=2.7cm]{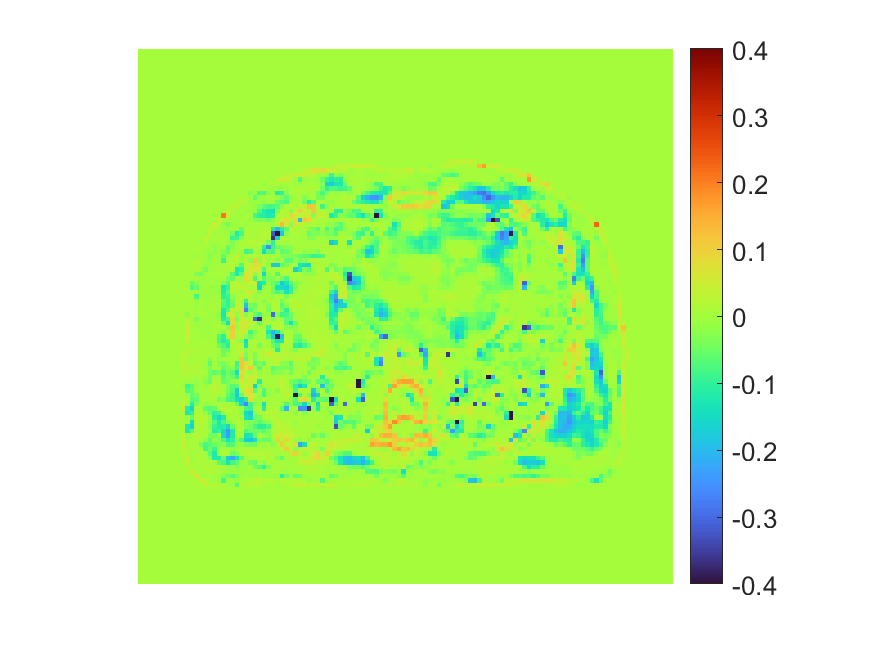}\\
			\end{minipage}%
		}%
		\subfloat[]{
			\begin{minipage}[t]{0.15\linewidth}
				\centering
				\includegraphics[trim=2.5cm 0.5cm 3.2cm 0cm, clip, height=2.7cm]{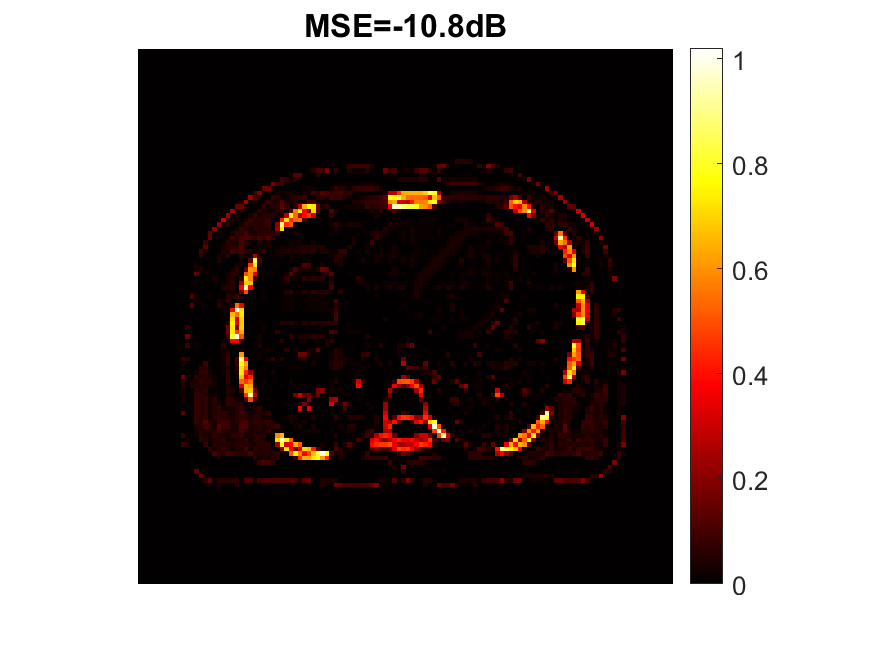}\\
				\includegraphics[trim=2.5cm 0.5cm 3.2cm 0cm, clip, height=2.7cm]{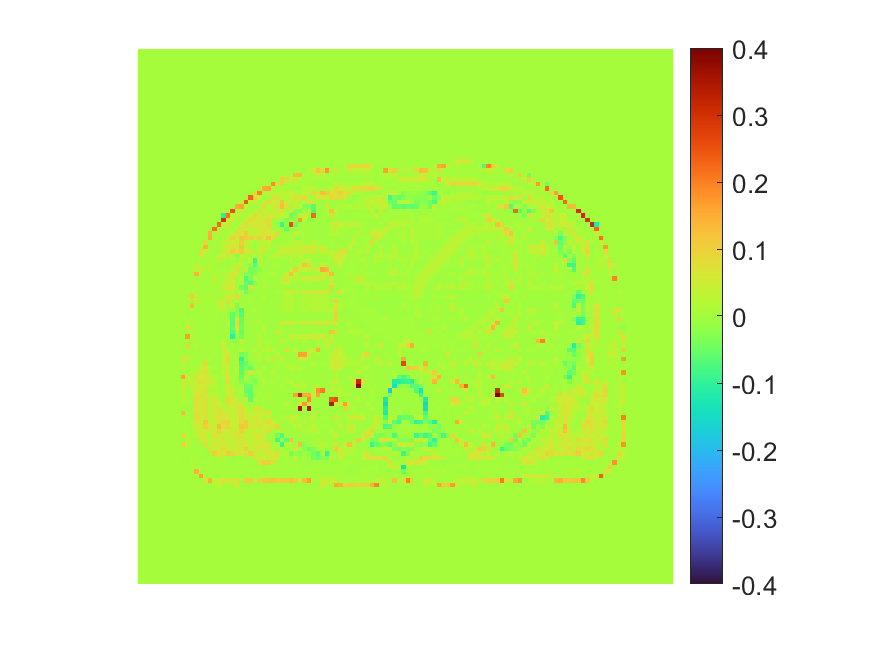}\\
				\includegraphics[trim=2.5cm 0.5cm 3.2cm 0cm, clip, height=2.7cm]{xcat_proj5m_soft_KMLAA.png}\\
				\includegraphics[trim=2.5cm 0.5cm 3.2cm 0cm, clip, height=2.7cm]{diff_water_Unet_kernel_MLAA.png}\\
			\end{minipage}%
		}%
		\subfloat[]{
			\begin{minipage}[t]{0.15\linewidth}
				\centering
				\includegraphics[trim=2.5cm 0.5cm 3.2cm 0cm, clip, height=2.7cm]{bone_Neural_MLAA_without_K.png}\\
				\includegraphics[trim=2.5cm 0.5cm 3.2cm 0cm, clip, height=2.7cm]{diff_bone_Neural_MLAA_without_kernel.png}\\
				\includegraphics[trim=2.5cm 0.5cm 3.2cm 0cm, clip, height=2.7cm]{soft_Neural_MLAA_without_K.png}\\
				\includegraphics[trim=2.5cm 0.5cm 3.2cm 0cm, clip, height=2.7cm]{diff_water_Neural_MLAA_without_kernel.png}\\
			\end{minipage}%
		}%
		\subfloat[]{
			\begin{minipage}[t]{0.15\linewidth}
				\centering
				\includegraphics[trim=2.5cm 0.5cm 3.2cm 0cm, clip, height=2.7cm]{xcat_proj5m_bone_Neural_MLAA_with_K.png}\\
				\includegraphics[trim=2.5cm 0.5cm 3.2cm 0cm, clip, height=2.7cm]{diff_bone_Neural_MLAA_with_kernel.png}\\
				\includegraphics[trim=2.5cm 0.5cm 3.2cm 0cm, clip, height=2.7cm]{xcat_proj5m_soft_Neural_MLAA_with_K.png}\\
				\includegraphics[trim=2.5cm 0.5cm 3.2cm 0cm, clip, height=2.7cm]{diff_water_Neural_MLAA_with_kernel.png}\\
			\end{minipage}%
		}%
		\caption{True and estimated fractional images of two basis materials using different reconstruction algorithms, as well as their corresponding error images: bone (top two rows) and soft tissue (bottom two rows). (a) Ground truth, (b) MLAA, (c) FBSTR-Net, (d) KAA, (e) \txtr{CDIP}, and (f) proposed neural KAA.}
		\label{MMD}
	\end{figure*}
	\subsection{FBSTR-Net Training}
	The unrolled FBSTR algorithm can be represented as a deep iterative neural network, FBSTR-Net $\phiv_{FBSTR}$, with $N$ blocks, as illustrated in Fig. \ref{Fig 21}. For each block, we have three steps: (1) the regularization update based on the previous gCT image estimate using a U-Net model; (2) MLTR from projection domain using the SPS (Eq. (\ref{SPS-update})); (3) pixel-by-pixel image fusion by Eq. (\ref{Fusion}). The trainable parameters of U-Net model are shared across all blocks. We formulate the training process between the output of FBSTR-Net ($\muv_s^{N}$) and the reference image ($\muv_s^{\rm{Ref}}$) using a mean-squared-error (MSE) loss function 
	\begin{equation}
	\hat{\tht} = \arg\min_{\tht} \frac{1}{N_s}\sum_{s=1}^{N_s}||\muv_s^{N} - \muv_s^{\rm{Ref}}||^2, \tag{53}
	\label{loss}
	\end{equation}
	with $\muv_s^{N} = \phiv_{FBSTR} (\tht|\y_s, \muv^0_s)$. $N_s$ is the number of training data. $\tht$ includes the trainable parameters in the U-Net model and the hyperparameter $\gamma$. $\y_s$ is the projection data and $\muv^0_s$ is the initial estimate.
	
	\subsection{Experiment Design and Implementation Details}
	Following the same simulation setting as described in Section V.A, we established a training dataset using ten other chest slices of the XCAT phantom, following a similar strategy as used in \cite{Lim2020}. The examples of gCT images for training are shown in Fig. \ref{gCT}.  For each case, we simulated ten noisy realizations so that total 100 training samples were included in our experiment. The ratio of the training set to the validation set is 2 to 8 \cite{Mehranian2020}. The simulated data in Section V.A was considered as the testing data for method comparison. The same U-Net structure used for the neural KAA and CDIP was used in the FBSTR-Net instead of the residual learning unit used in \cite{Mehranian2020} because of the U-Net's better performance. The block number $N$ and learning rate was selected as 50 and 0.01 respectively according to the performance of validation set. The initialization, $\muv^0$, was also the X-ray CT image-converted 511 keV attenuation map.
	
	Fig. \ref{loss fig} shows the plots of training loss and validation loss with the change of training epoch number. Here the loss value (Eq. (\ref{loss})) is converted into MSE measurement in dB. We thus chose the trained model at 150 epochs for testing.

	\subsection{Comparison for gCT Image Reconstruction}
	Fig. \ref{gCT image} shows the true and reconstructed 511-keV gCT images using different algorithms as well as the corresponding error images. Here, we mainly focus on the FBSTR-Net results. The image MSE value was also included for quantitative comparison. It can be seen that the gCT image derived from FBSTR-Net had a slightly worse MSE than KAA. Its error image demonstrated relatively larger bias around bone regions. \txty{The ensemble bias and SD for gCT quantification in a liver region and a bone region using different reconstruction methods are presented in Table \ref{tab:1}. In addition to FBSTR-Net, the quantification results of 400 iterations of other methods were also included for comparison. A larger bone bias was observed in the FBSTR-Net result, which can be reflected from the error image in Fig. \ref{gCT image}.}
	\begin{table}[h]
		\begin{threeparttable}
			\caption{Comparison of ROI bias and SD for gCT quantification}
			\label{tab:1}
			\setlength\tabcolsep{0pt} 
			\begin{tabular*}{\columnwidth}{@{\extracolsep{\fill}}cccccc}		
				\toprule
				ROI (\%) & MLAA & FBSTR-Net & KAA  & CDIP & Nerual KAA  \\
				\midrule
				Liver Bias & 1.52  & 1.1  & 1.09  & 1.43 & 0.82 \\
				Liver SD & 0.75 & 0.62 & 0.53 & 0.63 & 0.23 \\
				Bone Bias &9.94& 16.4 & 11.22 & 8.95 & 7.84 \\
				Bone SD & 0.39 & 0.33 & 0.23 & 0.43 & 0.21 \\
				\bottomrule
			\end{tabular*}
		\end{threeparttable}
	\end{table}
	\subsection{Comparison for Material Decomposition}
	Fig. \ref{MMD} shows the fractional basis images of bone and soft tissue obtained from MMD of the PET-enabled DECT images. The corresponding error images are also shown. The FBSTR-Net demonstrated a superior performance than CDIP and was comparable to KAA. \txty{While the CDIP had a better MSE than FBSTR-Net for gCT image comparison in Fig. \ref{gCT image}, in the case of fractional images in Fig. \ref{MMD}, the inferior MSE results of CDIP were attributable to significant artifacts in the uniform regions.} In comparison, the proposed neural KAA achieved the best result with the lowest MSE. Compared to KAA and FBSTR-Net, the improvement is due to the additional use of neural network as the deep coefficient prior ($\alp$). \txty{Similar to the Table \ref{tab:1}, the quantitative comparisons of ensemble bias and SD for ROI quantification on the soft tissue and bone fractional images are presented in Table \ref{tab:2}. While FBSTR-Net exhibited a bias in the soft tissue region comparable to that of KAA, it demonstrated a more pronounced bias in the bone region, which was propagated from the gCT reconstruction.}
	\begin{table}[h]
		\begin{threeparttable}
			\caption{Comparison of ROI quantification on soft tissue and bone fractional images}
			\label{tab:2}
			\setlength\tabcolsep{0pt} 
			\begin{tabular*}{\columnwidth}{@{\extracolsep{\fill}}cccccc}		
				\toprule
				ROI (\%) & MLAA & FBSTR-Net & KAA  & CDIP & Nerual KAA  \\
				\midrule
				Soft-tissue Bias & 14.53  & 6.39  & 6.12  & 7.83 & 3.92 \\
				Soft-tissue SD & 2.09 & 2.41 & 2.26 & 2.46 & 1.13 \\
				Bone Bias &15.53& 19.88 & 17.32 & 14.12 & 12.23 \\
				Bone SD & 0.7 & 0.64 & 0.48 & 0.8 & 0.49 \\
				\bottomrule
			\end{tabular*}
		\end{threeparttable}
	\end{table}
	\subsection{Discussion}
	By and large, the FBSTR-Net, an example of model-based deep reconstruction methods, suggested a promising direction for gCT image reconstruction. Once trained, it can be applied directly to other data. However, it is challenging to apply the method to patient scans directly given no training database has been built. In comparison, the single-subject deep learning reconstruction method is directly applicable to a patient scan. In the future, we will apply the population-based deep learning approach for the real data when a training dataset of PET-enabled DECT imaging becomes available and investigate its generalization performance on unseen PET emission data.
	
	\txty{The architecture of neural networks is crucial for training a model-based deep reconstruction method. In this preliminary study, we only compared the U-Net architecture and a general residual learning unit (no downsampling) \cite{Mehranian2020}. A better performance may be achieved by using a more advanced neural-network architecture. When applying such model-based methods, it will be important to also elaborately investigate the model hyperparameters (e.g., block number, learning rate, optimizer) and training strategies. A comprehensive investigation of these aspects for FBSTR-Net is beyond the scope of this paper but would be worth exploring in the future.}

\end{document}